\documentclass[journal,draftcls,onecolumn,12pt,twoside]{IEEEtran}
\usepackage{color}
\usepackage{mathrsfs}
\usepackage{amsfonts}
\usepackage{amsthm}
\theoremstyle{remark}

\newtheorem{corollary}{Corollary}
\newtheorem{theorem}{Theorem}
\newtheorem{lemma}{Lemma}
\newtheorem{remark}{Remark}
\newtheorem{example}{Example}
\usepackage{algorithm}
\usepackage{cite}
\usepackage{fancyhdr}
\usepackage{bbm}
\usepackage{datetime}

\newcommand{\tabincell}[2]{\begin{tabular}{@{}#1@{}}#2\end{tabular}}
\ifCLASSINFOpdf
   \usepackage[pdftex]{graphicx}
\else
   \usepackage[dvips]{graphicx}
   \graphicspath{{../eps/}}
   \DeclareGraphicsExtensions{.eps}
\fi

\usepackage[cmex10]{amsmath}
\usepackage{algorithmic}
\begin{document}

\title{\vspace{0.4cm}Graph Codes for Distributed Instant Message Collection in an Arbitrary Noisy Broadcast Network}
\author{
\IEEEauthorblockN{Yaoqing Yang, Soummya Kar and Pulkit Grover}
%
\thanks{This work was partially supported by the National Science Foundation under Grant CCF-1513936, NSF ECCS-1343324, NSF CCF-1350314, and by NSF grant ECCS-1306128.

Y. Yang, S. Kar and P. Grover are with the Department of Electrical and Computer Engineering, Carnegie Mellon University, Pittsburgh, PA, 15213, USA. Email: \{yyaoqing,soummyak,pgrover\}@andrew.cmu.edu}
}
\maketitle
\rfoot{}
\renewcommand{\headrulewidth}{0pt}


\vspace{-0.6in}

\begin{abstract}
We consider the problem of minimizing the number of broadcasts for collecting all sensor measurements at a sink node in a noisy broadcast sensor network. Focusing first on arbitrary network topologies, we  provide (i) fundamental limits on the required number of broadcasts of data gathering, and (ii) a general in-network computing strategy to achieve an upper bound within factor $\log N$ of the fundamental limits, where $N$ is the number of agents in the network. Next, focusing on two example networks, namely, \textcolor{black}{arbitrary geometric networks and random Erd\"os-R\'enyi networks}, we provide improved in-network computing schemes that are optimal in that they attain the fundamental limits, i.e., the lower and upper bounds are tight \textcolor{black}{in order sense}. Our main techniques are three distributed encoding techniques, called graph codes, which are designed respectively for the above-mentioned three scenarios. Our work thus extends and unifies previous works such as those of Gallager~\cite{Gal_TIT_88} and Karamchandani~\emph{et. al.}~\cite{Kara_TIT_11} on number of broadcasts for distributed function computation in special network topologies, while bringing in novel techniques, e.g., from error-control coding and noisy circuits, for both upper and lower bounds.
\end{abstract}

\section{Introduction}\label{intro}
\subsection{Motivations and Main Contributions}\label{motiv}
Distributed data collecting in a multi-agent sensor network~\cite{Gal_TIT_88} is crucial in many applications of \textcolor{black}{data processing and network control}. We focus on problems where there is one \emph{sink} node in the network that needs to collect all sensor measurements for further function computation tasks, e.g., SUM, MAX, Majority, Parity, Histogram, etc. Although distributed data processing in sensor networks is often studied from the perspective of distributed in-network function computation~\cite{Gir_CM_06,Dim_PI_10}, our focus here is on the computation of the most communication-intensive function: the identity function (see, e.g.~\cite{Gal_TIT_88,Goy_SJC_08,Kara_TIT_11,Li_TMC_13,Zheng_INFOCOM_07,Mos_IPSN_07,Joo_Allerton_08,Mar_IPSN_03}), where the goal is to collect all the measurements themselves at the sink node\footnote{Any other function computation will need only fewer number of transmissions because if the sink node can reliably compute the identity function, it can also compute any other function reliably.}. This problem is of practical importance: as discussed in \cite{Zheng_INFOCOM_07}, ``data gathering remains the primary service provided by wireless sensor networks''. Moreover, when the specific processing task of sensor measurements cannot be foreseen, collecting all sensor measurements is the safest strategy. Data gathering is also necessary in monitoring each agent in an emergency response system, for instance, the wearable wireless sensors that are connected with device-to-device links provide real-time monitoring signals for smart health care. \textcolor{black}{An interesting application is the optimization of the waste collector truck route based on the load levels of waste containers in a smart city, where binary bits that indicate whether load levels exceed thresholds are reported by a large number of wireless sensors~\cite{Zan_ITJ_14} to a remote data center.}

In the above-mentioned applications, data are often generated in sensors in the form of short and instant messages, and the number of sensors can be quite large. In this circumstance, communication throughput might not be the ultimate goal, since data are instant, instead of generated in streams. Following the seminal work of Gallager \cite{Gal_TIT_88}, we consider communication complexity~\cite{Kush_AC_97}, measured in number of broadcasts in bits, as the optimization goal. We assume, in each time slot, a network agent broadcasts a message bit to its neighborhood, and each other agent in this neighborhood receives an independent noisy copy\footnote{The assumption on noisy networks is suitable to model wireless sensor networks with limited transmission power and decoding capabilities.} of the broadcast message. Without loss of generality, we assume that each network agent has only one bit of information and the sink node needs to collect all these bits with some required accuracy and minimum number of broadcasts. The network consists of $(N+1)$ agents (also referred to as nodes in the sequel), among which one agent is assigned as the sink (arbitrarily but decided apriori). These $N+1$ agents may directly communicate with subsets of other agents through unidirectional or bidirectional noisy links as determined by a preassigned (but arbitrary, possibly sparse) inter-agent communication network. We model noisy links as binary symmetric channels (BSC) or binary erasure channels (BEC). Note that the results on communication complexity, which is the focus of this paper, are often obtained under specific assumptions on the network structure, including complete networks~\cite{Gal_TIT_88,Goy_SJC_08,Newman_CC_04,Kush_ACM_98,
Wang_ISIT_13}, grid networks \cite{Kara_TIT_11} and random geometric networks~\cite{Li_TMC_13,Zheng_INFOCOM_07,Ying_TIT_07,Dutta_ACM_08,Kano_ISIT_07,Kamath_TIT_14}. However, we seek to obtain results that are independent of the network topology. In other words, our goal is to characterize the communication complexity scaling in networks with arbitrary topologies. A similar problem is also considered in~\cite{Mos_IPSN_07,App_TIT_14,Kow_TIT_12,Jeon_ISIT_13}, but the problem of data gathering in a noisy network is not considered. Therefore, we believe that this work is the first to consider the minimum broadcasting complexity problem for data collecting in a distributed network with noisy links and arbitrary topologies. Interestingly, the communication complexity results in this paper coincide with many existing results obtained under specific graph topology. The comparison between our work and related works is discussed in detail in Section~\ref{related works}.

There are three major computation models in the field of in-network computing: one-shot computation \cite{Gal_TIT_88,Goy_SJC_08,Kara_TIT_11,Li_TMC_13,Newman_CC_04,Kush_ACM_98,Kano_ISIT_07,Kamath_TIT_14}, block computation \cite{Kow_TIT_12,Gir_CM_06} and pipelined computation \cite{Joo_Allerton_08,Mar_IPSN_03,Khu_TMC_08,App_TIT_14,Banerjee_ISIT_11}. We consider the one-time computation model, which means a one-time gathering of all the data, because each node only has a short message, e.g., one bit of information, to be sent as a separate data packet. This kind of communication problems with limited data is frequent~\cite{Gal_TIT_88} in distributed control of networks or a distributed monitoring system, where each sensor is required to report just a few bits to describe the state of the corresponding subsystem in a timely manner. Under the assumption of instant message collecting, applying classic error control coding to cope with noisy links is highly non-trivial, since it is impossible for each node to gather enough data to be encoded into blocks before being transmitted and distributed encoding is necessary\footnote{In fact, we use linear block codes with distributed encoding techniques in the paper. However, the encoding is instant in contrast to classical coding theoretic frameworks which operate on (large) blocks of data.}. This is also one of the main reasons why we explicitly consider noisy channels, rather than considering noiseless or effectively noiseless channels (on which noise-free communications can be achieved as long as the communication rate is below the channel capacity), as the classical notion of channel capacity is not generally applicable in scenarios involving instantaneous and distributed encoding. Rather, an effective computation (encoding) scheme in our context involves carefully designed in-network computations and inter-agent message exchanges (through neighborhood broadcasts).

\textcolor{black}{In~\cite{Gal_TIT_88}, Gallager considers the data gathering problem in a complete graph and obtains an upper bound $\mathcal{O}(N\log\log N)$ on the communication complexity. Here, we address the same problem in general graphs (possibly very sparse) and obtain a general upper bound. Specifically, we show that this upper bound reduces to $\mathcal{O}(N\log\log N)$ as long as the network diameter stays bounded as $N\rightarrow\infty$. The main technique that leads to the generalization of Gallager's result to arbitrary graph topologies is a distributed encoding scheme, called graph code, that extends error control coding to distributed in-network computations.} The graph codes constructed in this paper are conceptually different from the encoding scheme developed by Gallager for complete networks. We first consider a general network and design a general graph code for it. Then, we modify this code to improve its performance in more specific graph topologies. The formal definitions of graph codes will be given in Section~\ref{main_technique}. In the following, we briefly discuss the three graph codes that are used in this paper.

\subsubsection{$\mathcal{GC}$-1 Graph Code in General Graphs}
\textcolor{black}{In Section~\ref{Large_Diameter}, general graph topologies are considered and the $\mathcal{GC}$-1 graph code is provided. It is shown that in both BSC and BEC networks, the number of broadcasts required by the $\mathcal{GC}$-1 graph code is $\max\{\Theta(\bar{d}_\mathcal{G}N),\Theta(N\log N)\}$, where $\bar{d}_\mathcal{G}$ denotes the average distance from all agents to the sink. We also obtain a $\max\left\{\Theta(\bar{d}_\mathcal{G}N),\right.$ $\left.\Theta(N\log\log N)\right\}$ lower bound on the communication complexity through cut-set techniques in BSC networks, and a \textcolor{black}{$\Theta(\bar{d}_\mathcal{G}N)$ lower bound in BEC networks using the same techniques}. Note that there is a non-negligible gap between the above mentioned upper bound and the lower bound. When $\bar{d}_\mathcal{G}>\Theta(\log N)$, the upper bound coincides with the lower bound. When $\bar{d}_\mathcal{G}$ is small, there is at most a $\log N$ multiple between the two bounds. We also show a $\max\{\Theta(\bar{d}_\mathcal{G}N),\Theta(N\log N)\}$ lower bound in constant-degree networks with BEC channels, which implies that the $\mathcal{GC}$-1 graph code also achieves optimality in this scenario. To provide better intuition, we explain through examples how this gap gets introduced.}

\textcolor{black}{Compared with Gallager's result~\cite{Gal_TIT_88}, $\Theta(N\log\log N)$ complexity in a complete graph, $\bar{d}_\mathcal{G}N$ characterizes the cost due to possibly large graph diameter. Therefore, in general networks, we may need strictly more communications than the complete graph, so the gap between Gallager's and ours is not because our scheme is suboptimal. In fact, in Gallager's setup, $\bar{d}_\mathcal{G}=1$.}

\subsubsection{$\mathcal{GC}$-2 Graph Code in Geometric Graphs}
\textcolor{black}{Motivated by the above mismatch between the achievable result and the converse result, we reconsider the data gathering problem in general geometric graphs with BSC or BEC links (in contrast to random geometric graphs that have random node placement) in Section~\ref{geometric_graph}. For these graphs, we design a new in-network distributed encoding scheme, referred to as the $\mathcal{GC}$-2 code. \textcolor{black}{In a geometric graph, all nodes are placed within a 1-by-1 square, and each node is able to broadcast within a certain distance $r< 1$. We prove that, when $r$ is larger than a threshold with order $\Theta(\sqrt{\frac{\log N}{N}})$, the communication complexity upper bound achieved by the $\mathcal{GC}$-2 scheme is $\max\{\Theta(\bar{d}_\mathcal{G}N),\Theta(N\log\log N)\}$, }which coincides with the general lower bound in Section~\ref{Large_Diameter}.}

\textcolor{black}{The $\mathcal{GC}$-2 code utilizes this fact and each code bit calculated at a node $v$ is the parity of the bits held by a subset (possibly strict) of its neighbors in a local complete graph. Therefore, the $\mathcal{GC}$-2 utilizes local broadcasting among neighboring nodes as a means of reducing the distributed encoding cost. \textcolor{black}{Interestingly, we show that the $\mathcal{GC}$-2 code essentially reduces to the coding scheme in \cite[Section 7]{Goy_SJC_08} on complete graphs (see Remark \ref{remark_RGGtoGallager}). However, in contrast to the coding scheme developed in \cite[Section 7]{Goy_SJC_08} which applies to complete graphs only, the $\mathcal{GC}$-2 code is applicable to a much broader class of graphs (arbitrary connected geometric graphs) and achieves function computation using the same number (in the order sense) of broadcasts.}}

\subsubsection{$\mathcal{GC}$-3 Graph Code in Extended Erd\"os-R\'enyi Graphs}
\textcolor{black}{In Section~\ref{Upper_Bound}, we investigate the same problem in a low-diameter graph, because we can see from the previous discussion that large graph diameters lead to the $\bar{d}_\mathcal{G}N$ gap with respect to the $\Theta(N\log\log N)$ bound in~\cite{Gal_TIT_88}. Our motivation is to determine instances of non-complete graphs where it is possible to achieve the bounds in~\cite{Gal_TIT_88} for complete graphs. We find that an Erd\"os-R\'enyi random graph~\cite{Bol_Spr_98} suffices if two further assumptions are made:
\begin{itemize}
  \item \textcolor{black}{More links are added to the Erd\"os-R\'enyi graph such that} the multi-hop distance from each agent to the sink is bounded (e.g., when the sink is a central node and all other nodes have an extra directed link to it);
  \item \textcolor{black}{The noisy links are BEC} instead of BSC.
\end{itemize}
We call it the extended Erd\"os-R\'enyi graph. The technique utilized in the extended Erd\"os-R\'enyi graph is referred to as the $\mathcal{GC}$-3 code. Using the $\mathcal{GC}$-3 code, we show that the $\Theta(N\log\log N)$ upper bound can be achieved without the complete graph assumption. \textcolor{black}{The applicability of the $\mathcal{GC}$-3 code is however not limited to the extended Erd\"os-R\'enyi graphs. The $\mathcal{GC}$-3 code may outperform other types of graph codes in terms of error decay exponents in certain scenarios beyond the extended Erd\"os-R\'enyi case. For instance, we show that in a complete network with BEC channels (where both $\mathcal{GC}$-2, $\mathcal{GC}$-3 and the scheme in \cite[Section 7]{Goy_SJC_08} are applicable), the error exponent achieved by the $\mathcal{GC}$-3 code is better than that of the $\mathcal{GC}$-2 code (see Remark \ref{remark_GC3better}) which, in turn, is a generalization of the scheme in \cite[Section 7]{Goy_SJC_08} to general geometric graphs. This implies that although the $\mathcal{GC}$-2, $\mathcal{GC}$-3 and the scheme in \cite[Section 7]{Goy_SJC_08} all achieve a $\mathcal{O}(N\log\log N)$ complexity, $\mathcal{GC}$-3 outperforms the others in terms of error probability decay rates.}}

\textcolor{black}{The analysis of the error probability of $\mathcal{GC}$-3 code leads to, as by-products, new fundamental results in the design of erasure codes for point-to-point communications. In particular, we use the analyses for the $\mathcal{GC}$-3 code to show that there exist sparse erasure codes that can achieve diminishing error probability decaying polynomially with the code length.}

\textcolor{black}{We also borrow cut-set techniques of noisy circuits~\cite{Pip_FOC_85,Pip_TIT_91}, to derive a lower bound on the number of edges in the Erd\"os-R\'enyi graph, in order to determine the required sparseness of the underlying graph for reliable data collection. As by products, we show similar analysis techniques can be used to conclude that the number of ones in the generator matrix of an erasure code should be at least $\Omega(N\log N)$ in order to achieve decaying block error probability}. Note that $\Omega(N\log N)$ is in the same scale as LT codes (Luby transform codes)~\cite{Lub_FOCS_02}. In all, the $\mathcal{GC}$-3 code has strong a relevance to erasure codes, and techniques in the in-network computing problem can be applied to the analysis of erasure codes for the classical point-to-point communication setup.

The above mentioned three types of codes have the relationship $\mathcal{GC}$-3$\subset$$\mathcal{GC}$-2$\subset$$\mathcal{GC}$-1 according to the construction of each code bit. However, these codes have the encoding complexity relationship $\mathcal{GC}$-3$<$$\mathcal{GC}$-2$<$$\mathcal{GC}$-1. Therefore, the high-index codes are simple but \textcolor{black}{meant for specific graph scenarios}, while the low-index codes are complicated but suitable for general topologies. By studying different graph codes, our goal is to theoretically understand in-network computing and data aggregation under the assumptions of link noise and distributed data, with the aim of \textcolor{black}{minimizing} the number of communications. \textcolor{black}{Some of the major attributes of the three different types of graph codes are presented below.}

\textcolor{black}{\begin{center}\small
    \begin{tabular}{ | l | l | l | p{5.0cm} |}
    \hline
     & Applicable Networks & Analyzable in & Number of Broadcasts \\ \hline
     $\mathcal{GC}$-1 & Arbitrary connected networks & \tabincell{c}{BSC (Section V)\\ BEC (Section V-D)} & $\max\{\Theta(\bar{d}_\mathcal{G}N),\Theta(N\log N)\}$ \\ \hline
    $\mathcal{GC}$-2 & Arbitrary connected geometric networks & \tabincell{c}{BSC (Section VI)\\ BEC (Section VI-C)} & $\max\{\Theta(\bar{d}_\mathcal{G}N),\Theta(N\log\log N)\}$\\ \hline
    $\mathcal{GC}$-3 & Extended geometric networks & BEC (Section VII) & $\Theta(N\log\log N)$ \\
    \hline
    \end{tabular}
\end{center}}

\subsection{Related Works}\label{related works}
As mentioned earlier, although our problem of minimizing the broadcast complexity in an arbitrary topology is new, the scaling bounds obtained in this paper coincide with many existing ones under the assumptions of specific graph topologies. In what follows, we show how our results are related to existing results in the literature of in-network computing.

This work was initially inspired by the seminal work of Gallager~\cite{Gal_TIT_88}, where the minimum broadcast complexity problem in a noisy complete network is examined. If the naive approach of repetitive coding, which neglects the broadcast nature of the receptions, is used, the number of transmissions scales as $\mathcal{O}(N\log N)$. However, in \cite{Gal_TIT_88} a delicate broadcasting scheme is designed to achieve a complexity of $\Theta(N\log\log N)$ for the parity calculation problem and the identity calculation problem, i.e., data gathering. In~\cite{Goy_SJC_08}, this bound is proved tight for the identity calculation problem. For general graphs, Gallager's scheme is however, no longer applicable as it relies heavily on the complete graph structure. Nevertheless, this $\Theta(N\log\log N)$ bound still meets the upper bound obtained in Section~\ref{geometric_graph} and Section~\ref{Upper_Bound} when $\bar{d}_\mathcal{G}=\mathcal{O}(1)$. The proof technique for the $\mathcal{GC}$-3 code in Section~\ref{Upper_Bound} is based on rank analysis of random matrices, and can only be applied to a BEC, which is different from Gallager's original setting. Nonetheless, even if Gallager's algorithm is applied to a BEC in the complete network setting, the achieved order continues to be $\Theta(N\log\log N)$. \textcolor{black}{Moreover, the $\mathcal{GC}$-3 scheme is applicable for Gallager's complete setting with BEC links, but ends up using much fewer links in general.} Therefore, our result in an extended Erd\"os-R\'enyi random graph can be viewed as a generalization of prior results under weaker topology assumptions.

In~\cite{Kara_TIT_11}, data gathering in a grid network is studied. Theorem IV.1 and Theorem IV.2 in~\cite{Kara_TIT_11} state that, in an ${\sqrt{N}\times\sqrt{N}}$ grid broadcast network with a transmission radius $r$, the communication complexity for identity function computation is $\max\left\{\Theta(N^{3/2}/r),\right.$ $\left.\Theta(N \log\log N)\right\}$, which matches with the lower bounds in Section~\ref{Large_Diameter} and the upper bound in Section~\ref{geometric_graph}. In fact, the diameter for this network is in the order of $\sqrt{N}/r$.

In~\cite{Li_TMC_13}, the same problem in a random geometric graph is examined. The proposition 2 of~\cite{Li_TMC_13} claims that the communication complexity is upper bounded by $\mathcal{O}(N \sqrt{\frac{N}{\log N}})$, under the assumption that the diameter of the network is $\mathcal{O}(\sqrt{\frac{N}{\log N}})$. Therefore, their upper bound also matches our general upper bound for arbitrary geometric graphs. Although the communication protocol in~\cite{Li_TMC_13} also has a sink-tree-based multi-hop relay procedure similar to ours, their protocol requires nodes to be evenly distributed in the graph. In fact, many works on network-computing in random geometric graphs \cite{Khu_TMC_08,Zheng_INFOCOM_07,Kamath_TIT_14,Ying_TIT_07,Dutta_ACM_08,Li_TMC_13,Kano_ISIT_07} rely highly on the result that the graph can be clustered in an even manner into groups with $\Theta(\log N)$ nodes. \textcolor{black}{Then, codes with length $\Theta(\log N)$ are repeatedly used to facilitate hop-by-hop transmissions. This technique can also be viewed as distributedly encoding codes with low-density generator matrices which have a structure as shown in~\cite[Sec.IV]{Maz_JSAC_14}. However, in practical applications of network computation, the claim that nodes are evenly distributed might not hold~\cite{Alf_JSAC_09,Alf_ACM_10}.}

\textcolor{black}{From the perspective of coding theory, the proposed $\mathcal{GC}$-3 code is closely related to erasure codes that have low-density generator matrices (LDGM) \cite{Lub_FOCS_02,Dim_TON_06,Maz_JSAC_14,abadi2012capacity}. In fact, the graph code in this paper is equivalent to an LDGM erasure code with noisy encoding circuitry~\cite{Yang_All_14}, where the encoding noise is introduced by distributed encoding in the noisy inter-agent communication graph. Based on this observation, we show (in Corollary~\ref{coding_upb}) that our result directly leads to a known result in capacity-achieving LDGM codes. Similar results have been reported by \cite{Lub_FOCS_02} and \cite{abadi2012capacity} for communication and by \cite{Dim_TON_06} and \cite{Maz_JSAC_14} for distributed storage, both with noise-free encoding. Due to encoding noise, their analysis tools are not applicable for our problem. Moreover, our graph code achieves polynomially decaying error probability with increasing code length (which is 2$N$, twice the number of agents in the network), using only binary bits, rather than \textcolor{black}{polynomially} decaying error with expanding Galois field dimension. \textcolor{black}{We also obtain a lower bound on the number of ones in an erasure code generator matrix with techniques inspired by the analysis of graph codes. \textcolor{black}{Our work is also deeply related to network error control coding \cite{ahlswede2000network,koetter2008coding,lim2011noisy,jaggi2008resilient,silva2008rank}, but our work emphasizes more on the perspective of distributed encoding in noisy networks.}}}

\section{\textcolor{black}{Notation and Preliminary Results}}
The calligraphic letter $\mathcal{G}=(\mathcal{V},\mathcal{E})$ represents a graph with a node \textcolor{black}{(vertex)} set $\mathcal{V}$ and an edge set $\mathcal{E}$. In this paper, an edge is directed unless \textcolor{black}{otherwise} stated. Each graph with $N$ vertices has an $N$-by-$N$ adjacency matrix $\mathbf{A}=(A_{m,n})=(\mathbf{a}_1,\mathbf{a}_2,...,\mathbf{a}_N)$, which represents the \textcolor{black}{edges or} network connections\textcolor{black}{, i.e.,} $A_{m,n}=1$ if the node $v_m$ has a directed edge to the node $v_n$, or equivalently, $(v_m,v_n)\in \mathcal{E}$. Denote the one-hop out-neighbors of a node $v$ by $\mathcal{N}_v^{+}:=\{w\in\mathcal{V}|(v,w)\in\mathcal{E},w\neq v\}$. \textcolor{black}{Denote the one-hop in-neighbors of a node $v$ by $\mathcal{N}_v^{-}:=\{w\in\mathcal{V}|(w,v)\in\mathcal{E},w\neq v\}$. Note that the node $v_m \in \mathcal{N}^-(v_n)$, if and only if $A_{m,n}=1$.} If $A_{m,n}=A_{n,m}=1$, \textcolor{black}{we say that} $v_m$ and $v_n$ are \textcolor{black}{linked bidirectionally}. In an undirected graph, i.e., in which all edges are bidirectional, $v_m\in \mathcal{N}^{-}(v_n)$ is equivalent to $v_m\in \mathcal{N}^{+}(v_n)$. Thus, when the graph is undirected, we write $\mathcal{N}(v)$ for simplicity.

We will obtain scaling bounds on the communication complexity of discrete-time algorithms. Time is assumed to be discrete or slotted throughout the paper. The symbol $t$ denotes time. The \textcolor{black}{order} notations $f_1(N)=\mathcal{O}(f_2(N))$ and $f_1(N)=\Omega(f_2(N))$ respectively mean that $f_1(N)/f_2(N)\le C_1$ and $f_1(N)/f_2(N)\ge C_2$ for two positive constants $C_1$, $C_2$ and sufficiently large $N$. By $f_1(N)=\Theta(f_2(N))$ we mean that $f_1(N)=\mathcal{O}(f_2(N))$ and $f_1(N)=\Omega(f_2(N))$.

\textcolor{black}{By $\mathbb{F}_2$, we denote the binary field $\{0,1\}$.} We will use basic results \textcolor{black}{from} error control cod\textcolor{black}{ing, in particular, properties of} binary linear block codes. \textcolor{black}{A binary linear block code~\cite{Gal_Wil_68} with code length $N$ and rate $R<1$ is a set of $2^{NR}$ binary vectors (codewords) that form a linear subspace $\mathcal{C}\subset \mathbb{F}_2^N$. We always assume that $NR$ is an integer. Each codeword $c\in \mathcal{C}$ can be written as the product of a binary row vector $\mathbf{m}$ with length $NR$, called the message vector, and an $NR\times N$ binary matrix $\mathbf{G}$, called the generator matrix. If $\mathbf{G}=[\mathbf{I},\mathbf{A}]$, where $\mathbf{I}$ denotes the $NR\times NR$ identity matrix, we say that the code with the generator matrix $\mathbf{G}$ is systematic. }

A binary symmetric channel \textcolor{black}{(BSC) with crossover probability $\epsilon$ is a channel that flips a bit with probability $\epsilon$. A binary erasure channel (BEC) with erasure probability $\epsilon$ is a channel that outputs an erasure value `e' with probability $\epsilon$, no matter what value the input takes.} Finally, we state two useful results from the theory of reliable communication~\cite{Gal_Wil_68}. The first one \textcolor{black}{concerns repetition codes and the second one linear block codes for reliable message transmission over noisy communication channels.}
\begin{lemma}(\cite[Section 5.3]{Gal_Wil_68})\label{BSC_rept}
Suppose we have a BSC with crossover probability $\epsilon$. If one bit $x\in \mathbb{F}_2$ is repeatedly transmitted through the channel for $j$ times and the receiver uses the majority rule to \textcolor{black}{make a decision $\hat{x}$ the value of $x$}, then, the bit error probability is upper bounded by
\begin{equation}\label{BSC_rept_bd}
  P_e^{(bit)}=\Pr(\hat{x}\neq x)<[4\epsilon(1-\epsilon)]^{j/2}.
\end{equation}
\end{lemma}
\begin{remark}\label{Adaptive_Scheme}
Lemma~\ref{BSC_rept_bd} states that $\mathcal{O}(\frac{\log 1/P_e}{\log 1/\epsilon})$ repeated transmissions are sufficient to achieve an \textcolor{black}{error tolerance probability $P_e$ at the destination, when the point-to-point source to destination channel is a BSC.} One might consider using adaptive schemes, such as sequential detection~\cite{Wald_AMS_48}, to reduce the number of repetitions \textcolor{black}{to achieve the same level of $P_{e}$}. However, this does not change the number of transmissions in order sense.
\end{remark}

Binary linear block codes can be used to transmit binary vectors \textcolor{black}{over noisy channels}. Suppose we have a $K$-bit message vector $\mathbf{m}$ and a code $\mathcal{C}$ with length $N$ and \textcolor{black}{$R=\frac{K}{N}$}. Then, we can encode the message $\mathbf{m}$ into $N$ bits by multiplying $\mathbf{m}$ with the generator matrix $\mathbf{G}$, transmit these $N$ bits over a channel and decode the received bits. The block error probability is defined as the probability that the decoding result $\hat{\mathbf{m}}$ is different from the original $K$-bit message at least in one bit. The next lemma characterizes the performance of using binary linear codes over a BSC.
\begin{lemma}(\cite[Theorem 5.6.2]{Gal_Wil_68})\label{BSC_rm_cd}(Random Coding Theorem)
Suppose we have a $K$-bit message vector $\mathbf{m}$ to be transmitted through a BSC with crossover probability $\epsilon$. Then, \textcolor{black}{for each $R<C$}, where $C$ is the channel capacity, there exists a binary linear code with length $N_R$ and rate $R$, such that $K<N_R R$ and the $K$-bit message can be encoded into $N_R$ bits, transmitted through the BSC and decoded with block error probability upper bounded by
\begin{equation}\label{Gallager_Rm_Bd}
  P_e^{(blk)}=\Pr(\hat{\mathbf{m}}\neq \mathbf{m}) \le \exp[-KE_r(\epsilon,R)/R],
\end{equation}
where $E_r(\epsilon,R)>0$ is the random coding exponent.
\end{lemma}
\textcolor{black}{The random coding error exponent $E_r(\epsilon,R)$ for a BSC with crossover probability $\epsilon$ can be written as
\[
E_r(\epsilon,R)=\max_{0\le \rho\le 1} \left[-\rho R + E_0(\rho,\epsilon)\right],
\]
where
\[
E_0(\rho,\epsilon)=\rho \ln 2 -(1+\rho)\ln \left[\epsilon^{1/(1+\rho)}+(1-\epsilon)^{1/(1+\rho)}\right].
\]
The random coding error exponent $E_r(\epsilon,R)$ is always positive for coding rate $R<C=1-H(\epsilon)$.}

\section{System Model and Problem Formulations}\label{modeling}
\vspace{0.2cm}
\subsection{Data Gathering with Broadcasting}\label{Graph_Model}
Consider a network $\mathcal{G}=(\mathcal{V},\mathcal{E})$ with $N+1$ agents $\mathcal{V}=\{v_n\}_{n=0}^{N}$, \textcolor{black}{where $v_0$ is a preassigned sink node}. Each agent $v_n$ with $1\le n \le N$ has one bit of information $x_n\in\{0,1\}$ \textcolor{black}{distributed as $\text{Bernoulli}(\frac{1}{2})$}. This is called the \emph{self-information bit}. \textcolor{black}{All self-information bits are independent} \textcolor{black}{of} each other. Denote the vector of all self-information bits by $\mathbf{x}=(x_1,x_2,...,x_N)^\top$. The objective is to collect $\mathbf{x}$, in the sink $v_0$ with high accuracy.

Time is slotted. In the $t$-th slot, only one chosen node $v(t)$ is allowed to broadcast\footnote{The transmission scheduling is beyond the scope of this paper. This paper address a fundamental issue, the communication complexity, which \textcolor{black}{is minimum over all scheduling protocols in place.} Nevertheless, transmission scheduling indeed improves the network throughput~\cite{Gup_TIT_00}.} one bit of information in $\mathbb{F}_2$ to its out-neighborhood $\mathcal{N}^+(v(t))$. \textcolor{black}{The channel between any two connected nodes is assumed noisy. Since we consider different noise models, we make two assumptions for convenience of reference.}

\textcolor{black}{\noindent\textbf{(A.1a) BSC: }All channels or graph edges are BSCs with identical crossover probability $\epsilon\in (0,1/2)$. All channels are independent of each other.}

\textcolor{black}{\noindent\textbf{(A.1b) BEC: }All channels of graph edges are BECs with identical erasure probability $\epsilon$. All channels are independent of each other.}

A \emph{broadcast scheme} $\mathscr{S}=\{f_t\}_{t=1}^{\mathscr{C}_\mathscr{S}^\text{(N)}}$ is a \textcolor{black}{sequence of }Boolean functions, \textcolor{black}{such that at each time slot $t$ the broadcasting node $v(t)$ computes the function $f_{t}$ (whose arguments are to be made precise below) and broadcasts the computed output bit to its out-neighborhood}. The parameter $\mathscr{C}_\mathscr{S}^\text{(N)}$ is used to denote the total number of broadcasts in a broadcasting scheme $\mathscr{S}$ \textcolor{black}{which, in our setup, also corresponds to the time complexity or implementation time of $\mathscr{S}$, because in each time slot, only one node is allowed to broadcast}. \textcolor{black}{The minimum value of $\mathscr{C}_\mathscr{S}^\text{(N)}$ among all broadcast schemes is defined as the communication complexity of the data gathering problem, which is denoted as $\mathscr{C}^\text{(N)}$.} \textcolor{black}{The arguments of $f_t$ may consist of all the information that the broadcasting node $v(t)$ has up to time $t$}, including its self-information bit $x_{v(t)}$, randomly generated bits and information obtained from its in-neighborhood called the \emph{outer information}. We only consider oblivious transmission schemes, i.e., the \textcolor{black}{number of broadcasts $\mathscr{C}_\mathscr{S}^\text{(N)}$, all functions in $\mathscr{S}$ and the broadcasting order $\{v(t)\}_{t=1}^{\mathscr{C}_\mathscr{S}^\text{(N)}}$ are predetermined}. It also means that transmission by silence is not allowed, i.e., a node has to broadcast when it is required. \textcolor{black}{Further, we assume that a scheme terminates in finite time, i.e., $\mathscr{C}_\mathscr{S}^\text{(N)} < \infty$ for all $N$.} A scheme obviously has to \textcolor{black}{be feasible, meaning that all arguments of $f_t$ should be available in $v(t)$ before time $t$. Denote by $\mathcal{F}$ the set of all feasible oblivious schemes.} The final error probability is defined as $P_e^{(N)}=\Pr(\hat{\mathbf{x}}\neq\mathbf{x})$, \textcolor{black}{where $\hat{\mathbf{x}}$ denotes the final estimate of $\mathbf{x}$ at the sink $v_0$}. Usually it is required that the error probability is asymptotically bounded or $\lim\limits_{N\rightarrow \infty} P_e^{(N)}\le p_{\text{tar}}$ where $p_{\text{tar}}$ might be zero, which means that the error probability should be small even if the number of vertices in the network is large. Although \textcolor{black}{our objective does not involve convergence rate requirements}, in this paper, convergence rates are indeed given for all constructive results. The problem to be studied is therefore
\begin{equation}\label{op_problem}
\begin{split}
&\min_{\mathscr{S}\in\mathcal{F}}{\;\;\;\;}\mathscr{C}_\mathscr{S}^\text{(N)},\\
\text{s.t.}&{\;}\lim\limits_{N\rightarrow \infty} P_e^{(N)}\leq p_{\text{tar}}.
\end{split}
\end{equation}
\textcolor{black}{We call this problem the noisy broadcasting problem.} In this paper, we will consider both fixed graph topologies and random graph topologies, which will be clear in the next subsection. The above mentioned error probability $P_e^{(N)}$ \textcolor{black}{needs to be interpreted in the expected sense when dealing with random graph topologies. Specifically, for random topologies, denote by $P_e^{\mathcal{G}}$ the (conditional) error probability conditioned on an instance $\mathcal{G}$ of the communication graph. If the graph $\mathcal{G}$ involved is deterministic, $P_{e}^{\mathcal{G}}\equiv P_{e}^{N}$, otherwise, for random graph topologies, the conditional error probability $P_e^{\mathcal{G}}$ is itself a random variable and the error probability metric $P_{e}^{(N)}$ is defined as the expected error probability $P_e^{(N)}=\mathbb{E}_\mathcal{G} [P_e^{\mathcal{G}}]$ (When dealing with random graphs, the quantities $\mathbb{P}_{\mathcal{G}}(\cdot)$ and $\mathbb{E}_{\mathcal{G}}[\cdot]$ denote probability and expectation with respect to the distribution of the random graph ensemble.). The transmission scheme design problem for random graphs is the same as in~\eqref{op_problem}.}

\subsection{Network Models}\label{network_model}
\textcolor{black}{When working with deterministic (but arbitrary) graph topologies, we assume that the network is connected. Specifically, we impose the following connectivity assumption.}

\noindent\textbf{(A.2) Network Connectivity: }In the directed graph $\mathcal{G}=(\mathcal{V},\mathcal{E})$, \textcolor{black}{the sink node $v_{0}$ is reachable from each non-sink node $v\in\mathcal{V}\setminus\{v_0\}$ through a sequence $v\rightarrow v_{i_1}\rightarrow v_{i_2}\dots \rightarrow v_0$ of directed edges.}

In (A.2), recall that if the graph is undirected, each (bidirectional) edge corresponds to the two directed edges. Since network connectivity is necessary for data collection, we assume this assumption holds throughout.

We use $\mathcal{T}=(\mathcal{V},\mathcal{E}_{\mathcal{T}})$ to represent the breadth-first search (BFS) spanning tree~\cite{Cor_MIT_90} of $\mathcal{G}=(\mathcal{V},\mathcal{E})$ rooted at the sink $v_0$. The edge set $\mathcal{E}_{\mathcal{T}}$ is a subset of $\mathcal{E}$ and $|\mathcal{E}_{\mathcal{T}}|=|\mathcal{V}|-1$. A BFS tree can be constructed as follows:
\begin{itemize}
  \item Initialize: $\mathcal{V}_\mathcal{T}=\{v_0\}$, $\mathcal{E}_{\mathcal{T}}=\emptyset$.
  \item Find all directed edges $(u,v)\in \mathcal{E}$ such that $u\notin \mathcal{V}_\mathcal{T}$ and $v\in \mathcal{V}_\mathcal{T}$. Include $(u,v)$ in $\mathcal{E}_{\mathcal{T}}$ and include $u$ in $\mathcal{V}_{\mathcal{T}}$.
  \item Repeat the previous step until $\mathcal{V}_\mathcal{T}=\mathcal{V}$.
\end{itemize}
By assumption (A.2), the BFS tree exists. By $d(v,v_0)$, we denote the multi-hop distance from a node $v$ to the sink $v_0$. An obvious property of the breadth-first search spanning tree $\mathcal{T}$ is that the multi-hop distance $d(v,v_0)$ is the same in $\mathcal{T}$ as in the original graph $\mathcal{G}$. By the $l$-th layer $\mathcal{V}_l\subset\mathcal{V}$, we denote the set of nodes that have identical multi-hop distance $d(v,v_0)=l$. Denote the maximum distance from a node $v$ to the sink $v_0$ by $L_d$. We know that $\mathcal{V}=\bigcup\limits_{l=1}^{L_d}\mathcal{V}_l$ forms a layered partition of the node set. In the BFS tree, the parent-node $v_f$ of a node $v$ is defined to be the unique node such that there exists a directed edge $(v,v_f)$ in the BFS tree's edge set $\mathcal{E}_\mathcal{T}$. The descendants of a node $v$ is defined as the set $\mathcal{D}_{v}\subset\mathcal{V}$ that includes all nodes $w$ that are connected to $v$ through a sequence of directed edges in $\mathcal{E}_\mathcal{T}$.

In Section~\ref{Large_Diameter}, we consider the noisy broadcasting problem on a general graph. \textcolor{black}{The broadcasting scheme in~\cite{Gal_TIT_88} is designed for complete graphs and not directly applicable here.} Intuitively, the communication complexity is higher for general graphs, in contrast to complete graphs, because a non-negligible routing complexity might be incurred \textcolor{black}{due to the (possibly) large distances of some non-sink nodes to the sink $v_0$}.

In Section~\ref{geometric_graph}, we consider the noisy broadcasting problem in geometric graphs. \textcolor{black}{By geometric graph, we mean each node is connected to and can only communicate with nodes that are within a certain (specified) distance of itself}. \textcolor{black}{The formal definition is given in the assumption (A.3a). For comparison, we will cite a result on random geometric graphs~\cite{Li_TMC_13}. The definition of random geometric graphs is given in the assumption (A.3b). For arbitrary graphs and geometric graphs, we provide communication complexity results for both BECs and BSCs.}

\textcolor{black}{\noindent\textbf{(A.3a) Geometric Graph: }}The graph $\mathcal{G}=(\mathcal{V},\mathcal{E})$ is assumed to be a geometric graph, i.e., all $N+1$ nodes in $\mathcal{V}$ are located in a 1-by-1 square area, and any two nodes are connected bidirectionally if they are within a specified distance $r$. Further, we assume that $r>\sqrt{\frac{c_g\log N}{N}}$ where $c_g$ is a constant. \textcolor{black}{Finally, we assume that $\mathcal{G}$ is connected, which means that $\mathcal{G}$ satisfies the assumption (A.2)}\footnote{\textcolor{black}{This assumption is required because connectivity within distance $r$ does not necessarily ensure connectivity.}}.

\textcolor{black}{\noindent\textbf{(A.3b) Random Geometric Graph: }}The graph $\mathcal{G}=(\mathcal{V},\mathcal{E})$ is assumed to be a geometric graph which satisfies the assumption (A.3a)\footnote{Note that the connectivity assumption (A.2) here is still needed, although the random geometric graph is connected with high probability if $r$ is large enough.}. Moreover, each node in $\mathcal{V}$ is distributed uniformly in the 1-by-1 square area, independently of other nodes.

In Section~\ref{Upper_Bound}, we consider the noisy broadcasting problem in the extended Erd\"os-R\'enyi network, which is slightly different from the original Erd\"os-R\'enyi model in~\cite{Bol_Spr_98}. The definition is given in the assumption (A.4). In this model, \textcolor{black}{the connection probability $p=\Theta(\frac{\log N}{N})$ indicates that} the average node degree is $\Theta(\log N)$. We will also show that the minimum average node degree is at least $\Omega(\frac{\log N}{\log\log N})$, \textcolor{black}{if the error probability of data gathering is required to approach zero when the node number approaches infinity. This result states that $p=\Theta(\frac{\log N}{N})$ is minimum in the order sense except for a $\log\log N$ factor.} \textcolor{black}{A sink might be a base station and all agents have direct links to it}\footnote{\textcolor{black}{However, as long as these direct links are noisy, the communication complexity for \textcolor{black}{the} data gathering is in the order of $\Theta(N\log N)$~\cite{Gal_TIT_88} if \textcolor{black}{a naive scheme is used that aims to  transmit the self-bit of each node to the sink through the corresponding direct link}. We will prove that in-network computation makes this complexity smaller by utilizing more communications between non-sink nodes, i.e., information fusion. Moreover, these inter-node communications are usually cheaper than direct communications with the base station.}}. In this section, links are assumed to be BECs \textcolor{black}{as in the assumption (A.1b).} Furthermore, in the extended Erd\"os-R\'enyi network, the error probability $P_e^{(N)}$ in~\eqref{op_problem} \textcolor{black}{should be replaced by $\mathbb{E}_G(P_e^{(N)})$, where the expectation is taken over all random graph instances}. See Section~\ref{Upper_Bound} for more details.

\textcolor{black}{\noindent\textbf{(A.4) Extended Erd\"os-R\'enyi Graph: }The extended Erd\"os-R\'enyi graph is an ER graph with the minimal number of additional links that ensures that each non-source node has directed link to the sink. In the graph $\mathcal{G}=(\mathcal{V},\mathcal{E})$, all connections are independent of each other. Assume that $p$ satisfies $p=\frac{c\log N}{N}$, where $c$ is a constant. We further assume that each node in $\mathcal{V}$ has a direct link to the sink, in addition to the random connections between these nodes themselves. }

\textcolor{black}{Assumption \textbf{(A.4)} can be interpreted as follows: the edge set $\mathcal{E}$ can be decomposed into $\mathcal{E}=\mathcal{E}_1\cup\mathcal{E}_2$, where $\mathcal{E}_1$ is the set of directed edges connecting non-sink nodes, which form the edge set of a directed Erd\"os-R\'enyi network with connection probability $p=\frac{c\log N}{N}$, and $\mathcal{E}_2$ can be viewed as the minimum set of edges that is further added to the graph with edge set $\mathcal{E}_1$ so that each non-sink node has a directed link to the sink\footnote{\textcolor{black}{It can be shown that this assumption can be relaxed by assuming that the graph is an Erd\"os-R\'enyi graph with the minimum number of additional links to ensure bounded multi-hop distance $d_\text{max}$ from each non-sink node to the sink, That is, the edge set $\mathcal{E}_2$ can be viewed as the minimum set of edges that is further added to the graph with edge set $\mathcal{E}_1$ so that each non-sink node has a directed path of length smaller or equal to $d_\text{max}$ to the sink. This relaxation can be made because it does not affect the scaling bounds on the number of transmissions in the achievable scheme for the extended Erd\"os-R\'enyi Graph in order sense. More specifically, one bit can always be transmitted using a directed path to the sink node with $\mathcal{O}(1)$ transmissions to obtain an error probability $\epsilon$, which means that by increasing the number of transmissions by a constant multiple, the directed path of length $d_\text{max}$ between a non-sink node and the sink node can be viewed as a directed link with error probability $\epsilon$.}}. To be precise, in the standard Erd\"os-R\'enyi generation, there might already be some edges between sink and non-sink nodes. The set $\mathcal{E}_2$ is the additional set of source to non-source links not obtained through the Erd\"os-R\'enyi instantiation.}

\section{Main Techniques: Graph Codes}\label{main_technique}
Graph codes are distributed linear block codes which have generator matrices closely related to the network structure. Designing graph codes relies on the utilization of the network structure. We introduce three different types of graph codes in the following. \textcolor{black}{The following descriptions are informal, details and their usage will be made clear in the subsequent sections.}

A $\mathcal{GC}$-3 graph code is a rate-$\frac{1}{2}$ systematic code with a generator matrix $\mathbf{G}=[\mathbf{I},\mathbf{A}]$ with $\mathbf{A}$ being the graph adjacency matrix. The encoding of a $\mathcal{GC}$-3 graph code can be written as
\begin{equation}\label{3g_code}
  \mathbf{r}^\top=\mathbf{x}^\top \cdot\left[\mathbf{I},\mathbf{A}\right],
\end{equation}
\textcolor{black}{where $\mathbf{x}^\top$ denotes the message vector with length $N$ and $\mathbf{r}^\top$ denotes the encoding output with length $2N$.} \textcolor{black}{This means that the code bit calculated by a node $v$ is either its self-information bit $x_v$ or the parity of the self-information bits in its in-neighborhood $\mathcal{N}_v^-$. Therefore, $\mathcal{GC}$-3 codes are easy to encode with local communications and admit distributed implementations.} However, the decoding can be quite difficult \textcolor{black}{depending on the graph structure}. We only use $\mathcal{GC}$-3 codes for the extended Erd\"os-R\'enyi networks with BECs in Section~\ref{Upper_Bound}.

A $\mathcal{GC}$-2 graph code is also a rate-$\frac{1}{2}$ code with a generator matrix $\tilde{\mathbf{G}}=[\mathbf{I},\mathbf{\tilde{A}}]$. However, $\mathbf{\tilde{A}}$ is the adjacency matrix of a subgraph $\mathcal{\tilde{G}}=(\mathcal{V},\mathcal{\tilde{E}})$ of $\mathcal{G}=(\mathcal{V},\mathcal{E})$, where $\mathcal{\tilde{E}}\subset\mathcal{E}$. \textcolor{black}{Alternatively, a $\mathcal{GC}$-2 graph code may be viewed as} a generalization of the $\mathcal{GC}$-3 code when some edges in the original edge set $\mathcal{E}$ are removed. This code is much more flexible than the $\mathcal{GC}$-3 code and we will use it for geometric graphs in Section~\ref{geometric_graph}.

A $\mathcal{GC}$-1 graph code has no direct relationship with the adjacency matrix but the idea is similar to the previous codes. It assumes that each code bit calculated at a node $v$ is the parity of \textcolor{black}{a subset of} nodes that are within $\Theta(\log N)$ hops of $v$. This code is quite general and we will show that, \textcolor{black}{for arbitrary graph topologies}, a $\mathcal{GC}$-1 graph code can help achieve an upper bound on the communication complexity \textcolor{black}{of data gathering} which is at most a $\Theta(\log N)$ multiple of the lower bound.

The code length of the three graph codes are all in the order of $\Theta(N)$. Since the code length is in the same order as the number of nodes in the network, \textcolor{black}{and only one-shot computing of distributed encoding is required for the one-shot data gathering problem,} the average number of bits calculated by each node during the distributed encoding process is a constant (details follow in the subsequent sections). Furthermore, \textcolor{black}{all three types of graph codes are designed to possess a \textcolor{black}{sparseness} property: the number of ones in the generator matrix will be in the order of $\Theta(N \log N)$, because the studied graphs (either geometric graphs or extended Erd\"os-R\'enyi graphs have a \textcolor{black}{sparse} structure). This means that encoding each bit requires only $\Theta(\log N)$ self-information bits. Therefore, efficient distributed encoding with a small number of broadcasts becomes possible.}
\vspace{0.2cm}
\section{$\mathcal{GC}$-1 Graph Codes in a General Graph}\label{Large_Diameter}
In this section, we consider general connected network topologies. We first consider this problem on BSCs satisfying the assumption (A.1a), then we extend the results to BECs satisfying the assumption (A.1b). We design a general distributed in-network computing algorithm called the $\mathcal{GC}$-1 graph code. Recall that in the case of complete networks, as studied in~\cite{Gal_TIT_88,Goy_SJC_08}, \textcolor{black}{a lower bound on the} communication complexity \textcolor{black}{for data gathering} is $\Theta(N\log\log N)$. In what follows, we provide a lower bound for general networks. Then, we use the $\mathcal{GC}$-1 graph code to get an upper bound which, \textcolor{black}{we show,} is close to the lower bound when the graph diameter is small, and meets the lower bound when the diameter is large. We also give an intuitive example on why this upper bound can be achieved and why there is a small gap between the lower and upper bounds.
\subsection{\textcolor{black}{Communication Complexity Lower Bound in a General Graph}}
\begin{figure}
  \centering
  \includegraphics[scale=0.4]{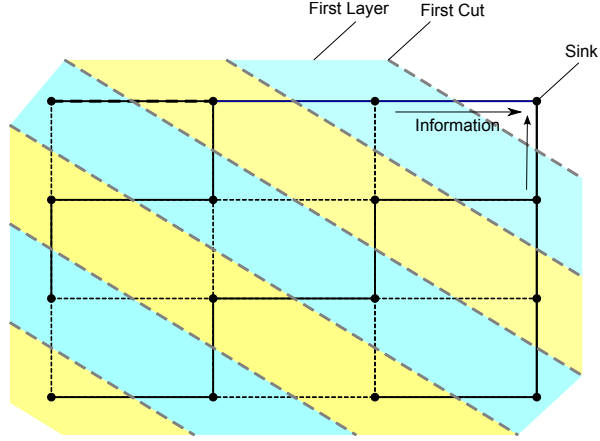}\\
  \caption{\emph{A grid network layered by the distance to the sink. The spanning tree rooted at the sink is represented by a solid line.} }\label{fig1}
\end{figure}

As shown in Fig.~\ref{fig1}, construct the breadth-first-search spanning tree of the network $\mathcal{G}=(\mathcal{V},\mathcal{E})$, and then, \textcolor{black}{construct the layered partition $\mathcal{V}=\bigcup\limits_{l=1}^{L_d}\mathcal{V}_l$ of the network} based on the multi-hop distance $d(v_n,v_0)$ from each node $v_n$ to the sink $v_0$, as defined in Section~\ref{network_model}. \textcolor{black}{Note that the} distance $d(v_n,v_0)$ in the tree is the same as in the original network.

\textcolor{black}{By definition of the BFS spanning tree and the associated layering}, we know that in the graph $\mathcal{G}$, no edges exist \textcolor{black}{between non-successive layers}, but edges connecting nodes in the same layer may exist. \textcolor{black}{By $l$-th cut, we denote the set of edges from the $l$-th layer $\mathcal{V}_l$ to the $(l-1)$-th layer $\mathcal{V}_{l-1}$.} We know that information can only be transmitted hop-by-hop \textcolor{black}{from the bottom layer $\mathcal{V}_{L_d}$ to the sink. Therefore, on each cut between two layers of the multi-layer BFS spanning tree, there is a certain amount of information that needs to be transmitted. The overall number of broadcasts can be lower bounded by the sum of information necessary to be transmitted on all of these disjoint cuts. This gives the basic lower bound for data gathering, i.e., transmitting the vector $\mathbf{x}=(x_1,x_2,...,x_N)^\top$ to $v_0$.}
\begin{theorem}\label{Lg_Dia_Lw_Bd}
\textcolor{black}{Suppose the communication links in the graph $\mathcal{G}$ satisfy the assumption (A.1a)}. Then, if all data are gathered at the sink $v_0$ with error probability $P_e^{(N)}$ by a feasible broadcasting scheme $\mathscr{S}$, the communication complexity is necessarily bounded below by
\begin{equation}\label{diameter_lw_bnd}
  \mathscr{C}^\text{(N)}\ge c_\epsilon\bar{d}_\mathcal{G}N,
\end{equation}
where $c_\epsilon=\frac{1-H(P_e^{(N)})}{1-H(\epsilon)}$ is a constant, \textcolor{black}{$N$ denotes the number of nodes in the graph and $\bar{d}_\mathcal{G}$ is the average distance to the sink, defined as}
\begin{equation}\label{ave_dis}
  \bar{d}_\mathcal{G}=\frac{1}{N}\sum\limits_{n=1}^N d(v_n,v_0).
\end{equation}
\end{theorem}
\begin{IEEEproof}\textcolor{black}{Denote by $\mathbf{z}_1$ the whole data received in the sink $v_0$ during the entire data gathering process, i.e., when the broadcasting scheme in place terminates. Then, for each bit $x_n$, we know that $x_n\rightarrow\mathbf{z}_1\rightarrow \hat{x}_n$ is a Markov chain, where $\hat{x}_n$ is the estimate of $x_n$ at $v_0$. Therefore, based on the data processing inequality} and Fano's inequality~\cite{Cover_Wiley_06}, it holds that
\begin{equation}\label{Fano_Inq}
\begin{split}
  H(x_n|\mathbf{z}_1)\le& H(x_n|\hat{x}_n) \le H(P_\text{bit})+P_\text{bit}\log(\left|\chi\right|-1)=H(P_\text{bit}),
\end{split}
\end{equation}
where $P_\text{bit}$ is the bit error probability $\Pr(x_n\neq \hat{x}_n)$ of estimating $x_n$ from $\mathbf{z}_1$ \textcolor{black}{and the second equality follows from} the fact that $\left|\chi\right|=|\mathbb{F}_2|=2$. \textcolor{black}{Since $x_1,x_2,\dots,x_N$ are assumed to be independent of each other, we know that $H(\mathbf{x})=N$.} Therefore, we have
\begin{equation}\label{Lw_Bd_Der1}
\begin{split}
  I(\mathbf{x};\mathbf{z}_1) = H(\mathbf{x})-H(\mathbf{x}|\mathbf{z}_1) \mathop  \ge \limits^{(a)}& N-\sum\limits_{n=1}^N H(x_n|\mathbf{z}_1)\ge N[1-H(P_\text{bit})],
\end{split}
\end{equation}
\textcolor{black}{where $(a)$ holds because
\[
H(\mathbf{x}|\mathbf{z}_1)=\sum_{n=1}^N H(x_n|\mathbf{z}_1,x_1,\ldots,x_{n-1})\le \sum_{n=1}^N H(x_n|\mathbf{z}_1).
\]}
\textcolor{black}{Since $v_{0}$ has no side information about $\mathbf{x}$ to start with, the amount of (mutual) information} $I(\mathbf{x};\mathbf{z}_1)$ needs to be broadcasted via the BSCs in the first cut. \textcolor{black}{Assume the number of broadcasts in the first cut is $\mathscr{C}_1$. Then, since the number of channel uses is $\mathscr{C}_1$ and each channel use has capacity $1-H(\epsilon)$, from the cut-set bound that $I(\mathbf{x};\mathbf{z}_1)<\mathscr{C}_1(1-H(\epsilon))$ we must have}
\begin{equation}\label{Lw_Bd_Der2}
  \mathscr{C}_1\mathop  \ge  \frac{I(\mathbf{x};\mathbf{z}_1)}{1-H(\epsilon)}\ge \frac{N(1-H(P_\text{bit}))}{1-H(\epsilon)}\mathop  \ge \limits^{(a)}  Nc_\epsilon,
\end{equation}
where step (a) follows from the fact that the \textcolor{black}{bit error probability} $P_\text{bit}$ is always smaller than total error probability $P_e^{(N)}$.

For each layer $l$, \textcolor{black}{denote by $\mathcal{S}_l$ the set of nodes in the union $\mathcal{V}_l\bigcup\mathcal{V}_{l+1}\bigcup\dots\bigcup\mathcal{V}_{L_d}$}. Denote \textcolor{black}{all self-information bits in $\mathcal{S}_l$ by $x_{\mathcal{S}_l}$}. Define $N_l=|\mathcal{S}_l|$. Similarly, we obtain
\begin{equation}\label{Lw_Bd_Der3}
  \mathscr{C}_l\ge\frac{I(\mathbf{x}_{\mathcal{S}_l};\mathbf{z}_l)}{1-H(\epsilon)}\ge \frac{\sum _{n\in \mathcal{S}_l}\left[1-H(x_n|\mathbf{z}_l)\right]}{1-H(\epsilon)}\mathop  \ge \limits^{(b)}  N_l c_\epsilon,
\end{equation}
where $\mathbf{z}_l$ is the data transmitted through the $l$-th cut and step (b) is obtained by combining~\eqref{Fano_Inq} with the data processing inequality
\begin{equation}\label{Dt_Prc}
  I(x_n;\mathbf{z}_l)\ge I(x_n;\mathbf{z_1}), \forall n\in \mathcal{S}_l.
\end{equation}
\textcolor{black}{To show that~\eqref{Dt_Prc} holds, we reason as follows:} \textcolor{black}{from the definition of layers and cuts, if $x_n$ is below the $\mathcal{V}_l$, the information of $x_n$ has to be routed through the $l$-th cut before it gets to the first cut, otherwise, $x_n$ must be in upper layers of the $l$-th cut\textcolor{black}{, a contradiction}. More formally, for $x_{n}\in\mathcal{S}_{l}$, we note that $\mathbf{z}_{1}$ is conditionally independent of $x_{n}$ given $\mathbf{z}_{l}$, and hence~\eqref{Dt_Prc} holds.} Summing over all $l$, we obtain the following lower bound on the (total) number of broadcasts:
\begin{equation}\label{Lw_Bd_Der4}
  \mathscr{C}_\mathscr{S}^\text{(N)}=\mathop\sum\limits_{l=1}^{L_d} \mathscr{C}_l\ge \mathop\sum\limits_{l=1}^{L_d} N_l c_\epsilon\mathop{=}\limits^{(c)}c_\epsilon N\bar{d}_\mathcal{G},
\end{equation}
where step (c) follows \textcolor{black}{from the definition of $\bar{d}_\mathcal{G}$ and the exchange of summation}. Thus, we obtain~\eqref{diameter_lw_bnd}. This lower bound holds for any broadcast scheme so~\eqref{Lw_Bd_Der4} is a lower bound on the communication complexity $\mathscr{C}^\text{(N)}$.
\end{IEEEproof}
\subsection{In-network Computing Algorithm}\label{Large_Diameter_Algorithm}
\begin{figure}
  \centering
  \includegraphics[scale=0.4]{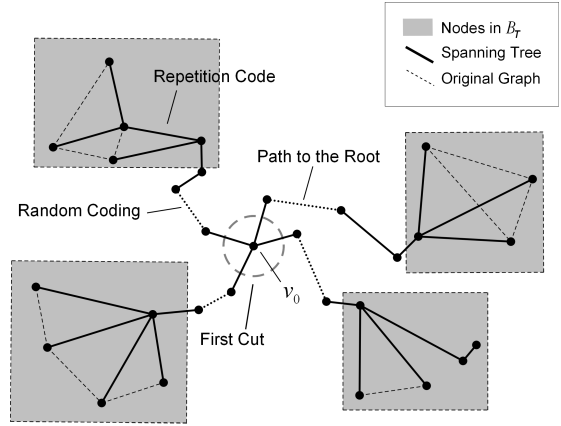}\\
  \caption{\emph{The in-network computing algorithm carried out on the spanning tree.} }\label{fig4}
\end{figure}

\vspace{0.2cm}
In this part we provide the $\mathcal{GC}$-1 in-network computing algorithm for gathering all data at $v_0$ in an arbitrary network. \textcolor{black}{Before we provide the algorithm, we provide some preparatory procedures as follows. First, we construct the BFS spanning tree $\mathcal{T}=(\mathcal{V},\mathcal{E}_{\mathcal{T}})$ rooted at the sink $v_0$, \textcolor{black}{as defined in Section~\ref{network_model}}. That is, in the layered network shown in Fig.~\ref{fig1}, we delete all edges in the same layer but reserve edges that span adjacent layers. The resulting network is like Fig.~\ref{fig4} and the edge set is denoted by $\mathcal{E}_{\mathcal{T}}$. As defined in Section~\ref{network_model}, denote all descendants of the node $v$ by $\mathcal{D}_v$. Define
\begin{equation}\label{B_T}
  \mathcal{B}_{\mathcal{T}}=\{v\in \mathcal{V}: \left|\mathcal{D}_v\right| < \gamma\log N\},
\end{equation}
where $\gamma$ is a constant. Define $\mathcal{A}_{\mathcal{T}}=\mathcal{V}\setminus \mathcal{B}_{\mathcal{T}}$. It is obvious that each path from a leaf-node $v_n$ to the root $v_0$ is constituted by a series of nodes in $\mathcal{B}_{\mathcal{T}}$, followed by another series of nodes in $\mathcal{A}_{\mathcal{T}}$ (as shown in Fig.~\ref{fig4}).}

\textcolor{black}{Then, we propose the $\mathcal{GC}$-1 algorithm, as shown in Algorithm~\ref{alg1}. The basic idea is: Each $v\in \mathcal{V}$ gathers all self-information bits from its descendants in $\mathcal{D}_v$. Then, it sends all the information in $\mathcal{D}_v \cup \{v\}$, including bits from its descendants and its own self-information bit, to its parent-node. In order to make this scheme a feasible in-network computing scheme, each node $v$ has to start transmitting after all of its children nodes complete transmitting.}

\textcolor{black}{All nodes use linear block codes to encode the information that it needs to transmit. Nodes with small descendant size ($|\mathcal{D}_v|<\gamma \log N$) has to insert zeros (dummy bits) to the message vector before encoding. The performance guarantee of this algorithm is shown in Theorem~\ref{Lg_Dia_Up_Bd}. The intuition underlying why the error probability is small is put in Remark~\ref{alg1_intuition}.}
\begin{algorithm}\caption{$\mathcal{GC}$-1 algorithm}\label{alg1}
\textcolor{black}{\noindent\textbf{Initialization}: Construct the BFS spanning tree $\mathcal{T}=(\mathcal{V},\mathcal{E}_{\mathcal{T}})$ rooted at the sink $v_0$.}

\textcolor{black}{\noindent\textbf{Step 1}: Each leaf-node $v$ encodes the binary vector $(x_v,0,\dots,0)$ with length $\gamma\log N$ using random coding with rate $R$ and transmits the codeword to its parent-node.}

\textcolor{black}{\noindent\textbf{Step 2}: Each non-leaf node $v$, from its children-nodes, receives the self-information bits of its entire set of descendants $\mathcal{D}_v$. After all of its children-nodes finish transmitting, the node $v$ relays the self-information bits of all of its descendants and its own self-information bit $x_v$ to its parent-node, using error control codes. Depending on if $v$ is in $\mathcal{B}_{\mathcal{T}}$ or $\mathcal{A}_{\mathcal{T}}$, the coding schemes differ. The coding details are shown below.}
\begin{itemize}
  \item \textbf{Actions in $\mathcal{B}_{\mathcal{T}}$:} Each $v\in \mathcal{B}_{\mathcal{T}}$ decodes the self-information bits from $\mathcal{D}_v$ and form a binary vector with length $\mathcal{D}_v+1$ with its own self-information bit. Then the node $v$ inserts $\gamma\log N-1-|\mathcal{D}_v|$ zeros to the vector to make the length $\gamma\log N$ and uses random coding to encode this vector. Finally, it sends the whole $\left\lceil {(|\mathcal{D}_v| + 1)/R} \right\rceil$ bits to its parent-node, where $R$ is the coding rate.
  \item \textbf{Actions in $\mathcal{A}_{\mathcal{T}}$:} Each $v\in \mathcal{A}_{\mathcal{T}}$ decodes the self-information bits from $\mathcal{D}_v$, and uses random coding to encode these bits and its own self-information. Finally, it sends the whole $\left\lceil {(|\mathcal{D}_v| + 1)/R} \right\rceil$ bits to its parent-node, where $R$ is the coding rate.
\end{itemize}
\end{algorithm}

\begin{theorem}\label{Lg_Dia_Up_Bd}
\textcolor{black}{Suppose the communication links in the graph $\mathcal{G}$ satisfy the assumption (A.1a)}. Then, \textcolor{black}{for each tuple of constants $(R,\gamma)$ satisfying}
\begin{equation}\label{Rate_Cond}
  R<\gamma E_r(\epsilon,R),
\end{equation}
where $E_r(\epsilon,R)$ is the random coding error exponent from~\eqref{Gallager_Rm_Bd}, \textcolor{black}{the number of broadcasts that the scheme $\mathscr{S}$ provided in Algorithm~\ref{alg1} incurs is upper bounded by}
\begin{equation}\label{diameter_up_bnd}
\begin{split}
  \mathscr{C}_\mathscr{S}^\text{(N)}< &N(\frac{\bar{d}_\mathcal{G}}{R}+1)+N(\gamma\log N/R+1)=\max\{\Theta(\bar{d}_\mathcal{G}N),\Theta(N\log N)\},
\end{split}
\end{equation}
\textcolor{black}{where $N$ denotes the number of nodes in the graph and $\bar{d}_\mathcal{G}$ is the average distance to the sink, which is defined in~\eqref{ave_dis}}. Moreover, as $N\rightarrow\infty$, the error probability $P_{e}^{(N)}$ decreases polynomially as
\begin{equation}\label{error_polynomial}
  P_e^{(N)}<N^{-(\frac{\gamma E_r(\epsilon,R)}{R}-1)}\cdot \left(1+\exp[-E_r(\epsilon,R)/R]\right),
\end{equation}
\textcolor{black}{and, in particular, achieves $\lim_{N\rightarrow\infty}P_{e}^{(N)}=0$.}
\end{theorem}
\vspace{0.2cm}
\begin{IEEEproof}
\textcolor{black}{In what follows, we show how to obtain the upper bound on the number of broadcasts in~\eqref{diameter_up_bnd}, while the error probability analysis of~\eqref{error_polynomial} is put in the Appendix~\ref{PofT2}.} Each node $v\in \mathcal{B}_{\mathcal{T}}$ (including leaf-nodes) transmits a codeword of size $\lceil\gamma \log N/R\rceil$, so the number of broadcasts at each node $v\in \mathcal{B}_{\mathcal{T}}$ satisfies
\begin{equation}\label{CC_Each_Node_B}
  \mathscr{C}_v<\gamma\log N/R+1.
\end{equation}
The number of broadcasts at each node $v\in \mathcal{A}_{\mathcal{T}}$ is
\begin{equation}\label{CC_Each_Node_A}
  \mathscr{C}_v=\lceil(\mathcal{D}_v+1)/R\rceil<(\mathcal{D}_v+1)/R+1.
\end{equation}
Therefore, the final number of broadcasts is
\begin{equation}\label{Large_d_cmplx}
\begin{split}
  \mathscr{C}_\mathscr{S}^\text{(N)}=&\mathop\sum\limits_{v\in \mathcal{A}_{\mathcal{T}}}\mathscr{C}_v+\mathop\sum\limits_{v\in \mathcal{B}_{\mathcal{T}}}\mathscr{C}_v <\mathop\sum\limits_{v\in \mathcal{V}}[(\mathcal{D}_v+1)/R+1]+\mathop\sum\limits_{v\in \mathcal{V}}(\gamma\log N/R+1)\\
  =&N(\frac{\bar{d}_\mathcal{G}}{R}+1)+N(\gamma\log N/R+1).
\end{split}
\end{equation}
\textcolor{black}{In Appendix~\ref{PofT2} the remaining part of the theorem, i.e., Eq.~\eqref{error_polynomial}, is proved in detail.}
\end{IEEEproof}

\begin{remark}\label{alg1_intuition}
\textcolor{black}{The nodes in $\mathcal{B}_{\mathcal{T}}$ all have a descendent size $\left|\mathcal{D}_v\right| < \gamma\log N$, and hence they do not have enough data to use powerful error control codes with large code length, unless dummy bits are inserted. The code length $\gamma \log N$ is to ensure that, the probability that all transmissions in $\mathcal{B}_{\mathcal{T}}$ are reliable, decays polynomially with $N$ under the union bound. The nodes in $\mathcal{A}_{\mathcal{T}}$ all have large descendent size, so they can use powerful error control codes to carry out block transmissions with low error probability.}
\end{remark}
\subsection{Comparison between the Upper Bound and the Lower Bound}
Clearly, when the average distance $\bar{d}_\mathcal{G}$ to the sink is large and \textcolor{black}{grows polynomially with $N$}, the first term in the RHS of~\eqref{diameter_up_bnd} dominates. Thus, the upper bound is \textcolor{black}{the same order as} the lower bound in Theorem~\ref{Lg_Dia_Lw_Bd} when the average multi-hop distance $\bar{d}_\mathcal{G}$ is large. In this section, we make a summary of results both in this paper and~\cite{Goy_SJC_08} and discuss the tightness of the obtained scaling results in different cases.
\begin{corollary}\label{corollary1}
Suppose the communication links in the graph $\mathcal{G}$ satisfy the assumption (A.1a). Then, the communication complexity $\mathscr{C}^\text{(N)}$ of data gathering has an upper bound $\overline{\mathscr{C}^\text{(N)}}$ and an lower bound $\underline{\mathscr{C}^\text{(N)}}$, satisfying
\begin{eqnarray}
  &\overline{\mathscr{C}^\text{(N)}}=\max\{\Theta(\bar{d}_\mathcal{G}N),\Theta(N\log N)\}, \label{UpB}\\
  &\underline{\mathscr{C}^\text{(N)}}=\max\{\Theta(\bar{d}_\mathcal{G}N),\Theta(N\log\log N)\}. \label{LowerB}
\end{eqnarray}
\end{corollary}
\begin{IEEEproof}
Considering~\eqref{diameter_lw_bnd} and \eqref{diameter_up_bnd}, to prove~\eqref{LowerB}, it suffices to show $\mathscr{C}^\text{(N)}=\Omega(N \log\log N)$. \textcolor{black}{In fact, it is stated in Theorem 1 in [5] that if the number of noisy broadcasts is
\[\mathscr{C}^\text{(N)}=\beta(N)N,\]
the error probability $P_e^{(N)}$ that the receiver does not output all self-information bits satisfies
\begin{equation}\label{der1}
1-P_e^{(N)}<\sqrt{\frac{1}{N}}+\frac{48\beta^2\log(1/\epsilon)}{\epsilon^{4\beta}\log N}.
\end{equation}
Then, we have
\[\begin{split}
\text{Inequality \eqref{der1}}\Longleftrightarrow &\left(1-P_e^{(N)}-\sqrt{\frac{1}{N}}\right)\frac{\log N}{48\log(1/\epsilon)}<\frac{\beta^2}{\epsilon^{4\beta}}\\
\Longleftrightarrow & \log\log N +\log \left(1-P_e^{(N)}-\sqrt{\frac{1}{N}}\right)-\log \left(48\log\left(\frac{1}{\epsilon}\right)\right)\\
&<2\log \beta+4\beta \log\left(\frac{1}{\epsilon}\right).\\
\end{split}
\]
Dividing both the LHS and the RHS with $4\log\left(\frac{1}{\epsilon}\right)$, we have
\begin{equation}\label{Saks_beta}
\begin{split}
  \beta+\frac{\log \beta}{2\log \frac{1}{\epsilon}}>&\frac{\log\log N}{4\log\frac{1}{\epsilon}} +\frac{\log(1-P_e^{(N)}-\sqrt{\frac{1}{N}})-\log(48\log(\frac{1}{\epsilon}))}{4\log\frac{1}{\epsilon}}\\
  =&\Omega(\log\log N).
\end{split}
\end{equation}
From \eqref{Saks_beta}, we immediately have $\beta(N)=\Omega(\log\log N)$.}
\end{IEEEproof}

The lower bound is tight in the order sense in many cases. An example to support this claim in the low-diameter regime is the $\Theta(N\log\log N)$ communication complexity upper bound obtained in complete graphs in~\cite{Gal_TIT_88}. An example in the high-diameter regime is the grid network studied in~\cite{Kara_TIT_11}. Theorem IV.1 and Theorem IV.2 in~\cite{Kara_TIT_11} prove that in a ${\sqrt{n}\times\sqrt{n}}$ grid broadcast network with a transmission radius $r$, the communication complexity for data gathering is $\max\{\Theta(N^{3/2}/r),\Theta(N \log\log N)\}$, which matches the lower bound in this section, if the fact that the typical diameter for this network is $\sqrt{N}/r$ is considered.

\textcolor{black}{However, the upper bound obtained by the $\mathcal{GC}$-1 algorithm might not be tight in all occasions.} For example, it is apparently loose when $\bar{d}_\mathcal{G}<2$. To show this claim, plug in $\bar{d}_\mathcal{G}<2$ into~\eqref{UpB} and~\eqref{LowerB}. Then, we know that $\overline{\mathscr{C}^\text{(N)}}=\Theta(N\log N)$, $\underline{\mathscr{C}^\text{(N)}}=\Theta(N\log\log N)$. This mismatch is because the $\mathcal{GC}$-1 algorithm is designed for general graph topologies and is not adaptive in specific graph topologies. However, the $\mathcal{GC}$-1 algorithm can be improved in specific graph topologies to meet the lower bound. We use the following example to show a basic topology structure that helps achieve the lower bound, which motivates the geometric graph in the next section.

\begin{example}
Consider the examples shown in Fig.~\ref{fig2}. We abuse the terminologies, and use `heavy-tail' to describe the case of (a), and use `light-tail' to describe the case of (b). Suppose in the heavy-tail star network (a), there are $\gamma\log N$ nodes that form a clique (complete graph) on the end of each tail, where $\gamma$ satisfies the condition \eqref{Rate_Cond} in Theorem~\ref{Lg_Dia_Up_Bd}. All nodes in these cliques form the set $\mathcal{B}_{\mathcal{T}}$ defined in~\eqref{B_T}. Then, we modify Algorithm~\ref{alg1} by letting each $v\in \mathcal{B}_{\mathcal{T}}$ broadcast $j_t=\lceil\frac{2\log(\gamma \log N/p_{\text{ch}})}{\log[1/4\epsilon(1-\epsilon)]}\rceil$ times to all other nodes in the clique that $v$ lies in, where $p_{\text{ch}}<1/2$ is a constant. This modification changes the number of broadcasts at each node $v\in \mathcal{B}_{\mathcal{T}}$ from~\eqref{CC_Each_Node_B} to $\mathscr{C}_v=j_t=\Theta(\log\log N)$. Therefore, the total number of broadcasts is changed from~\eqref{Large_d_cmplx} to
\begin{equation}\label{Net1_CC}
\begin{split}
  \mathscr{C}_\mathscr{S}^\text{(N)}<&j_t |\mathcal{B}_{\mathcal{T}}|+\mathop\sum\limits_{v\in \mathcal{A}_\mathcal{T}}[(\mathcal{D}_v+1)/R+1]=\max\{\Theta(N\log\log N),\Theta(\bar{d}_\mathcal{G}N)\},
\end{split}
\end{equation}
which achieves the lower bound~\eqref{LowerB}. It can be shown that the overall probability of getting an error in the broadcasts in all cliques decays polynomially with $N$ (see Section~\ref{geo_net_analysis_sec}).

However, for the light-tail network shown in Fig.~\ref{fig2} (b), there is no convenient structure to be utilized for a broadcast. When the length of each tail is greater than $\gamma \log N$, we can use error control coding for the nodes in $\mathcal{A}_{\mathcal{T}}$, but the nodes in $\mathcal{B}_{\mathcal{T}}$ (nodes that are close to the tail ends) can only insert dummy bits to obtain large code length in order to ensure reliability. This issue limits the number of broadcasts to scale as $\mathscr{C}_\mathscr{S}^\text{(N)}=\max\{\Theta(N\log N),\Theta(\bar{d}_\mathcal{G}N)\}$. Further, when the length of each tail is smaller than $\gamma \log N$, we can only use error control coding with dummy bits and length $\Theta(\log N)$ at all nodes, since all nodes are in $\mathcal{B}_{\mathcal{T}}$. This limits the number of broadcasts to scale as $\mathscr{C}_\mathscr{S}^\text{(N)}=\Theta(N\log N)$. Therefore, the total number of broadcasts \textcolor{black}{has the same form as~\eqref{UpB}. and does not reach the lower bound.}
\begin{figure}
  \centering
  \includegraphics[scale=0.3]{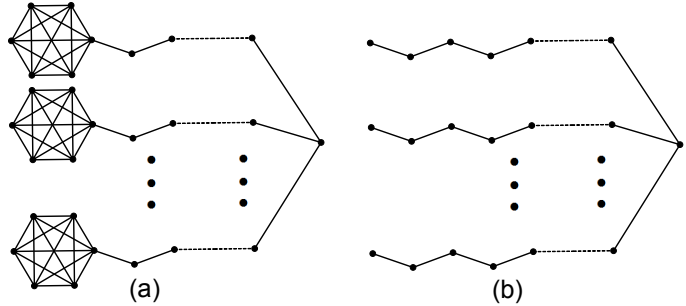}\\
  \caption{\emph{(a) A heavy-tail star network that achieves the lower bound~\eqref{LowerB}; (b) A light-tail star network that cannot achieve the lower bound.}}\label{fig2}
\end{figure}
\end{example}

\begin{remark}\label{Ex_achv_lb}
The heavy-tail structure in Fig.~\ref{fig2} (a) is the basic structure that achieves the $\Theta(\bar{d}_\mathcal{G}N)$ upper bound by using error control coding. This is essentially the structure considered in~\cite[Theorem IV.2]{Kara_TIT_11}. In the following section, we consider a general geometric graph, which is essentially a generalization of this heavy-tail network structure.
\end{remark}

\subsection{Extension to Binary Erasure Channels}
The conclusion of the previous section can be easily generalized to BECs.
\begin{corollary}\label{corollary_BEC}
Suppose the communication links in the graph $\mathcal{G}$ satisfy the assumption (A.1b). Suppose the parameters of Algorithm~\ref{alg1} is the same as in Theorem~\ref{Lg_Dia_Up_Bd}, except that all the error probability $\epsilon$ is changed into $\epsilon/2$. Then, using Algorithm~\ref{alg1}, we can achieve polynomially decaying error probability. The number of broadcasts $\mathscr{C}_\mathscr{S}^\text{(N)}$ has an upper bound that is the same with~\eqref{diameter_up_bnd}.
\end{corollary}
\begin{IEEEproof}
Note that when a bit is erased by the BEC, we can always flip a fair coin and assign a random binary value to this bit on the receiver side. The equivalent channel of combining a BEC with erasure probability $\epsilon$ and a fair coin flip is a BSC with crossover probability $\epsilon/2$. Thus, all conclusions of Theorem~\ref{Lg_Dia_Up_Bd} holds, after the crossover probability of all BSCs are changed to $\epsilon/2$.
\end{IEEEproof}

\textcolor{black}{
\begin{corollary}\label{corollary_BEC_lb}
Suppose the communication links in the graph $\mathcal{G}$ satisfy the assumption (A.1b). Then, if all data are gathered at the sink $v_0$ with error probability $P_e^{(N)}$ by a feasible broadcasting scheme $\mathscr{S}$, the communication complexity is necessarily bounded below by
\begin{equation}
  \mathscr{C}^\text{(N)}\ge c_\epsilon\bar{d}_\mathcal{G}N,
\end{equation}
where $c_\epsilon=\frac{1-H(P_e^{(N)})}{1-\epsilon}$ is a constant, $N$ denotes the number of nodes in the graph and $\bar{d}_\mathcal{G}$ is the average distance to the sink, defined as
\begin{equation}
  \bar{d}_\mathcal{G}=\frac{1}{N}\sum\limits_{n=1}^N d(v_n,v_0).
\end{equation}
\end{corollary}
\begin{IEEEproof}
The proof is almost exactly the same as the one of Theorem~\ref{Lg_Dia_Lw_Bd}. The only difference is that the channel capacity of each BEC link is $1-\epsilon$ instead of $1-H(\epsilon)$.
\end{IEEEproof}}

The upper bound that we obtained in Corollary \ref{corollary_BEC} is $\max\{\Theta ( \bar{d}_\mathcal{G} N), \Theta(N \log N)\}$. Therefore, the two bounds meet with each other when $\bar{d}_\mathcal{G}=\Omega(log N)$, which holds in many networks, such as a square grid network of size $\sqrt{N}\times \sqrt{N}$.

In fact, one can show that in some other types of graphs, the lower bound $\Omega(N\log N)$ is also valid for BEC models. We will show in the following that for a network with constant degree, $\Omega(N\log N)$ is a valid lower bound on the number of broadcasts for the noisy broadcast problem. Therefore, for this particular type of networks, $\max\{\Theta ( \bar{d}_\mathcal{G} N), \Theta(N \log N)\}$ is both the upper and lower bound on the number of broadcasts.

\begin{lemma}
Suppose each node $v\in\mathcal{V}$ in the graph $\mathcal{G}=(\mathcal{V},\mathcal{E})$ satisfies $\text{deg}(v)\le D$ and $D$ is a constant. Then, for any scheme $\mathscr{S}$ to obtain an output $\hat{\mathbf{x}}$ with constant error probability $\Pr(\hat{\mathbf{x}}\neq \mathbf{x})<\delta$ in the noisy broadcast problem, the number of broadcasts satisfies $\mathscr{C}_\mathscr{S}^{(N)}=\Omega(N\log N)$.
\end{lemma}

\begin{IEEEproof}
For an arbitrary node $v\in\mathcal{V}$, suppose the number of broadcasts by $v$ is $\mathscr{C}_{v}$. Then, since the degree of $v$ satisfies $\text{deg}(v)\le D$, the probability $p_v$ that all broadcasts made by $v$ are erased is lower bounded by
\begin{equation}
p_v\ge \epsilon^{D\mathscr{C}_{v}},
\end{equation}
where recall that $\epsilon$ is the erasure probability. If all broadcasts from one particular node are erased, the sink can never recover the entire input bits $\mathbf{x}$. Therefore,
\begin{equation}
P_e^{(N)}=\Pr(\hat{\mathbf{x}}\neq \mathbf{x})\ge 1-\prod_{v\in\mathcal{V}}(1-p_v)\ge 1-\prod_{v\in\mathcal{V}}(1-\epsilon^{D\mathscr{C}_{v}}).
\end{equation}
This implies that
\begin{equation}
\begin{split}
1-\delta<1-P_e^{(N)}\le \prod_{v\in\mathcal{V}}(1-\epsilon^{D\mathscr{C}_{v}})\le \left[\frac{1}{N}\sum_{v\in\mathcal{V}}(1-\epsilon^{D\mathscr{C}_{v}})\right]^N=\left[1-\frac{1}{N}\sum_{v\in\mathcal{V}}\epsilon^{D\mathscr{C}_{v}}\right]^N\\
\le {\left( {1 -{{\epsilon}^{\frac{1}{N}\sum_{v\in\mathcal{V}} {D\mathscr{C}_{v}} }}} \right)^N}
 \overset{(a)}{\le} \exp {\left( { - N \cdot {{\epsilon}^{\frac{1}{N}\sum_{v\in\mathcal{V}} {D\mathscr{C}_{v}} }}} \right)},
\end{split}
\end{equation}
where (a) is from $1-x\le \exp(-x)$. Rearranging the terms in the above inequality gives
\begin{equation}
\sum_{v\in\mathcal{V}} {\mathscr{C}_{v}}>\frac{N}{D}\cdot\frac{\log N-\log\log(1/(1-\delta))}{\log(1/\epsilon)}=\Omega(N\log N).
\end{equation}
\end{IEEEproof}
\section{$\mathcal{GC}$-2 Graph Codes in a Geometric Graph}\label{geometric_graph}
In the previous section, we considered the communication complexity problem in a general graph where the upper and lower bound has a $\Theta(\log N)$ gap. In the following sections, we consider graphs where the communication complexity lower bound $\max\{\Theta(\bar{d}_\mathcal{G}N),\Theta(N\log\log N)\}$ can indeed be achieved. The in-network computing algorithm utilized in this section is the $\mathcal{GC}$-2 graph code. \textcolor{black}{Our $\mathcal{GC}$-2 graph code in this section is based on cell partitioning in geometric graphs and node replication. In particular, we partition all nodes in the network into cells based on geographic location. If we can partition nodes into groups of $\Theta(\log N)$ and each group forms a local complete graph, we can then use similar ideas from \cite{Goy_SJC_08} to aggregate data, i.e., we can use a short code of length $\Theta(\log N)$ to aggregate data reliably in a local complete graph. However, for general geometric graphs, this partitioning does not apply directly. Thus, we introduce ``dummy nodes'', so that the number of nodes in each cell always exceeds $\Theta(\log N)$. For the geometric graph with connection distance $r>\sqrt{\frac{c_g\log N}{N}}$ (see Assumption (A.3a)) that we consider, the introduction of dummy nodes does not change the number of broadcasts in order sense.}
\begin{figure}
  \centering
  \includegraphics[scale=0.6]{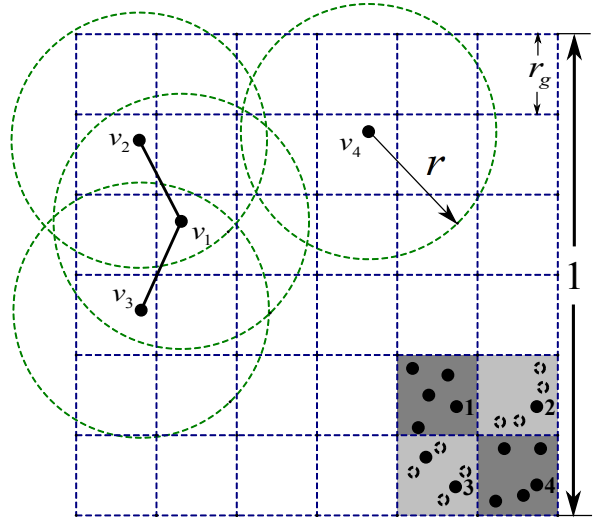}\\
  \caption{\emph{A geometric graph where each node can only broadcast within a certain distance. The node $v_1$ can broadcast to $v_2$ and $v_3$ but cannot broadcast to $v_4$. Cells 1 and 4 are grouped into dense set $\mathcal{S}_{d}$ while cells 2 and 3 are grouped into sparse set $\mathcal{S}_{s}$. Dashed circles around a solid node represents the replications of this node, i.e., dummy nodes.}}\label{fig6}
\end{figure}

We first consider geometric graphs \textcolor{black}{$\mathcal{G}=(\mathcal{V},\mathcal{E})$ that satisfy the connectivity assumption (A.2), the topology assumption (A.3a) and the channel assumption (A.1a). Extensions to random geometric graphs with assumption (A.3b) and BECs with (A.1b) are given in Section~\ref{RG_BEC}.} As shown in Fig~\ref{fig6}, we use a square tessellation scheme to partition the 1-by-1 area into $B^2=\lceil \frac{\sqrt{2}}{r}\rceil^2$ small squares, such that each square has length $r_g<\frac{r}{\sqrt{2}}$. We call each small square a \emph{cell}. By \textcolor{black}{the assumption of} a geometric graph, each node is connected to all other nodes in the same cell since the diagonal of each cell is smaller than $r$. \textcolor{black}{Therefore, for any two nodes $v$ and $v'$ in the same cell, the difference between the multi-hop distances to the sink $v_0$ satisfies $|d(v,v_0)-d(v',v_0)|\le 1$. According to the topology assumption (A.3a), $r>\sqrt{\frac{c_g\log N}{N}}$, and hence
\begin{equation}\label{Bsquare}
  B^2<\left(\sqrt{\frac{2N}{c_g \log N}}+1\right)^2.
\end{equation}
\textcolor{black}{Denote by $\mathcal{V}_l$ the nodes that belong to a particular cell indexed by $l$. Then, the node set $\mathcal{V}$ is divided into $B^2$ subsets, which is written as}
\begin{equation}\label{Set_div}
  \mathcal{V}=\mathop\bigcup\limits_{l=1}^{B^2} \mathcal{V}_l.
\end{equation}
We use $\mathcal{S}=\{1,2,\dots,B^2\}$ to denote the cell index set. In each cell, call the node with the minimum distance to the sink $v_0$ the \emph{cell head}. }Separate $\mathcal{S}$ into two parts $\mathcal{S}=\mathcal{S}_{d}\cup\mathcal{S}_{s}$, where
\begin{equation}\label{dense_s}
  \mathcal{S}_{d}=\{l\in \mathcal{S}||\mathcal{V}_l|>\rho \log N\},
\end{equation}
and $\mathcal{S}_{s}=\mathcal{S}\setminus \mathcal{S}_{d}$, \textcolor{black}{where $\rho$ is a constant. It is clear that} $\mathcal{S}_{d}$ denotes the cells where the nodes are dense and $\mathcal{S}_{s}$ denotes the opposite (see Fig.~\ref{fig6}).

\textcolor{black}{
For each cell $l$ in the set of dense cells $\mathcal{S}_{d}$, partition $\mathcal{V}_l$ into groups, such that the number of nodes in each group ranges between $\rho \log N$ and $2\rho \log N$. Each group is located in the same cell, and hence forms a local complete graph. For each cell $l$ in the set of sparse cells $\mathcal{S}_{d}$, if $\mathcal{V}_l\neq \emptyset$, replicate each node, together with the held self-information bit, for $\lceil \frac{\rho \log N}{|\mathcal{V}_l|}\rceil$ times. Therefore, in each cell, there are $|\mathcal{V}_l|\lceil \frac{\rho \log N}{|\mathcal{V}_l|}\rceil\in [\rho \log N,2\rho \log N]$ replicated \emph{dummy nodes} (see Fig.~\ref{fig6}). Each dummy node holds one \emph{dummy self-information bit} that replicates the original self-information bit. We assume that all the dummy nodes in a cell $l$ form a single group. In the following sections, we provide the algorithm for the graph with dummy nodes. But it should always be clear that all actions taken by a dummy node is actually implemented by the original physical node. By introducing dummy nodes, all nodes in $\mathcal{V}$ are partitioned into groups of $\Theta(\rho\log N)$ nodes, and all nodes in one group form a local complete graph. The reason to partition $\mathcal{V}$ into approximately even groups of size $\Theta(\log N)$ is to ensure the polynomial decay of the final error probability with $N$, which will be explained soon in detail. Suppose the total number of groups is $D$. Denote by $v_\mathcal{A}$ the cell head of the corresponding cell that the group $\mathcal{A}$ is located in.
}
\subsection{\textcolor{black}{In-network Computing Algorithm}}\label{geo_net_alg_sec}
The algorithm for data gathering in a geometric graph has two steps. In the first step, in each cell, all self-information bits are gathered in the cell head, using a $\mathcal{GC}$-2 code. In the second step, a backbone network constituted by cell heads is constructed, in order to route all information to the sink $v_0$.

We first design the generator matrix of the $\mathcal{GC}$-2 graph code. As defined in Section~\ref{main_technique}, a $\mathcal{GC}$-2 graph code is a binary linear block code with a generator matrix $\tilde{\mathbf{G}}=[\mathbf{I},\mathbf{\tilde{A}}]$, where $\mathbf{\tilde{A}}$ is the adjacency matrix of a subgraph of the original graph. The structure of this generator matrix ensures that each code bit can be calculated by local information exchanges. Here we design the generator matrix to be $\tilde{\mathbf{G}}=[\mathbf{I},\mathbf{\tilde{A}}]$, such that $\mathbf{\tilde{A}}$ is a block diagonal matrix written as
\begin{equation}\label{block_diagonal}
  \mathbf{\tilde{A}}=\text{Diag}\{\tilde{A}_1,\tilde{A}_2,\dots,\tilde{A}_{D}\},
\end{equation}
where $D$ is the number of groups, and the $m$-th block $\tilde{A}_m$ is a square matrix that has the same size as the $m$-th group. This definition is always valid, because each group forms a local complete graph, and since a sub-graph of a complete graph can have arbitrary topology, we know that each block $\tilde{A}_m$ can be arbitrary, as long as it is symmetric and its size is the same as the size of the $m$-th group. In the algorithm to be shown, we require each group to distributedly encode all of its self-information bits with a generator matrix $\tilde{G}_m=[I,\tilde{A}_m]$, which means that we are decoupling the encoding with matrix $\tilde{\mathbf{G}}$ into local computations in each group (and equivalently, in each cell) based on the block diagonal structure of $\mathbf{\tilde{A}}$. This distributed encoding can clearly be done with local information exchange. It should be noted that the matrix $\mathbf{\tilde{A}}$ is not actually the adjacency matrix of a subgraph of the original graph $\mathcal{G}$, but of the augmented graph with replicated dummy nodes in sparse cells.

Apart from to be block diagonal, we further require each block $\tilde{A}_m$ of $\mathbf{\tilde{A}}$ to satisfy the property that a systematic code with the generator matrix $\tilde{G}_m=[I,\tilde{A}_m]$ achieves the random coding exponent in Lemma~\ref{BSC_rm_cd}, which ensures that each local encoding process yields a codeword with powerful error correcting capabilities. The formal description of the local encoding scheme, or the local gathering scheme, is given in the `Local Computing' part of Algorithm~\ref{alg2}. In each group $\mathcal{A}_m$, as mentioned above, a rate-$\frac{1}{2}$ graph code with the generator matrix $\tilde{G}_m=[I,\tilde{A}_m]$ and code length $2|\mathcal{A}_m|>2\rho\log N$ is utilized to distributedly encode all data in this group and ensure reliable decoding in the cell head. In particular, denote by $\mathbf{x}_m$ the vector that contains all self-information bits in group $\mathcal{A}_m$. Then, the encoding yields $\mathbf{x}_m^\top\tilde{G}_m=[\mathbf{x}_m^\top,\mathbf{x}_m^\top\tilde{A}_m]$. Thus, each code bit can be calculated using local broadcasts, since each code bit is either a self-information bit, or the parity of some self-information bits in a local complete graph.

It should be noted that in a sparse cell, although the actual actions are taken by the physical nodes, these actions can be viewed as being performed by the dummy nodes without changing the statistical properties of the distributed encoding scheme. \textcolor{black}{To be specific, let each physical node broadcast its self-information bit for $j_g\cdot\lceil \frac{\rho \log N}{|\mathcal{B}|}\rceil$ times (here $\mathcal{B}$ is the only group in the sparse cell and is the same as $\mathcal{V}_l$ in \eqref{dense_s}), which is the same with letting each dummy node broadcast $j_g$ times.} Then, each dummy node receives all other bits, and computes one code bit signified by the local graph code generator matrix $\tilde{G}_m$. Finally, all these code bits and all self-information bits (all the dummy bits) are transmitted to the cell head to be decoded. The broadcast channel between two replications (dummy nodes) of the same physical node is actually a perfect channel, which only incurs less errors.
\begin{figure}
  \centering
  \includegraphics[scale=0.6]{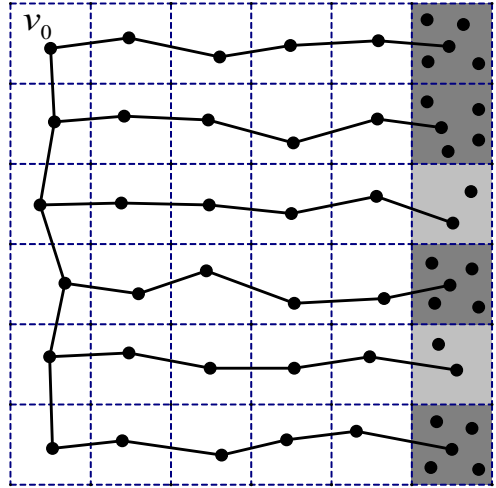}\\
  \caption{\emph{A geometric graph routing backbone. Both sparse and dense cells introduce dummy bits to facilitate routing.}}\label{geo_net_backbone}
\end{figure}

In the routing step, all self-information bits are routed along a backbone network. The algorithm is given in the `Backbone Routing' part in Algorithm~\ref{alg2}. As explained in the algorithm, if two cell heads $v_i$ and $v_j$ may interact with a path $v_i\rightarrow v_i'\rightarrow v_j'\rightarrow v_j$, they are defined to be connected in the backbone network. Since the underlying network $\mathcal{G}=(\mathcal{V},\mathcal{E})$ is connected, the backbone network is also connected. Each node in the spanning tree $\mathcal{T}$ of the backbone network is a cell head and has the task of forwarding all the self-information bits in the corresponding cell. Furthermore, each node in the backbone network has to relay all information bits from its children-nodes (direct descendants) in $\mathcal{T}$ as well. This decode-and-forward routing is carried on along the entire spanning tree $\mathcal{T}$ of the backbone network, until the sink $v_0$ receives all the data, i.e., all the self information bits in the whole network. The number of dummy bits introduced in each cell head $v$ is smaller than $\rho\log N$. As shown in the proof, the reason to append $\mathcal{O}(\rho\log N)$ dummy bits in each node $v\in \mathcal{T}$ at the end of the `Backbone Routing' part of Algorithm~\ref{alg2} is to ensure the polynomial decay of error probability with $N$. Moreover, appending $\mathcal{O}(\rho\log N)$ dummy bits in each cell does not change the number of broadcasts in the order sense.

\begin{algorithm}\caption{\textcolor{black}{Identity-Function Computation in a Geometric Graph}}\label{alg2}
\begin{itemize}
  \item \textbf{Initialization:} For all cell $l\in \mathcal{S}_{d}$, partition $\mathcal{V}_l$ into groups where each group has $\rho \log N$ to $2\rho \log N$ nodes. For all cell $l\in \mathcal{S}_{s}$, replicate each node $\lceil \frac{\rho \log N}{|\mathcal{V}_l|}\rceil$ times and form one group in this cell.
  \item \textbf{Local Computing:} For each group $\mathcal{A}_m$, first let each node in $\mathcal{A}_m$ broadcast its self-information bit for $j_g$ times where
   \begin{equation}\label{dense_j}
      j_g=\lceil\frac{2\log(2\rho \log N/p_{\text{ch}})}{\log[1/4\epsilon(1-\epsilon)]}\rceil,
   \end{equation}
   and $p_{\text{ch}}<1/2$ is a constant.

   Secondly, each node computes one code bit using the corresponding column in $\tilde{A}_m$, the $m$-th sub-matrix of the generator matrix $\tilde{\mathbf{G}}$. For example, the $i$-th node $v_{m,i}$ in the $m$-th group calculates
   \begin{equation}\label{local_encoding_each_node}
     y_{m,i}=\hat{\mathbf{x}}_{m,i}\tilde{\mathbf{a}}_{m,i},
   \end{equation}
   where $\tilde{\mathbf{a}}_{m,i}$ is the $i$-th column of $\tilde{A}_m$, and $\hat{\mathbf{x}}_{m,i}$ is the majority-rule-based estimate of $\mathbf{x}_{m,i}$ at $v_{m,i}$ based on information received by $v_{m,i}$ from the first step of local computing, where $\mathbf{x}_{m,i}$ is the self-information bits of the nodes in group $\mathcal{A}_m$. The summation is in the sense of modulo-2. Then, all these code bits are transmitted to the node $v_\mathcal{A}$, the cell head.

   Thirdly, each node in the group transmits its own self-information to $v_\mathcal{A}$.

   Finally, $v_\mathcal{A}$ performs decoding on the received bits to recover all self-information bits in group $\mathcal{A}$.
  \item \textbf{Backbone Routing:} Construct the backbone network constituted by all cell heads in the following way: two cell heads $v_i$ and $v_j$ in cell $i$ and cell $j$ are connected if there exist one node $v_i'$ in cell $i$ and $v_j'$ in cell $j$ such that the pair $(v_i',v_j')$ is in the original edge set $\mathcal{E}$.

      Construct the breadth-first-search spanning tree $\mathcal{T}$ of the backbone network rooted at the sink $v_0$. Each node relays all information bits from itself and its descendants in $\mathcal{T}$ to its parent-node.

      Suppose a node $v\in \mathcal{T}$ needs to route a binary vector $\mathbf{x}$. Then, $v$ partitions $\mathbf{x}$ into blocks with length $\rho\log N$. If the length of $\mathbf{x}$ is not a multiple of $\rho\log N$, some dummy bits are appended into $\mathbf{x}$. After that, $v$ encodes each block with a rate-$R$ block code and transmits the codeword to the parent-node.
\end{itemize}
\end{algorithm}

\subsection{Upper Bounds on the Error Probability and the Number of Broadcasts}\label{geo_net_analysis_sec}
In the following lemma, we analyze the error probability and the number of broadcasts in the local computing step.
\begin{lemma}\label{local_gathering}
Assume $\mathcal{G}=(\mathcal{V},\mathcal{E})$ satisfies the topology assumption (A.3a) and the channel assumption (A.1a). Further assume that
\begin{equation}\label{local_condition}
  4\rho E_r(\epsilon+p_\text{ch},\frac{1}{2})>1,
\end{equation}
where $\epsilon$ is the channel crossover probability, $p_\text{ch}<1/2$ is a constant defined in~\eqref{dense_j}, $\rho$ is the constant defined in~\eqref{dense_s} and $E_r(\cdot)$ is the random coding exponent for BSCs. Then, using the local computing step in Algorithm~\ref{alg2}, i.e., the $\mathcal{GC}$-2 code, and using $\Theta(N\log\log N)$ number of broadcasts, all cell heads learn all the self-information bits in their own cells with high accuracy, that is, the total error probability in the local computing step, $P_{e,\text{local}}=\Pr(\exists v\in\mathcal{T},v$\text{ has a wrong decoding output}$)$, eventually decays polynomially with $N$.
\end{lemma}
\begin{IEEEproof}
According to Lemma~\ref{BSC_rept}, after transmitting each bit $x$ for $j_g$ times (defined in \eqref{dense_j}), the bit $x$ is erroneous with error probability
\begin{equation}\label{Geo_Der_1}
  P_e<[4\epsilon(1-\epsilon)]^{\frac{j_g}{2}}<\frac{p_\text{ch}}{2\rho\log N}.
\end{equation}
Since each code bit calculated at a node $v_{m,i}$ is the XOR of at most $2\rho\log N$ self-information bit (defined in~\eqref{local_encoding_each_node}), by the union bound, each code bit of the $\mathcal{GC}$-2 code is encoded incorrectly with probability
\begin{equation}\label{Geo_Der_2}
  P_{e,v}=\Pr(\hat{\mathbf{x}}_{m,i}\neq \mathbf{x}_{m,i})<\frac{p_\text{ch}}{2\rho\log N}\cdot 2\rho\log N=p_\text{ch}.
\end{equation}
By Lemma~\ref{BSC_rm_cd}, this makes the error probability of recovering all self-information bits in group $\mathcal{A}$ at the cell head $v_{\mathcal{A}}$ be upper bounded by
\begin{equation}\label{Geo_Der_3}
  P_{e,\text{local}}^{\mathcal{A}}<\exp[-4\rho\log N E_r(\epsilon+p_\text{ch},\frac{1}{2})]=N^{-4\rho E_r(\epsilon+p_\text{ch},\frac{1}{2})}.
\end{equation}
The error probability upper bound is the same for sparse and dense cells. Therefore, the total error probability is
\begin{equation}\label{Pe_local}
\begin{split}
  P_{e,\text{local}}=&\mathop\sum\limits_{\mathcal{A}\text{ is in a dense cell}}P_{e,\text{local}}^{\mathcal{A}}+
  \mathop\sum\limits_{\mathcal{B}\text{ is in a sparse cell}}P_{e,\text{local}}^{\mathcal{B}}\\
  \overset{(a)}{<}&N^{-4\rho E_r(\epsilon+p_\text{ch},\frac{1}{2})} \left(\mathop\sum\limits_{\mathcal{A}\text{ is in a dense cell}} 1+\mathop\sum\limits_{\mathcal{B}\text{ is in a sparse cell}} 1\right)\\
  \overset{(b)}{<}& N^{-4\rho E_r(\epsilon+p_\text{ch},\frac{1}{2})}\left[\frac{N}{\rho\log N}+\left(\sqrt{\frac{2N}{c_g \log N}}+1\right)^2\right],
\end{split}
\end{equation}
which eventually decays polynomially with $N$ when $4\rho E_r(\epsilon+p_\text{ch},\frac{1}{2})>1$. Note that step (a) follows from~\eqref{Geo_Der_3} and step (b) follows from the fact that the number of groups in dense cells is upper bounded by $\frac{N}{\rho\log N}$ and the number of groups in sparse cells is upper bounded by the total number of cells $B^2$ in~\eqref{Bsquare}.

The number of broadcasts consumed by group $\mathcal{A}$ in a dense cell is
\begin{equation}\label{Cmpl_A_local}
  \mathscr{C}^{\mathcal{A}}_{\text{local}}=(j_g+2)|\mathcal{A}|,
\end{equation}
where $|\mathcal{A}|$ denotes the number of nodes in group $\mathcal{A}$, and the constant $2$ is because each node, apart from broadcasting its own self-information bit for $j_g$ times, has to transmit a code bit and its own self-information bit to the cell head. Similarly, consider the fact that each node is replicated into $\lceil \frac{\rho \log N}{|\mathcal{B}|}\rceil$ dummy nodes, we know that the number of broadcasts consumed by group $\mathcal{B}$ in a sparse cell is
\begin{equation}\label{Cmpl_B_local}
  \mathscr{C}^{\mathcal{B}}_{\text{local}}=(j_g+2)\cdot |\mathcal{B}|\lceil \frac{\rho \log N}{|\mathcal{B}|}\rceil.
\end{equation}
Since
\begin{equation}
  |\mathcal{B}|\lceil \frac{\rho \log N}{|\mathcal{B}|}\rceil<|\mathcal{B}|( \frac{\rho \log N}{|\mathcal{B}|}+1)=|\mathcal{B}|+\rho\log N<2\rho\log N,
\end{equation}
we have
\begin{equation}
  \mathscr{C}^{\mathcal{B}}_{\text{local}}<(j_g+2)\cdot2\rho\log N.
\end{equation}
Thus, the total number of broadcasts is
\begin{equation}\label{Cmpl_local}
\begin{split}
  \mathscr{C}_{\text{local}}=&\mathop\sum\limits_{\mathcal{A}}\mathscr{C}^{\mathcal{A}}_{\text{local}}+
    \mathop\sum\limits_{\mathcal{B}}\mathscr{C}^{\mathcal{B}}_{\text{local}}
  <(j_g+2)(\mathop\sum\limits_{\mathcal{A}}|\mathcal{A}|+\mathop\sum\limits_{\mathcal{B}}2\rho\log N)\\
  <&(j_g+2)\left[N+2\rho\log N \left(\sqrt{\frac{2N}{c_g \log N}}+1\right)^2\right]
  =\Theta(N\log\log N),
\end{split}
\end{equation}
where we have used $j_g=\Theta(\log\log N)$ in~\eqref{dense_j} and~\eqref{Bsquare}.
\end{IEEEproof}

Lemma~\ref{local_gathering} states that $\Theta(\log\log N)$ broadcasts suffice to make all cell heads successfully gather all local information. \textcolor{black}{After that, cell heads form a backbone network and all local information is routed to the sink $v_0$. The analysis of the whole Algorithm~\ref{alg2} is given in the following.}
\begin{theorem}\label{geo_thm}
Suppose $\mathcal{G}=(\mathcal{V},\mathcal{E})$ satisfies the topology assumption (A.3a) and the channel assumption (A.1a). Suppose the parameters of Algorithm~\ref{alg2} satisfy~\eqref{local_condition} and
\begin{equation}
  \frac{\rho}{R} E_r(\epsilon,R)>3/2,
\end{equation}
where the parameters $\rho$ and $\epsilon$ are defined the same as in Lemma~\ref{local_gathering}, $R$ is the code rate of backbone routing, and $E_r(\cdot)$ is the random coding exponent for BSCs, as defined in~\eqref{Gallager_Rm_Bd}. Then, using the in-network computing scheme defined in Algorithm~\ref{alg2}, in which the number of broadcasts scales as $\max\{\Theta(\bar{d}_\mathcal{G}N),\Theta(N\log\log N)\}$, the final error probability eventually decays polynomially with $N$.
\end{theorem}
\begin{IEEEproof}
We respectively analyze the total error probability and the total number of broadcasts. As defined in the backbone routing step of Algorithm~\ref{alg2}, each transmission relays a block of $\rho \log N$ bits with a rate-$R$ code. Thus, the error probability of each block on one transmission is bounded by
\begin{equation}\label{Geo_Der_3_2}
  P_{e,\text{routing}}^{B}<\exp(-\frac{\rho\log N}{R} E_r(\epsilon,R))=N^{-\frac{\rho}{R} E_r(\epsilon,R)}.
\end{equation}
In all, the number of appended dummy bits is at most $\rho\log N\cdot B^2$ (at most $\rho \log N$ dummy bits in each cell and $B^2$ cells), and hence the number of blocks is at most $\frac{N+\rho\log N\cdot B^2}{\rho\log N}$. Each block is transmitted along at most $3\cdot2B$ hops, where the multiple $3$ is because each path $v_i\rightarrow v_i'\rightarrow v_j'\rightarrow v_j$ between two cell heads is constituted by at most three hops in the underlying graph $\mathcal{G}$, and the multiple $2B$ is the longest multi-hop distance to the sink on a $B\times B$ grid. Using the union bound, the error probability that the sink $v_0$ gets a wrong version of all information bits is bounded from above by
\begin{equation}\label{routing_error}
  P_{e,\text{routing}}<6B\cdot\frac{N+\rho\log N\cdot B^2}{\rho\log N}N^{-\frac{\rho}{R} E_r(\epsilon,R)}=6B\cdot(\frac{N}{\rho \log N}+B^2)N^{-\frac{\rho}{R} E_r(\epsilon,R)}.
\end{equation}
Using the fact that $B=\mathcal{O}(N^{\frac{1}{2}})$, we know that the total routing error probability decays polynomially if $\frac{\rho}{R} E_r(\epsilon,R)>3/2$.

As for the number of broadcasts in the backbone routing phase, each bit from a cell head $v$ is now routed along a path on the backbone network, the length of which is at most 3 times the original distance $d(v,v_0)$. All appended dummy bits are only transmitted for one hop on the backbone network, or equivalently, at most three hops on the original network. Therefore, the total number of broadcasts for routing is
\begin{equation}\label{routing_cost}
\begin{split}
  \mathscr{C}_{\text{routing}}<&3\bar{d}_\mathcal{G}N+3\rho\log N\cdot B^2<
  3\bar{d}_\mathcal{G}N+3\rho\log N (\sqrt{\frac{2N}{c_g \log N}}+1)^2=\Theta(\bar{d}_\mathcal{G}N).
\end{split}
\end{equation}
Combining~\eqref{Pe_local}\eqref{Cmpl_local}\eqref{routing_error}\eqref{routing_cost}, we know that the overall error probability decays polynomially with $N$ and the number of broadcasts scales as $\max\left\{\Theta(\bar{d}_\mathcal{G}N),\right.$ $\left.\Theta(N\log\log N)\right\}$. Therefore, the proof is completed.
\end{IEEEproof}
\begin{remark}\label{remark_RGGtoGallager}
The proposed $\mathcal{GC}$-2 code can be viewed as an extension of the coding scheme in \cite[Section 7]{Goy_SJC_08} in complete graphs to arbitrary connected geometric graphs. In a complete graph, we can partition all nodes into non-overlapping cells of size $\Theta(\log N)$. Then, nodes in each cell form a complete graph of size $\Theta(\log N)$, which means that all nodes in the graph are in dense cells. In that case, we do not need to define dummy nodes. We do not need to construct the backbone network either, because all nodes in the network have direct links to the sink node. Therefore, the number of broadcasts in a complete network is $\Theta(N\log\log N)$.
\end{remark}

\subsection{\textcolor{black}{Extension to Random Geometric Graphs and BECs}}\label{RG_BEC}
A counterpart of Theorem~\ref{geo_thm} in random geometric graphs is the following corollary. This result generalizes the Theorem 2 in~\cite{Li_TMC_13} to cases when the connectivity range is larger than $\Theta(\sqrt{\frac{\log N}{N}})$. Note that in random graphs, we only care about the expected error probability $P_e^{(N)}=\mathbb{E}_\mathcal{G} [P_e^{\mathcal{G}}]$, which has been discussed at the end of Section~\ref{op_problem}.

\begin{corollary}\label{geo_thm_rm}(\cite[Theorem 2]{Li_TMC_13})Suppose $\mathcal{G}=(\mathcal{V},\mathcal{E})$ satisfies the topology assumption (A.3b) and the channel assumption (A.1a). Suppose the parameters of Algorithm~\ref{alg2} satisfy the same conditions as in Theorem~\ref{geo_thm}. Further assume that $c_g>\frac{1}{\pi}$. Then, using the in-network computing scheme in Algorithm~\ref{alg2}, we can obtain the identify function at the sink with high probability, and the number of broadcasts scales as $\max\{\Theta(\bar{d}_\mathcal{G}N),\Theta(N\log\log N)\}$. That is, the expected error probability $P_e^{(N)}=\mathbb{E}_\mathcal{G} [P_e^{\mathcal{G}}]$ goes down polynomially with $N$.
\end{corollary}
\begin{IEEEproof}
One possible way to prove this corollary is to use the same idea in~\cite{Li_TMC_13}, which relies on the result that in a random geometric graph satisfying the assumption (A.3b), after the tessellation step as shown in Fig.~\ref{fig6}, each cell has $\Theta(\log N)$ nodes with high probability. However, we present a different proof.

\begin{equation}\label{error_decoupling}
\begin{split}
  P_e^{(N)}=\mathbb{E}_\mathcal{G} [P_e^{\mathcal{G}}]<&\Pr(\mathcal{G}\text{ is connected})\mathbb{E}_\mathcal{G} [P_e^{\mathcal{G}}|\mathcal{G}\text{ is connected}]
  +[1-\Pr(\mathcal{G}\text{ is connected})].
\end{split}
\end{equation}
According to the conclusion of Theorem~\ref{geo_thm}, we know that, as long as the randomly generated graph $\mathcal{G}$ is connected, $P_e^{\mathcal{G}}$ decays polynomially with $N$. Moreover, we know from~\cite{Gir_CM_06} that the random geometric graph is connected with polynomially decaying probability as long as $c_g>\frac{1}{\pi}$. Thus, we obtain our claim.
\end{IEEEproof}

\begin{remark}
  The proof technique can be generalized easily to other extended random geometric graph distributions, if the connectivity assumption is satisfied with high probability. This is advantageous over the strict assumptions in~\cite{Li_TMC_13}, that nodes are all uniformly distributed .
\end{remark}

A counterpart of Theorem~\ref{geo_thm} with BECs can also be obtained.
\begin{corollary}\label{geo_thm_BEC}
Suppose $\mathcal{G}=(\mathcal{V},\mathcal{E})$ satisfies the topology assumption (A.3a) and the channel assumption (A.1b). Suppose the parameters of Algorithm~\ref{alg2} satisfy the same conditions as in Theorem~\ref{geo_thm},i.e., $\frac{\rho}{R} E_r(\epsilon,R)>\frac{3}{2}$, where $E_r(\epsilon,R)$ is the random coding error exponent of a BEC channel with rate $R$. Then, using the in-network computing scheme in Algorithm~\ref{alg2}, we can obtain the identity function at the sink with high probability, and the number of broadcasts scales as $\max\{\Theta(\bar{d}_\mathcal{G}N),\Theta(N\log\log N)\}$. That is, the error probability $P_e^{(N)}=\mathcal{O}(N^{-\frac{\rho}{R} E_r(\epsilon,R)+\frac{3}{2}})$.
\end{corollary}
\begin{IEEEproof}
The proof is exactly the same as the proof of Theorem~\ref{geo_thm}, except that the random coding error exponent is now of BEC channels instead of BSC channels.
\end{IEEEproof}

\section{$\mathcal{GC}$-3 Codes in a Low-Diameter Graph}\label{Upper_Bound}
In this section, we provide an in-network computing scheme when the graph diameter is low \textcolor{black}{(in particular, when the average multi-hop distance $\bar{d}_\mathcal{G}$ is a constant)} and the graph topologies are random, i.e., specifically, when the graph $\mathcal{G}$ satisfies the topology assumption (A.4) and the channel assumption (A.1b). \textcolor{black}{In this in-network computing scheme}, the number of broadcasts meets the general lower bound~\eqref{LowerB}\footnote{Note that the lower bound~\eqref{LowerB} is for BSCs and the techniques we use here are for BECs. However, even if the algorithm in~\cite{Gal_TIT_88} is applied to a complete graph with BECs, the number of broadcasts still scales as $\Theta(N \log\log N)$. Thus, our result is still better in that we allow non-complete graph topologies.}, with the assumption that $\bar{d}_\mathcal{G}$ has order $\mathcal{O}(1)$. \textcolor{black}{As noted in Section~\ref{Graph_Model}, since we are dealing with random graph instances in this section, there are two error probabilities associated with an in-network computation scheme: the conditional error probability $P_{e}^{\mathcal{G}}$ conditioned on a given graph instance and the expected $P_{e}^{N}$ over the ensemble. Note that, there might be cases in which the graph instance is simply not connected and $P_{e}^{\mathcal{G}}$ is bounded to be one. In what follows, we will use the expected error probability  $P_{e}^{N}$ over all random graph instances as the evaluation metric.}

\textcolor{black}{We recall the assumption (A.4) of extended Erd\"os-R\'enyi-type graphs in Section~\ref{Graph_Model}.} We assume there are $N$ nodes $\mathcal{V}=\{v_n\}_{n=1}^N$. The graph $\mathcal{G}=(\mathcal{V},\mathcal{E})$ is obtained as follows: each node pair $(v_i,v_j)\in \mathcal{V}^2$ is connected with a directed link with probability $p_N=\frac{c\log N}{N}$, where $c>0$ is a constant. All connections are assumed to be independent of each other. Link $(v_i,v_j)$ and $(v_j,v_i)$ are connected independently as well. Note that we allow self-loops, because each node can certainly broadcasts information to itself. Furthermore, there is a unique sink node $v_0$ and each node is assumed to have a directed link to it, so that the sink can hear all the broadcasted information.\footnote{\textcolor{black}{This assumption has been discussed in Section~\ref{Graph_Model}. In fact, we only require each node to have a bounded distance to the sink, which ensures that transmitting one bit to the sink has an erasure probability strictly less than 1 and the number of broadcasts required is $\mathcal{O}(1)$. However, for conciseness, we only consider cases when direct links are present.}} \textcolor{black}{Each link is assumed to be a BEC with erasure probability $\epsilon$. That is, if one bit is erased, the receiver knows explicitly the erasure position.}

Note that since each node is connected directly to the sink, there is a naive scheme to achieve polynomially decaying error probability with $N$, i.e., each node transmits the self-information bit to the sink for $\Theta(\log N)$ times. However, this naive scheme can only provide a solution in which the number of broadcasts scales as $\Theta(N\log N)$. \textcolor{black}{This scheme is also feasible in complete graphs, but since it does not achieve the lower bound, even in complete graph settings, a more involved scheme was required in~\cite{Gal_TIT_88}. As shown in~\cite{Goy_SJC_08}, the data gathering problem in a complete noisy broadcast network has a communication complexity lower bound which scales as $\Omega(N \log \log N)$, in order to achieve a constant error probability, even in a complete network where each node pair is connected. In what follows, we show that our proposed $\mathcal{GC}-3$ coding based in-network computing scheme achieves polynomial decay (in $N$) of the error probability in the above mentioned random graph settings and requires $\Theta(N\log\log N)$ broadcasts. Therefore, our broadcasting scheme can indeed achieve the broadcasting communication complexity lower bound in order sense, and, moreover, in sparser graph settings.}

\subsection{In-network Computing Algorithm}\label{Algorithm}
In this section, an in-network computing algorithm with two steps is provided. During the first step, let each node broadcast its self-information bit to its out-neighborhood $\mathcal{N}^+(v)$ for $t$ times, where
\begin{equation}\label{Broadcasting_Times}
  t=\frac{\log(\frac{c\log N}{p_{\text{ch}}})}{\log(1/\epsilon)},
\end{equation}
and $p_{\text{ch}}>0$ is a predetermined constant smaller than $1/2$. Then, each node estimates each self-information bit from its in-neighbors. The next lemma provides the probability of a certain bit being erased when transmitted from a node $v$ to one of its out-neighbors. This lemma is a counterpart result of Lemma~\ref{BSC_rept} in BEC.
\begin{figure}
  \centering
 \includegraphics[scale=0.3]{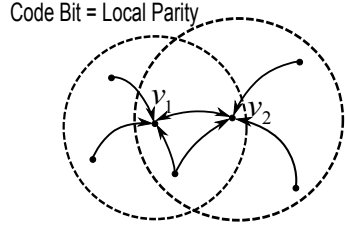}\\
  \caption{\emph{Each code bit is the parity of all one-hop in-neighbors of a specific node. Some edges might be bi-directional.}}\label{L_parity}
\end{figure}
\begin{lemma}\label{one_bit_error}
Suppose we have a BEC with erasure probability $\epsilon$. Then, the erasure probability of a bit that is repeatedly transmitted for $t$ times on this channel is
\begin{equation}\label{one_bit_error_equation}
  P_e=\epsilon^t = \frac{p_{\text{ch}}}{c\log N}.
\end{equation}
\end{lemma}
\begin{IEEEproof}The proof follows immediately by substituting in~\eqref{Broadcasting_Times}.\end{IEEEproof}

After estimating each bit, each $v_n$ calculates the local parity. Suppose node $v_n$ receives the self-information bits from its in-neighborhood $\mathcal{N}^-(v_n)$ and if all information bits are sent successfully, $v_n$ can calculate
\begin{equation}\label{Local_Parity}
  y_n=\mathop\sum \limits_{v_m\in \mathcal{N}^-(v_n)}x_m=\mathbf{x}^\top\mathbf{a}_n,
\end{equation}
where $\mathbf{a}_n$ is the $n$-th column of the adjacency matrix $\mathbf{A}$, and the summation is in the sense of modulo-2. If any bit $x_m$ is not sent successfully, i.e., erased for $t$ times, the local parity cannot be calculated. In this case, $y_n$ is assumed to take the value `$e$'. We denote the vector of all local parity bits by $\mathbf{y}=[y_1,y_2,...,y_N]^\top$. If all nodes could successfully receive all information from their in-neighborhood, we would have
\begin{equation}\label{Local_Parity_Encoding}
  \mathbf{y}^\top=\mathbf{x}^\top \mathbf{A},
\end{equation}
where $A$ is the adjacency matrix of the graph $\mathcal{G}$, and particularly, a random matrix in this section.

During the second step, each node $v_n$ transmits its self-information bit $x_n$ and the local parity $y_n$ in its in-neighborhood back to the sink exactly once. Denote the received version of the bit $x_n$ at the sink by $\tilde{x}_n$. Denote the vector of all self-information bits at the sink by $\tilde{\mathbf{x}}=[\tilde{x}_1,\tilde{x}_2,...,\tilde{x}_N]^\top$. There might be `$e$'s in this vector. Apart from self-information bits, the sink also gets a (possibly erased) version of all local parities. We denote all information gathered at the sink by
\begin{equation}\label{All_information}
  \mathbf{r}=[\tilde{x}_1,...,\tilde{x}_N,\tilde{y}_1,...,\tilde{y}_N]=[\tilde{\mathbf{x}}^\top,\tilde{\mathbf{y}}^\top],
\end{equation}
where $[\tilde{y}_1,...,\tilde{y}_N]$ is the received version (with possible erasures) of all local parity bits $\mathbf{y}$. That is, there might be some bits in $\mathbf{y}$ changed into value `$e$' during the second step. If the channels were perfect, the received information could be written as
\begin{equation}\label{All_info_noiseless}
\mathbf{r}^\top=\mathbf{x}^\top\cdot [\mathbf{I},\mathbf{A}],
\end{equation}
which is exactly a channel control code with rate $1/2$ and a generator matrix $\mathbf{G}=[\mathbf{I},\mathbf{A}]$. However, the received version is possibly with erasures, so the sink carries out the Gaussian elimination algorithm to recover all information bits, using all non-erased information. If there are too many bits erased, leading to more than one possible decoded values $\hat{\mathbf{x}}^\top$, the sink claims an error.

In all, the number of broadcasts is
\begin{equation}\label{Low_Diam_Up_Bd}
  \mathscr{C}_\mathscr{S}^\text{(N)}=N\cdot t+2N=N(2+\frac{\log(\frac{c\log N}{p_{\text{ch}}})}{\log(1/\epsilon)})=\Theta(N\log\log N),
\end{equation}
where $t$ is defined in~\eqref{Broadcasting_Times}, and the constant $2$ is introduced in the second step of the in-network computing algorithm, when the self-information bit and the local parity are transmitted directly to the sink.
\begin{remark}
\textcolor{black}{Note that in the proposed network-computing algorithm, the sink only uses received information in the second step for reconstructing the identify function (all data). However, based on our assumption, all broadcasts during the first step reach the sink as well. Thus, effectively, the sink does not (directly) take into account the bits or erasures received in the first step for the purpose of decoding. This indicates that our algorithm could be additionally advantageous in application scenarios where inter-sensor broadcasts (broadcasts between among non-sink nodes) are cheap, and direct communications between sensors and the sink are expensive, as the number of inter-sensor broadcasts required by the in-network computation algorithm is $\Theta(N\log\log N)$, whereas, the number of direct communications between sensors and the sink is only $2N$.}
\end{remark}
\subsection{An Upper Bound on the Error Probability}\label{Analysis}
In this subsection, we analyze the expected error probability of the previous algorithm. As defined in Section~\ref{Graph_Model}, denote by $P_e^{\mathcal{G}}(\mathbf{x})$ the conditional error probability in gathering all data at the sink conditioned on a graph instance $\mathcal{G}$ and self-information bit vector $\mathbf{x}$. The expected error probability is defined to be $P_e^{(N)}(\mathbf{x})=\mathbb{E}_\mathcal{G} [P_e^{\mathcal{G}}(\mathbf{x})]$. In this section, we prove that $P_e^{(N)}(\mathbf{x})$ converges to zero as $N\rightarrow \infty$ for all $\mathbf{x}$.

From Section~\ref{Algorithm}, we know that an error occurs when there exist more than one feasible solutions that satisfy the version with possible erasures of~\eqref{All_info_noiseless}. That is to say, when all positions with erasures are eliminated from the received vector, there are at least two solutions to the remaining linear equations. Denote by $\mathbf{x}_1$ and $\mathbf{x}_2$ two different vectors of self-information bits. We say that \emph{$\mathbf{x}_1$ is confused with $\mathbf{x}_2$} if the true vector of self-information bits is $\mathbf{x}_1$ but $\mathbf{x}_2$ also satisfies the possibly erased version of~\eqref{All_info_noiseless}, in which case $x_{1}$ is indistinguishable from $\mathbf{x}_2$. Denote by $P_e^{\mathcal{G}}(\mathbf{x}_1\rightarrow \mathbf{x}_2)$ the probability that $\mathbf{x}_1$ is confused with $\mathbf{x}_2$.

The Lemma~\ref{identical_error} in the following states that $P_e^{\mathcal{G}}(\mathbf{x})$ is upper bounded by an expression which is independent of the argument $\mathbf{x}$ (self-information bits).

\begin{lemma}\label{identical_error}
The error probability $P_e^{\mathcal{G}}$ can be upper-bounded by
\begin{equation}\label{identical_error_equation}
  P_e^{\mathcal{G}}(\mathbf{x})\le \mathop \sum \limits_{\mathbf{x}_0^\top \in \{0,1\}^N\setminus \{\mathbf{0}_N\}} P_e^{\mathcal{G}}(\mathbf{x}_0\rightarrow \mathbf{0}_N),
\end{equation}
where $\mathbf{0}_N$ is the $N$-dimensional zero vector.
\end{lemma}
\begin{IEEEproof}
See Appendix~\ref{PofLconfuse}.
\end{IEEEproof}

Each term on the RHS of~\eqref{identical_error_equation} can be interpreted as the probability of the existence of a non-zero vector input $\mathbf{x}_0^\top$ that is confused with the all-zero vector after all the non-zero entries of $\mathbf{x}_0^\top\cdot[\mathbf{I},\mathbf{A}]$ are erased, in which case $\mathbf{x}_{0}^{\top}$ is indistinguishable from the all zero channel input. \textcolor{black}{For example, suppose the code length is $2N=6$ and the codeword $\mathbf{x}_0^\top\cdot[\mathbf{I},\mathbf{A}]=[x_1,0,0,x_4,x_5,x_6]$ is sent and the output happens to be $\mathbf{r}^\top=[e,0,0,e,e,e]$. In this case, we cannot distinguish between the input vector $\mathbf{x}_0^\top$ and the all-zero vector $\mathbf{0}^\top_N$ based on the channel output.}

The Lemma~\ref{error_bound_big} in the following states that the expected error of the error event discussed above can be upper-bounded. This upper bound is obtained by decomposing the error event into the union of three error events on each bit.

\begin{lemma}\label{error_bound_big}
Define $\varepsilon_0=(\frac{2}{1-1/e}+1)p_{\text{ch}}+\epsilon$, where $\epsilon$ is the erasure probability of the BECs and $p_{\text{ch}}$ is a constant defined in~\eqref{Broadcasting_Times}. Then, the expected error probability $P_e^{(N)}(\mathbf{x})=\mathbb{E}_\mathcal{G} [P_e^{\mathcal{G}}(\mathbf{x})]$ can be upper-bounded by
\begin{equation}\label{Error_Middle}
  P_e^{(N)}(\mathbf{x})=\mathbb{E}_\mathcal{G} [P_e^{\mathcal{G}}(\mathbf{x})]\le \mathop\sum\limits_{k=1}^N\binom{N}{k}\epsilon^k \left[\varepsilon_0+(1-\varepsilon_0)\cdot\frac{1+(1-2p)^k}{2}\right]^N.
\end{equation}
\end{lemma}
\begin{IEEEproof}
We will first show how to decompose the error event mentioned in the above example to obtain an upper bound on the conditional error probability $P_e^{\mathcal{G}}(\mathbf{x})$. Then, we show how to obtain an upper bound on the expected error probability $P_e^{(N)}(\mathbf{x})=\mathbb{E}_\mathcal{G} [P_e^{\mathcal{G}}(\mathbf{x})]$. Finally, we compute the expected error probability upper bound using random graph theory.

\subsubsection{\textbf{Decomposing the error event conditioned on $\mathcal{G}$}}
The ambiguity event mentioned above, i.e., a non-zero vector of self-information bits being confused with the all-zero vector $\mathbf{0}_{N}$, happens if and only if each entry of the received vector $\mathbf{r}^\top$ is either zero or `$e$'. When $\mathbf{x}_0^\top$ and the graph $\mathcal{G}$ are both fixed, different entries in $\mathbf{r}^\top$ are independent of each other. Thus, the ambiguity probability $P_e^{\mathcal{G}}(\mathbf{x}_0\rightarrow \mathbf{0}_N)$ for a fixed non-zero input $\mathbf{x}_0^\top$ and a fixed graph instance $\mathcal{G}$ is the product of the corresponding ambiguity probability of each entry in $\mathbf{r}^\top$ (being a zero or a `$e$').

\textcolor{black}{The ambiguity event of each entry may occur due to structural deficiencies in the graph topology as well as due to erasures.} In particular, three events contribute to the error at the $i$-th entry of $\mathbf{r}^\top$: the product of $\mathbf{x}_0^\top$ and the $i$-th column of $[\mathbf{I},\mathbf{A}]$ is zero; the $i$-th entry of $\mathbf{r}^\top$ is `$e$' due to erasures in the first step; the $i$-th entry is `$e$' due to an erasure in the second step. We denote these three events respectively by $A_{1}^{(i)}(\mathbf{x}_0^\top)$, $A_{2}^{(i)}(\mathbf{x}_0^\top)$ and $A_{3}^{(i)}(\mathbf{x}_0^\top)$, where the superscript $i$ and the argument $\mathbf{x}_0^\top$ mean that the events are for the $i$-th entry and conditioned on a fixed message vector $\mathbf{x}_0^\top$. The ambiguity event on the $i$-th entry is the union of the above three events. \textcolor{black}{Note that the first event is due to structural deficiency, while the second and the third events are due to erasures.} Therefore, by applying the union bound over all possible inputs, the error probability $P_e^{\mathcal{G}}(\mathbf{x})$ can be upper bounded by
\begin{equation}\label{Error_pre}
  P_e^{\mathcal{G}}(\mathbf{x})\le \mathop \sum \limits_{\mathbf{x}_0^\top \in \{0,1\}^N\setminus\{\mathbf{0}^N\}} \mathop\prod \limits_{i=1}^{2N} \Pr[A_{1}^{(i)}(\mathbf{x}_0^\top)\cup A_{2}^{(i)}(\mathbf{x}_0^\top)\cup A_{3}^{(i)}(\mathbf{x}_0^\top)|\mathcal{G}],
\end{equation}
In this expression, $\mathcal{G}$ is a random graph. The randomness of $\mathcal{G}$ lies in the random edge connections.
\subsubsection{\textbf{Decomposing the unconditioned error event}}
We will further show that
\begin{equation}\label{6142}
  P_e^{(N)}(\mathbf{x})=\mathbb{E}_\mathcal{G} [P_e^{\mathcal{G}}(\mathbf{x})]\le\mathop \sum \limits_{\mathbf{x}_0^\top \in \{0,1\}^N\setminus\{\mathbf{0}^N\}}
  \mathop\prod \limits_{i=1}^{2N} \Pr[A_{1}^{(i)}(\mathbf{x}_0^\top)\cup A_{2}^{(i)}(\mathbf{x}_0^\top)\cup A_{3}^{(i)}(\mathbf{x}_0^\top)],
\end{equation}

We use a set of random binary indicators $\{E_{mn}\}_{m,n=1}^N$ to denote these edges, i.e., $E_{mn}=1$ if there is a directed edge from node $v_m$ to $v_n$. Note that we allow self-loops, because each node can certainly broadcasts information to itself. By Assumption (A.4), all random variables in $\{E_{mn}\}_{m,n=1}^N$ are mutually independent. Since in the in-network computing algorithm, the self-information bit $x_i$ and the local parity bit $y_i$ is only calculated based on the in-edges of $v_i$, i.e., the edge set $\mathcal{E}_i^{\text{in}}=\{E_{ni}|1\le n\le N\}$, we obtain
\[  \Pr[A_{1}^{(i)}(\mathbf{x}_0^\top)\cup A_{2}^{(i)}(\mathbf{x}_0^\top)\cup A_{3}^{(i)}(\mathbf{x}_0^\top)|\mathcal{G}]=\Pr[A_{1}^{(i)}(\mathbf{x}_0^\top)\cup A_{2}^{(i)}(\mathbf{x}_0^\top)\cup A_{3}^{(i)}(\mathbf{x}_0^\top)|E_{ni},1\le n\le N].\]
Thus
\begin{equation}
\begin{split}
  &\mathop\prod \limits_{i=1}^{2N} \Pr[A_{1}^{(i)}(\mathbf{x}_0^\top)\cup A_{2}^{(i)}(\mathbf{x}_0^\top)\cup A_{3}^{(i)}(\mathbf{x}_0^\top)|\mathcal{G}]\\
  =&\mathop\prod \limits_{i=1}^{2N} \Pr[A_{1}^{(i)}(\mathbf{x}_0^\top)\cup A_{2}^{(i)}(\mathbf{x}_0^\top)\cup A_{3}^{(i)}(\mathbf{x}_0^\top)|E_{ni},1\le n\le N].
\end{split}
\end{equation}
Note a bidirectional edge in the current setting corresponds to two independently generated directional edges. Therefore
\begin{equation}
\begin{split}
  P_e^{(N)}(\mathbf{x})=&\mathbb{E}_\mathcal{G} [P_e^{\mathcal{G}}(\mathbf{x})]\\
  \le&\mathop \sum \limits_{\mathbf{x}_0^\top \in \{0,1\}^N\setminus\{\mathbf{0}^N\}} \mathbb{E}_\mathcal{G} \left[\mathop\prod \limits_{i=1}^{2N} \Pr[A_{1}^{(i)}(\mathbf{x}_0^\top)\cup A_{2}^{(i)}(\mathbf{x}_0^\top)\cup A_{3}^{(i)}(\mathbf{x}_0^\top)|\mathcal{G}]\right]\\
  \overset{(a)}{=}&\mathop \sum \limits_{\mathbf{x}_0^\top \in \{0,1\}^N\setminus\{\mathbf{0}^N\}}
  \mathop\prod \limits_{i=1}^{2N} \mathbb{E}_\mathcal{G}\left[\Pr[A_{1}^{(i)}(\mathbf{x}_0^\top)\cup A_{2}^{(i)}(\mathbf{x}_0^\top)\cup A_{3}^{(i)}(\mathbf{x}_0^\top)|E_{ni},1\le n\le N]\right]\\
  =&\mathop \sum \limits_{\mathbf{x}_0^\top \in \{0,1\}^N\setminus\{\mathbf{0}^N\}}
  \mathop\prod \limits_{i=1}^{2N} \Pr[A_{1}^{(i)}(\mathbf{x}_0^\top)\cup A_{2}^{(i)}(\mathbf{x}_0^\top)\cup A_{3}^{(i)}(\mathbf{x}_0^\top)],
\end{split}
\end{equation}
where the equality (a) follows from the fact that the sets $\{E_{ni}\}_{1\leq n\leq N}$ and $\{E_{nj}\}_{1\leq n\leq N}$ are independent (by the link generation hypothesis) for any pair $(i,j)$ with $i\neq j$.

\subsubsection{\textbf{Computing the expected error upper bound using random graph theory}}

\begin{lemma}\label{Lemma1}
Define $k$ as the number of ones in $\mathbf{x}_0^\top$ and $\varepsilon_0=(\frac{2}{1-1/e}+1)p_{\text{ch}}+\epsilon$, where $\epsilon$ is the erasure probability of the BECs and $p_{\text{ch}}$ is a constant defined in~\eqref{Broadcasting_Times}. Further suppose $c\log N>1$. Then, for $1\le i\le N$, it holds that
\begin{equation}\label{Each_error_decompose_2}
  \mathop\prod \limits_{i=1}^{N}\Pr[A_1^{(i)}(\mathbf{x}_0^\top)\cup A_2^{(i)}(\mathbf{x}_0^\top)\cup A_3^{(i)}(\mathbf{x}_0^\top)]= \epsilon^k.
\end{equation}
For $N+1\le i\le 2N$, it holds that
\begin{equation}\label{Each_error_decompose}
  \Pr[A_1^{(i)}(\mathbf{x}_0^\top)\cup A_2^{(i)}(\mathbf{x}_0^\top)\cup A_3^{(i)}(\mathbf{x}_0^\top)]\le \varepsilon_0+(1-\varepsilon_0)\cdot\frac{1+(1-2p)^k}{2},
\end{equation}
where $p$ is the connection probability defined in Assumption (A.4).
\end{lemma}

\begin{IEEEproof}
See Appendix \ref{PofL1}.
\end{IEEEproof}

Based on Lemma~\ref{Lemma1} and simple counting arguments, note that~\eqref{6142} may be bounded as
\begin{equation}
  P_e^{(N)}(\mathbf{x})\le \mathop\sum\limits_{k=1}^N\binom{N}{k}\epsilon^k \left[\varepsilon_0+(1-\varepsilon_0)\cdot\frac{1+(1-2p)^k}{2}\right]^N,
\end{equation}
where the binomial expression $\binom{N}{k}$ is from the fact that there are $\binom{N}{k}$ codewords $\mathbf{x}_0$ with $k$ ones. Thus, we conclude the proof.
\end{IEEEproof}

By respectively analyzing the upper bound in Lemma \ref{error_bound_big} for $k=o\left(\frac{N}{\log N}\right)$
and $k=\Omega\left(\frac{N}{\log N}\right)$, we obtain the final error bound as follows.

\begin{theorem}\label{Theorem1}\label{ER_main_thm}
Suppose the graph $\mathcal{G}$ satisfies the topology assumption (A.4) and the channel assumption (A.1b). Suppose $\delta>0$ is a constant, $p_{\text{ch}}\in (0,\frac{1}{2})$ is a constant, $\epsilon$ is the channel erasure probability and $\varepsilon_0=(\frac{2}{1-1/e}+1)p_{\text{ch}}+\epsilon$. Assume $c\log N>1$. Define
\begin{equation}\label{b}
  b_{\delta} = \frac{1}{2}(1 - {\varepsilon _0})(1 - \frac{{1 - {e^{ - 2c\delta }}}}{2}),
\end{equation}
and assume
\begin{equation}\label{N_big_enough}
  \epsilon<b_\delta.
\end{equation}
Then, for the transmission scheme in Section~\ref{Algorithm}, we have
\begin{equation}\label{error_exponent}
P_e^{(N)} \le \left\{{(1 - b_{\delta})^N}{\rm{ + }}\delta e \epsilon \frac{{{N^{2 - c(1 - {\varepsilon _0})(1 - c\delta )}}}}{{\log N}}\right\}.
\end{equation}
That is to say, if $2 < c(1 - {\varepsilon _0})(1 - c\delta )$, the error probability eventually decreases polynomially with $N$. The rate of decrease can be maximized over all $\delta$ that satisfies~\eqref{N_big_enough}.
\end{theorem}
\begin{IEEEproof}
See Appendix~\ref{PofT1}.
\end{IEEEproof}

\begin{remark}\label{remark_GC3better}
\textcolor{black}{The $\mathcal{GC}$-3 code is ``capacity achieving'' in some sense, in that this code has rate $\frac{1}{2}$, and this code can be used even when the erasure probability $\epsilon\approx\frac{1}{2}$. Consider the case when $\epsilon=\frac{1}{2}-\Delta$, where $\Delta$ is a small constant. In Theorem 4, choose $\delta=\frac{\Delta}{2c}$ and $p_\text{ch}=\frac{\Delta}{2(\frac{2}{1-1/e}+1)}$. In this case, the constants in Theorem 4 satisfy $\varepsilon_0=\epsilon+\frac{\Delta}{2}=\frac{1}{2}-\frac{\Delta}{2}$, and $2b_\delta\ge(1-\varepsilon_0)(1-c\delta)\ge1-\varepsilon_0-c\delta=\frac{1}{2}$. Then, the error probability upper bound in Theorem 4 can be simplified to
\begin{equation}
\begin{split}
  P_e^{(N)} \le& {(1-(\frac{1}{2}-(\frac{1}{2}-\Delta)))^N}{\rm{ + }}\frac{e\Delta}{2c}(\frac{1}{2}-\Delta) \frac{{{N^{2 - c(\frac{1}{2}+\frac{\Delta}{2})(1-\frac{\Delta}{2})}}}}{{\log N}}\\
  \le &{(1-\Delta)^N}{\rm{ + }}\frac{e\Delta}{4c} \frac{{{N^{2 - c(\frac{1}{2}+\frac{\Delta}{4})}}}}{{\log N}},
\end{split}
\end{equation}
which decays polynomially with $N$ for all small $\Delta>0$ and $c>4$.}

However, consider using $\mathcal{GC}$-2 code in a complete graph with BEC channels. \textcolor{black}{From Corollary \ref{geo_thm_BEC}, the error probability of $\mathcal{GC}$-2 code in a complete graph can be shown to be $P_e^{(N)}=\mathcal{O}(N^{-2\rho  E_r(\epsilon,\frac{1}{2})+\frac{3}{2}})$, where $E_r(\epsilon,\frac{1}{2})$ is the random coding exponent for a BEC with erasure probability $\epsilon$ and code rate $\frac{1}{2}$. In the case that $\epsilon\to\frac{1}{2}$, i.e., the capacity achieving limit, $E_r(\epsilon,\frac{1}{2})$ vanishes, and hence the $\mathcal{GC}$-2 code requires a much denser network (it requires $\rho> \frac{3}{4E_r(\epsilon,\frac{1}{2})}$) than the $\mathcal{GC}$-3 code (it only requires $c>4$)}.
\end{remark}

Interestingly, the result of Theorem~\ref{Theorem1} implies a more fundamental result for erasure codes.
\begin{corollary}\label{coding_upb}
For a discrete memoryless \textcolor{black}{point-to-point} BEC with erasure probability $\epsilon$, there exists a systematic linear code with rate-$1/2$ and an $N\times 2N$ generator matrix $\mathbf{G}=[\mathbf{I},\mathbf{A}]$ such that the block error probability decreases polynomially with $N$. Moreover, the generator matrix is sparse: the number of ones in $\mathbf{A}$ is $\mathcal{O}(N\log N)$.
\end{corollary}
\begin{IEEEproof}
The proof relies on building the relation between the $\mathcal{GC}$-3 graph code and an ordinary error control code. We construct the error control code as follows:
\begin{itemize}
  \item Construct a directed Erd\"os-R\'enyi network $\mathcal{G}=(\mathcal{V},\mathcal{E})$ with $N$ nodes and connection probability $p=\frac{c\log N}{N}$, where $c$ is a constant which will be defined later.
  \item Construct a linear code with the generated matrix $\mathbf{G}=[\mathbf{I},\mathbf{A}]$, where $\mathbf{A}_{N\times N}$ is the adjacency matrix of the directed network in the previous step, i.e., the entry $A_{m,n}=1$ if and only if $v_m$ is connected to $v_n$.
\end{itemize}
The number of edges in $\mathcal{E}$ is a binomial random variable distributed according to $\text{Binomial}(N^2,p)$. Using the Chernoff bound~\cite{Che_AMS_52}, we obtain
\begin{equation}\label{Bin_large_deviation}
  \Pr(|\mathcal{E}|>2pN^2)<\exp(-\frac{p^2}{2}N^2)=(\frac{1}{N})^{\frac{c^2}{2}\log N}.
\end{equation}
Then we use the code constructed above to encode $N$ binary bits and transmit the encoded bits via $2N$ parallel BECs to the receiver. Denote by $A_e^{(N)}$ the event of a block error on the receiver side. \textcolor{black}{Define $P_e^{(N)}=\Pr(A_e^{(N)})$ as the block error probability. Note that
\begin{equation}
  P_e^{(N)}=\mathbb{E}\left[P_e^{\mathcal{G}}\right],
\end{equation}
where $P_e^{\mathcal{G}}=\Pr\left(A_e^{(N)}\mid\mathcal{G}\right)$ is the block error probability conditioned on the graph instance $\mathcal{G}$. In other words, $P_e^{(N)}$ is the expected block error probability of an ensemble of codes constructed based on directed Erd\"os-R\'enyi networks.}

Clearly, this point-to-point transmitting scheme is the same as carrying out the in-network computing algorithm in Section~\ref{Algorithm}, except that the encoding step in the point-to-point case is centralized instead of being distributed. This is equivalent to the in-network computing scheme when channels between neighboring sensor nodes are without erasures and erasures happen only when communicating over the channels to the decoder (compare with the second step of the in-network computing algorithm). Since erasure events constitute a strict subset of those encountered in the in-network computing scheme, the upper bound on the error probability in Theorem~\ref{Theorem1} still holds, which means that the expected block error probability $P_e^{(N)}$ goes down polynomially when the constant $c$ designed for the connection probability $p=\frac{c\log N}{N}$ satisfies the same condition in Theorem~\ref{Theorem1}. Note that
\begin{equation}\label{rm_cd_argu}
\begin{split}
  P_e^{(N)}= \Pr(A_e^{(N)})=&\Pr(|\mathcal{E}|>2pN^2)\Pr\left(A_e^{(N)}\mid|\mathcal{E}|>2pN^2\right)\\
  &+\Pr(|\mathcal{E}|<2pN^2)\Pr\left(A_e^{(N)}\mid|\mathcal{E}|<2pN^2\right).
\end{split}
\end{equation}
Thus, combining~\eqref{rm_cd_argu} with~\eqref{Bin_large_deviation} and~\eqref{error_exponent}, we conclude that the block error probability conditioned on $|\mathcal{E}|<2pN^2$, or equivalently $\Pr(A_e^{(N)}||\mathcal{E}|<2pN^2)$, decreases polynomially with $N$. This means that, by expurgating the code ensemble and eliminating the codes that have more than $2pN^2=\mathcal{O}(N\log N)$ ones in their generator matrices, we obtain a sparse code ensemble, of which the expected error probability decreases polynomially with $N$. Therefore, there exists a series of sparse codes which obtains polynomially decaying error probability with $N$.
\end{IEEEproof}

\begin{remark}
\textcolor{black}{In fact, the $\mathcal{GC}$-2 code also satisfies all required properties in this theorem. This fact is mentioned implicitly in~\cite{Maz_JSAC_14}. Therefore, the $\mathcal{GC}$-3 code can serve as another instance of sparse codes that satisfy these properties.}
\end{remark}

\textcolor{black}{We simulate the $\mathcal{GC}$-3 code with different code lengths in an extended Erd\"os-R\'enyi network. The ratio of successful identity function computing at the sink node is compared with the number of broadcasts during the entire in-network function computing scheme (see Section \ref{Algorithm} for details), including $t$ in-network broadcasts in the first phase and 2 transmissions to the sink node in the second phase. We can see from the simulation result that the number of broadcasts at each node required for successful identity function computing almost does not change for different network size. This is because the required number of broadcasts is $\mathcal{O}(\log\log N)$ at each node, and hence it increases very slowly with the code length or the number of nodes in the network.}

\begin{figure}
\centering
\includegraphics[scale=0.7]{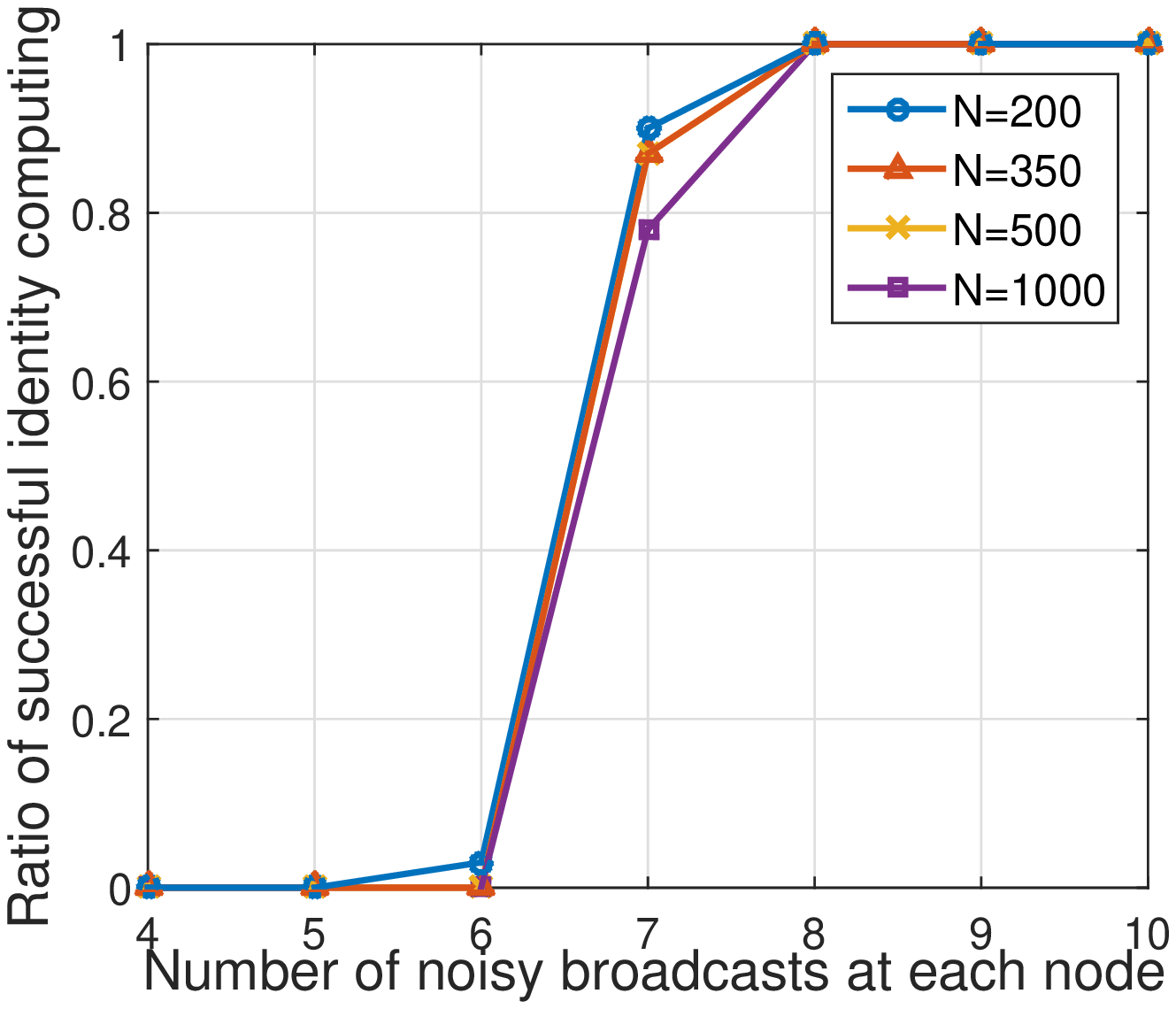}

\end{figure}

\subsection{The Degree Lower Bound for the $\mathcal{GC}$-3 Graph Code}\label{Best_Degree}
In this part, we prove that $p=\Theta(\frac{\log N}{N})$ is the minimum connection probability that gives the polynomial decay of error probability in Theorem~\ref{ER_main_thm}. In fact, we will prove a worst-case result for the total number of edges in the computation graph $\mathcal{G}$: the number of edges in the network must be $\Omega(\frac{N\log N}{\log \log N})$. This result suggests that, despite a negligible ratio $\frac{1}{\log \log N}$, the connection probability $p=\frac{c\log N}{N}$ is optimal in terms of sparseness. Since the worst-case result is for a fixed graph, we require the connectivity assumption (A.2).
\begin{theorem}\label{best_degree}
Suppose the channel assumption (A.1b) holds. Suppose the algorithm in Section~\ref{Algorithm} is carried out. Then, if $\lim\limits_{N\rightarrow\infty}P_e^{(N)}=0$, it holds that
\begin{equation}\label{edge_lower_bound}
  |\mathcal{E}|=\Omega(\frac{N \log (N/P_e^{(N)})}{\log \log N}),
\end{equation}
where $|\mathcal{E}|$ denotes the number of all directed edges in the edge set $\mathcal{E}$.
\end{theorem}
\begin{IEEEproof}During the first step of the algorithm in Section~\ref{Algorithm}, each self-information bit is broadcasted for $t$ times. Therefore, for a node $v_n$, the total number of possibly erased versions of $x_n$ is $d_nt$ where $d_n=\mathop\sum\limits_{m=1}^N \mathbf{1}_{\{v_n\in \mathcal{N}^-(v_m)\}}$. Each directed edge is counted once, so we have
\begin{equation}\label{num_edge}
  \mathop\sum\limits_{n=1}^Nd_n=|\mathcal{E}|.
\end{equation}
During the second step of the algorithm, each self-information bit $x_n$ is transmitted to the sink once. For any $x_n$, the probability that all $d_nt+1$ copies of $x_n$ are erased is
\begin{equation}\label{one_bit_erase}
  p_\text{n}=\epsilon^{d_nt+1}.
\end{equation}
If this event happens for any $x_n$, the identity function cannot be computed reliably, because at least all possible information about $x_n$ has been erased. Thus, we have
\begin{equation}\label{Ege_Low_der1}
  P_e^{(N)}>1-\mathop\prod\limits_{n=1}^N (1-p_\text{n}).
\end{equation}
\begin{figure*}
  \centering
  \includegraphics[scale=0.32]{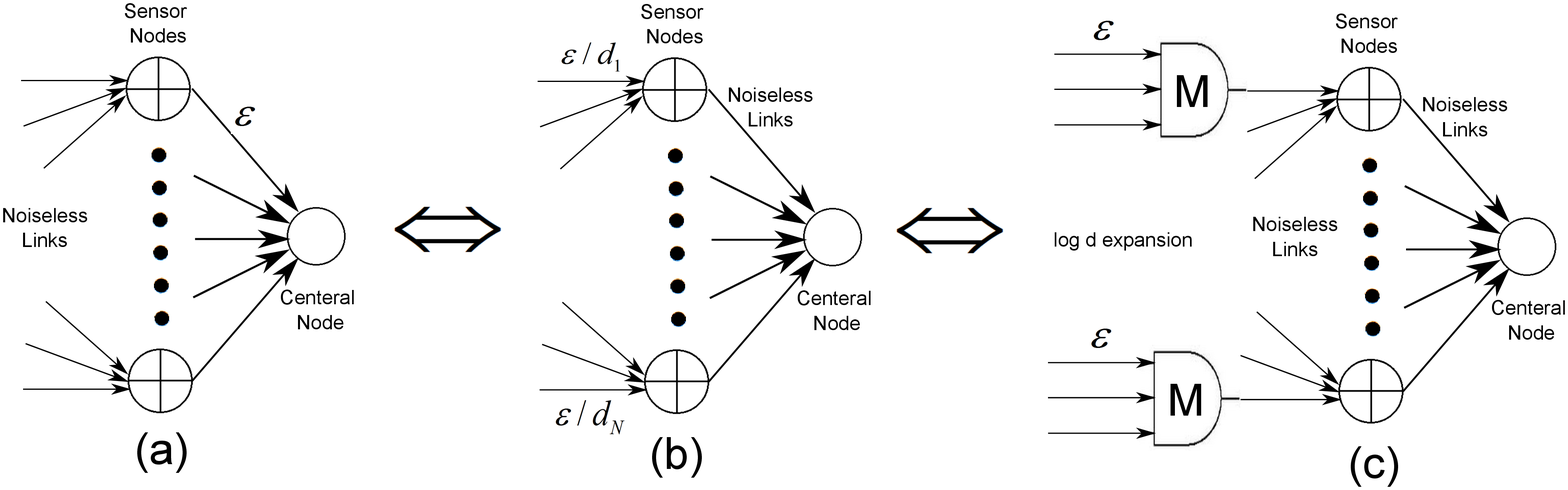}\\
  \caption{Network transformations that relate coding theory to noisy broadcast networks.}\label{it_convert}
\end{figure*}
Based on $1-x\le\exp(-x)$ and the fact that arithmetic mean is no less than geometric mean, we have
\begin{equation}\label{Ege_Low_der2}
\begin{split}
  1-P_e^{(N)}<&\mathop\prod\limits_{n=1}^N (1-p_\text{n})\le {\left[ {\frac{1}{N}\sum\limits_{n = 1}^N {(1 - p_n)} } \right]^N}= {\left( {1 - \frac{1}{N}\sum\limits_{n = 1}^N p_n } \right)^N}\\
  \le & {\left( {1 -{{\epsilon}^{\frac{1}{N}\sum\limits_{n = 1}^N {{d_n}t+1} }}} \right)^N}
 \le \exp {\left( { - N \cdot {{\epsilon}^{\frac{1}{N}\sum\limits_{n = 1}^N {{d_n}t+1} }}} \right)},
\end{split}
\end{equation}
which can be translated into
\begin{equation}\label{Ege_Low_der3}
  \sum\limits_{n = 1}^N {(t{d_n}+1)}  \ge N \cdot \frac{{\log N - \log \log (1/(1 - {P_{e}^{(N)}}))}}{{\log (1/\epsilon )}}.
\end{equation}
When $\lim\limits_{N\rightarrow\infty}P_e^{(N)}=0$,  it holds that $-\log \log (1/(1 - {P_{e}^{(N)}}))=\Theta(\log\frac{1}{P_{e}^{(N)}})$.
Therefore, jointly considering~\eqref{Broadcasting_Times}, we get
\begin{equation}\label{edge_low_bound_final}
  |\mathcal{E}|=\sum\limits_{n = 1}^N {{d_n}}=\Omega(\frac{N\log (N/P_e^{(N)})}{\log (c \log N/p_{\text{ch}})}).
\end{equation}
\end{IEEEproof}
\begin{remark}
\textcolor{black}{Note that the lower bound~\eqref{edge_lower_bound} holds for individual graph instances with arbitrary graph topologies, instead of holding for certain ensemble average.}
\end{remark}

Similar with Theorem~\ref{Theorem1} and Corollary~\ref{coding_upb}, Theorem~\ref{best_degree} also implies a result in \textcolor{black}{point-to-point} coding theory, but the proof is not obtained by directly applying Theorem~\ref{best_degree}. We have to carry out a series of network transforms, as shown in Fig.~\ref{it_convert}.
\begin{corollary}\label{coding_lowb}
For a rate-$1/2$ linear block code with an $N\times2N$ generator matrix $\mathbf{G}=[\mathbf{I},\mathbf{A}]$, if there are $d_n$ ones in the $n$-th column of $\mathbf{A}$, then, the code is asymptotically good for a \textcolor{black}{point-to-point} discrete memoryless BEC with erasure probability $\epsilon$, i.e., the block error probability $\lim\limits_{N\rightarrow\infty}P_e^{(N)}=0$, only if
\begin{equation}\label{IT_low_bound}
  \sum\limits_{n = 1}^N d_n\log d_n=\Omega(N\log (N/P_e^{(N)})).
\end{equation}
\end{corollary}
\begin{IEEEproof}Suppose we have a code $\mathbf{G}=[\mathbf{I},\mathbf{A}]$ that satisfies the conditions in this corollary. As shown in Fig~\ref{it_convert}(a), construct a directed graph $\mathcal{\mathcal{G}}=(\mathcal{V},\mathcal{E})$ with the following procedures
\begin{itemize}
  \item Set $|\mathcal{V}|=N$;
  \item Connect a directed edge from the node $v_m$ to the node $v_n$ if $A_{m,n}=1$, where $m$ can be equal to $n$, in which case a directed self loop is constructed;
  \item Assume each edge is a noiseless channel.
\end{itemize}
After constructing the graph, construct an extra node $v_0$ to be the sink, and connect each node to the sink. The links to the sink are all assumed to be discrete memoryless BECs with identical erasure probability $\epsilon$. Suppose in the network constructed above, each node $v_n\in \mathcal{V}$ carries a self-information bit $x_n$. Then, we can use the in-network computing algorithm in Section~\ref{Algorithm} to gather all sensor measurements at the sink $v_0$. Clearly, what the algorithm does is encoding the information vector $\mathbf{x}$ with the generator matrix $\mathbf{G}=[\mathbf{I},\mathbf{A}]$ (see~\eqref{All_info_noiseless}) and sending the encoded message through 2$N$ parallel BECs to the sink. Until now, the inter-sensor edges in $\mathcal{E}$ are all noiseless. The only noisy edges are from sensors to the sink, which means in the first step of the in-network computing algorithm, instead of broadcasting each self-information bit for $t$ times (as defined in~\eqref{Broadcasting_Times}), each node only needs to broadcast once. Therefore, the in-network gathering of all data in the constructed network is equivalent to the encode-and-decode procedure with the block code $\mathbf{G}=[\mathbf{I},\mathbf{A}]$ on a point-to-point link, and hence they have the same error probability $P_e^{(N)}$.

Now, modify the constructed network by assuming that links from all sensor nodes to the sink are noiseless when transmitting the parity bits. That is, in the second step of the in-network computing algorithm, these sensor-to-sink links are only noisy when self-information bits are transmitted. However, assume that the links between sensors are noisy, as shown in Fig~\ref{it_convert}(b). Specifically, for each node $v_n$, assume that all the directed links from the in-neighborhood $\mathcal{N}_{v_n}^-$ are changed into BECs with identical erasure probability $\epsilon/d_n$, where $d_n=|\mathcal{N}^-(v_n)|$. Now that the local parity that $v_n$ sends to the sink is erased with probability $1-(1-\frac{\epsilon}{d_n})^{d_n}<\epsilon$, therefore, if the original network can gather all data with error probability $P_e^{(N)}$, the transformed network can compute it with error probability strictly less than $P_e^{(N)}$.

Now make a further change as shown in Fig~\ref{it_convert}(c), which consists of substituting each sensor-to-sensor link with erasure probability $\epsilon/d_n$ to a set of $\lceil1+\frac{\log d_n}{\log(1/\epsilon)}\rceil$ parallel links with erasure probability $\epsilon$ connected to a merging gate. This gate claims an `erasure' only if all bits in the incoming edges are erased. This transform is exactly the same as repeatedly transmitting $t$ times of the same bit as defined in~\eqref{Broadcasting_Times}. After this transform, the erasure probability changes to $\epsilon^{1+\frac{\log d_n}{\log(1/\epsilon)}}<\epsilon/d_n$. Similarly, if the original network can reliably gather all data with error probability $P_e^{(N)}$, the new network can also compute it with lower error probability.

Therefore, if the block code $\mathbf{G}=[\mathbf{I},\mathbf{A}]$ can be used to successfully transmit all bits on a point-to-point BEC with error probability $P_e^{(N)}$, data gathering in the transformed network shown in Fig~\ref{it_convert}(c) can be reliably completed with lower error probability. By Theorem~\ref{best_degree}, to achieve error probability $P_e^{(N)}$, the degree of the transformed network should satisfy
\begin{equation}\label{IT_Lw_bd}
  \sum\limits_{n = 1}^N d_n\lceil1+\frac{\log d_n}{\log(1/\epsilon)}\rceil>N \cdot \frac{{\log N - \log \log (1/(1 - {P_{e}^{(N)}}))}}{{\log (1/\epsilon )}}.
\end{equation}
This implies that~\eqref{IT_low_bound} holds.
\end{IEEEproof}

This corollary suggests that, if one wants to find a sparse linear block code for BECs, then~\eqref{IT_low_bound} can serve as a lower bound on `sparseness'. Moreover, if the matrix $\mathbf{A}$ has the same number of ones in each column, then, there are $\Omega(\frac{\log N}{\log\log N})$ ones in each column, in order for~\eqref{IT_low_bound} to hold. \textcolor{black}{A similar result was obtained in~\cite{Maz_JSAC_14}, which states that $\underset{1\le n\le N}{\max}{\;}d_n$ is at least $\Omega(\log n)$, in order to achieve an error probability strictly less than 1. However, this result is obtained only for the maximum value $d_n$, which does not capture the total sparseness. Moreover, our result is in coding theory but relates to distributed encoding as well.}

\section{Conclusions}
In this paper, we obtain both upper and lower scaling bounds on the communication complexity of data gathering in arbitrary noisy broadcast networks. In particular, using different graph-based distributed encoding schemes, which we call graph codes, we find two special graph topologies, i.e., geometric graphs and extended random Erd\"os-R\'enyi graphs, in which the upper bounds on the number of broadcasts obtained by graph codes meet with the general lower bound in order sense. Furthermore, the analysis techniques of the third graph code is used to construct a sparse erasure code that is used in point-to-point communications. We also use cut-set techniques to show that the obtained code is almost optimal in terms of sparseness (with minimum number of ones in the generator matrix) except for a $\log\log N$ multiple gap, such that the block error probability approaches zero in the limit of large code length $N$. \textcolor{black}{However, quite a few open questions worthy of further research remain. For instance, an issue with the $\mathcal{GC}$-3 code proposed in this paper is that it can be analyzed only in BEC networks. The technical difficulty is that the ``effective channel noise'' is determined by the graph structure and hence is dependent of the code structure itself. The analysis of $\mathcal{GC}$-3 code in BEC networks as achieved in this paper is feasible because the upper bound on error probability can be decomposed as the product of the error probability of each particular bit. However, this decomposition cannot be readily obtained for BSC channels (and networks). A meaningful direction is to understand and characterize this effective channel noise for analyzing $\mathcal{GC}$-3 codes in BSCs. The focus of this paper has been primarily on the design of codes that minimize the broadcast complexity, i.e., the number of broadcasts required to achieve function computation. Other practical metrics such as the energy of broadcast (which, depending on the network structure, is somewhat indirectly related to the number of broadcasts) may be of interest in applications too. An extension of $\mathcal{GC}$-3 codes from an energy minimization perspective is provided in a follow up conference paper \cite{7541796}.}

\appendices
\section{Proof of \eqref{error_polynomial} in Theorem~\ref{Lg_Dia_Up_Bd}}\label{PofT2}
Since the code length at each node $v\in \mathcal{B}_{\mathcal{T}}$ is $\gamma \log N$, according to Lemma~\ref{BSC_rm_cd}, the decoding error probability is
\begin{equation}\label{Up_Each_Node_B}
  P_{e,v}<\exp[-(\gamma \log N+1)E_r(\epsilon,R)/R]=\exp[-E_r(\epsilon,R)/R] N^{-\gamma E_r(\epsilon,R) /R}.
\end{equation}
Similarly, the decoding error probability at a node $v\in \mathcal{A}_{\mathcal{T}}$ is
\begin{equation}\label{Up_Each_Node_A}
\begin{split}
  P_{e,v}< &\exp[-(\mathcal{D}_v+1)E_r(\epsilon,R)/R]
  <\exp[-\frac{\gamma}{R}\log N E_r(\epsilon,R)]
  <N^{-\gamma E_r(\epsilon,R) /R},
\end{split}
\end{equation}
where we used the fact that the message size $\mathcal{D}_v$ in $v$ is greater than or equal to $\gamma \log N$, and hence we can find a code with length $\lceil(\mathcal{D}_v+1)/R\rceil>\frac{\gamma}{R}\log N$.

Combining~\eqref{Up_Each_Node_B} and \eqref{Up_Each_Node_A} and using the union bound, the error probability is bounded as follows
\begin{equation}\label{Large_d_error_bd}
\begin{split}
  P_e^{(N)}<&\mathop\sum\limits_{v\in \mathcal{A}_{\mathcal{T}}}P_{e,v}+\mathop\sum\limits_{v\in \mathcal{B}_{\mathcal{T}}}P_{e,v}
  <N\cdot N^{-\gamma E_r(\epsilon,R) /R}+N\cdot\exp[-E_r(\epsilon,R)/R] N^{-\gamma E_r(\epsilon,R) /R}\\
  =&N^{-(\frac{\gamma E_r(\epsilon,R)}{R}-1)}\cdot \left(1+\exp[-E_r(\epsilon,R)/R]\right).
\end{split}
\end{equation}
When the condition $R<\gamma E_r(\epsilon,R)$ is satisfied, the error probability in~\eqref{Large_d_error_bd} satisfies the property that $\lim_{N\rightarrow\infty} P_e^{(N)}=0$ and the convergence rate is polynomial. This concludes the proof.
\section{Proof of Lemma~\ref{identical_error}}\label{PofLconfuse}
We know from the union bound that
\begin{equation}
\begin{split}
  P_e^{\mathcal{G}}(\mathbf{x})\le \mathop \sum \limits_{\mathbf{x}_1^\top \in \{0,1\}^N \setminus \{\mathbf{x}\}} P_e^{\mathcal{G}}(\mathbf{x}\rightarrow \mathbf{x}_1).
\end{split}
\end{equation}

\begin{lemma}\label{all_zero_vector}
The probability that $\mathbf{x}_1$ is confused with $\mathbf{x}_2$ equals the probability that $\mathbf{x}_1-\mathbf{x}_2$ is confused with the $N$-dimensional zero vector $\mathbf{0}_N$, i.e.,
\begin{equation}\label{confusion_identical}
  P_e^{\mathcal{G}}(\mathbf{x}_1\rightarrow \mathbf{x}_2)=P_e^{\mathcal{G}}(\mathbf{x}_1-\mathbf{x}_2\rightarrow \mathbf{0}_N).
\end{equation}
\end{lemma}
\begin{IEEEproof}
We define an \emph{erasure matrix} $\mathbf{E}$ as a $2N$-by-$2N$ diagonal matrix in which each diagonal entry is either an `$e$' or a $1$. Define an extended binary multiplication operation with `$e$', which has the rule that $ae=e,a\in \{0,1\}$. The intuition is that both $0$ and $1$ become an erasure after being erased. Under this definition, the event that $\mathbf{x}_1$ is confused with $\mathbf{x}_2$ can be written as
\begin{equation}
  \mathbf{x}_1^\top\cdot [\mathbf{I},\mathbf{A}] \cdot \mathbf{E}=\mathbf{x}_2^\top\cdot [\mathbf{I},\mathbf{A}] \cdot \mathbf{E},
\end{equation}
\textcolor{black}{where a diagonal entry in $\mathbf{E}$ being `$e$' corresponds to erasure/removal of the corresponding linear equation.} We know that if the erasure matrix $\mathbf{E}$ remains the same, we can arrange the two terms and write
\begin{equation}
  (\mathbf{x}_1^\top-\mathbf{x}_2^\top)\cdot [\mathbf{I},\mathbf{A}] \cdot \mathbf{E}=0_N^\top\cdot [\mathbf{I},\mathbf{A}] \cdot \mathbf{E}.
\end{equation}
That is to say, if $\mathbf{x}_1$ is confused with $\mathbf{x}_2$, then, if all the erasure events are the same and the self-information bits are changed to $\mathbf{x}_1-\mathbf{x}_2$, they will be confused with the all zero vector $\mathbf{0}_N$ and vice-versa. Thus, in order to prove~\eqref{confusion_identical}, we only need to show that the probability of having particular erasure events remains the same with different self-information bits. This claim is satisfied, because by the BEC assumption the erasure events are independent of the channel inputs and identically distributed.
\end{IEEEproof}
Thus, using the result from Lemma~\ref{all_zero_vector}, we obtain
\begin{equation}
\begin{split}
P_e^{\mathcal{G}}(\mathbf{x})\le \mathop \sum \limits_{\mathbf{x}_1^\top \in \{0,1\}^N \setminus \{\mathbf{x}\}} P_e^{\mathcal{G}}(\mathbf{x}-\mathbf{x}_1\rightarrow \mathbf{0}_N),
\end{split}
\end{equation}
and hence,~\eqref{identical_error_equation} holds.

\section{Proof of Lemma~\ref{Lemma1}}\label{PofL1}

First, we notice that for $1\le i\le N$, the vector $\tilde{\mathbf{x}}^\top$ received is the noisy version of $\mathbf{x}_0^\top$. Since, according to the in-network computing algorithm in Section~\ref{Algorithm}, the vector $\tilde{\mathbf{x}}^\top$ is obtained in the second step, the event $A_3^{(i)}(\mathbf{x}_0^\top)$ is the only ambiguity event. Moreover, if the $i$-th entry of $\mathbf{x}_0^\top$ is zero, it does not matter whether an erasure happens to this entry. Thus, the error probability can be calculated by considering all the $k$ non-zero entries, which means
\[
\mathop\prod \limits_{i=1}^{N}\Pr[A_1^{(i)}(\mathbf{x}_0^\top)\cup A_2^{(i)}(\mathbf{x}_0^\top)\cup A_3^{(i)}(\mathbf{x}_0^\top)]= \epsilon^k.
\]
For $N+1\le i\le 2N$, $A_3^{(i)}(\mathbf{x}_0^\top)$ is the erasure event during the second step and is independent from the previous two events $A_1^{(i)}(\mathbf{x}_0^\top)$ and $A_2^{(i)}(\mathbf{x}_0^\top)$. Therefore
\begin{equation}\label{Derive_1}
  \begin{split}
&\Pr\left[A_1^{(i)}(\mathbf{x}_{0}^{\top})\cup A_2^{(i)}(\mathbf{x}_{0}^{\top})\cup A_3^{(i)}(\mathbf{x}_{0}^{\top})\right]\\
\le& \Pr\left[(A_3^{(i)}(\mathbf{x}_{0}^{\top}))^C\right]+\Pr\left[A_3^{(i)}(\mathbf{x}_{0}^{\top})\right]\Pr\left[A_1^{(i)}(\mathbf{x}_{0}^{\top})\cup A_2^{(i)}(\mathbf{x}_{0}^{\top})\right]\\
=& 1-\epsilon+\epsilon\Pr\left[A_1^{(i)}(\mathbf{x}_{0}^{\top})\cup A_2^{(i)}(\mathbf{x}_{0}^{\top})\right]\\
=& 1-\epsilon+\epsilon\left(\Pr\left[A_1^{(i)}(\mathbf{x}_{0}^{\top})\right]+\Pr\left[(A_1^{(i)}(\mathbf{x}_{0}^{\top}))^C\cap A_2^{(i)}(\mathbf{x}_{0}^{\top})\right]\right).
\end{split}
\end{equation}
The event $A_1^{(i)}(\mathbf{x}_0^\top)$ happens when the local parity $\mathbf{x}_0^\top \mathbf{a}_{i}$ equals zero, i.e., in the $k$ locations of non-zero entries in $\mathbf{x}_0^\top$, there are an even number of ones in the corresponding entries in $\mathbf{a}_{i}$, the $i$-th column of the graph adjacency matrix $\mathbf{A}$. Denote by $l$ the number of ones in these $k$ corresponding entries in $\mathbf{a}_{i}$. Since each entry of $\mathbf{a}_{i}$ takes value 1 independently with probability $p$, the probability that an even number of entries are 1 in these $k$ locations is
\begin{equation}\label{Derive_2}
  \begin{split}
\Pr[A_1^{(i)}(\mathbf{x}_{0}^{\top})]=&\Pr[l\text{ is even}]=\mathop\sum\limits_{l\text{ is even}}p^l(1-p)^{k-l}=\frac{1+(1-2p)^k}{2}.
\end{split}
\end{equation}
The event $(A_1^{(i)}(\mathbf{x}_{0}^{\top}))^C\cap A_2^{(i)}(\mathbf{x}_{0}^{\top})$ indicates that $l$ is odd and at least one entry of all non-zero entries in $\mathbf{x}_0^\top$ is erased. Suppose in the remaining $N-k$ entries in $\mathbf{a}_{i}$, $j$ entries take the value 1 and hence there are $(l+j)$ 1's in $\mathbf{a}_{i}$. Therefore, for a fixed $l$, we have
\[\begin{split}
\Pr[(A_1^{(i)}(\mathbf{x}_{0}^{\top}))^C\cap A_2^{(i)}(\mathbf{x}_{0}^{\top})|l]
=&\mathop\sum\limits_{j=0}^{N-k}\binom{N-k}{j}p^j (1-p)^{N-k-j}\cdot[1-(1-p_e)^{l+j}]\\
\le & \mathop\sum\limits_{j=0}^{N-k}\binom{N-k}{j}p^j (1-p)^{N-k-j}(l+j)p_e,
\end{split}\]
where $p$ is the edge connection probability and $p_e$ is the probability that a certain bit in $\mathbf{x}_0$ is erased for $t=\frac{\log(\frac{c\log N}{p_{\text{ch}}})}{\log(1/\epsilon)}$ times when transmitted to $v_i$ from one of its neighbors during the first step of the algorithm. Combining the above inequality with Lemma~\ref{one_bit_error}, we get
\[\begin{split}
&\Pr[(A_1^{(i)})^C\cap A_2^{(i)}(l)]\le \mathop\sum\limits_{j=0}^{N-k}\binom{N-k}{j}p^j (1-p)^{N-k-j}(l+j)\frac{p_{\text{ch}}}{c\log N}\\
=&l\frac{p_{\text{ch}}}{c\log N}\mathop\sum\limits_{j=0}^{N-k}\binom{N-k}{j}p^j (1-p)^{N-k-j}
+\frac{p_{\text{ch}}}{c\log N}\mathop\sum\limits_{j=1}^{N-k}j\binom{N-k}{j}p^j (1-p)^{N-k-j}\\
\overset{(a)}{=}&l\frac{p_{\text{ch}}}{c\log N}
+\frac{p_{\text{ch}}p}{c\log N}\mathop\sum\limits_{j=1}^{N-k}(N-k)\binom{N-k-1}{j-1}p^{j-1} (1-p)^{N-k-j}\\
=&l\frac{p_{\text{ch}}}{c\log N}+\frac{p_{\text{ch}}(N-k)}{N}\mathop\sum\limits_{j=1}^{N-k}\binom{N-k-1}{j-1}p^{j-1} (1-p)^{N-k-j}\\
=&l\frac{p_{\text{ch}}}{c\log N}+p_{\text{ch}}\cdot\frac{N-k}{N},
\end{split}\]
where step (a) follows from $j\binom{N-k}{j}=(N-k)\binom{N-k-1}{j-1}$. Therefore
\[\begin{split}
&\Pr[(A_1^{(i)})^C\cap A_2^{(i)}]\\
=&\mathop\sum\limits_{l\text{ is odd}}\binom{k}{l}p^l (1-p)^{k-l}\Pr[(A_1^{(i)})^C\cap A_2^{(i)}(l)]\\
\le&\mathop\sum\limits_{l\text{ is odd}}\binom{k}{l}p^l (1-p)^{k-l} (l\frac{p_{\text{ch}}}{c\log N}+p_{\text{ch}}\cdot\frac{N-k}{N})\\
=&\sum\limits_{l\text{ is odd}}{\binom{k}{l}{{p}^{l}}{{(1-p)}^{k-l}}{{p}_{\text{ch}}}\cdot \frac{N-k}{N}}+\sum\limits_{l\text{ is odd}}{l\binom{k}{l}{{p}^{l}}{{(1-p)}^{k-l}}\frac{{{p}_{\text{ch}}}}{c\log N}}\\
=&{{p}_{\text{ch}}}\cdot \frac{N-k}{N}\sum\limits_{l\text{ is odd}}{\binom{k}{l}}{{p}^{l}}{{(1-p)}^{k-l}}+\frac{kp{{p}_{\text{ch}}}}{c\log N}\sum\limits_{l\text{ is odd}}{\binom{k-1}{l-1}}{{p}^{l-1}}{{(1-p)}^{k-l}}\\
=&p_{\text{ch}}\cdot\frac{N-k}{N}\frac{1-(1-2p)^k}{2}+p_{\text{ch}}\cdot\frac{k}{N}\frac{1+(1-2p)^{k-1}}{2}\\
\overset{(a)}{\le}& Lp_{\text{ch}}\frac{1-(1-2p)^k}{2},
\end{split}\]
where the constant $L$ in step (a) is to be determined. Now we show that $L=\frac{2}{1-1/e}+1$ suffices to ensure that (a) holds. In fact, we only need to prove
\[\frac{N-k}{N}\frac{1-{{(1-2p)}^{k}}}{2}+\frac{k}{N}\frac{1+{{(1-2p)}^{k-1}}}{2}\le L\frac{1-{{(1-2p)}^{k}}}{2}.\]
Since $\frac{N-k}{N}<1$, it suffices to show that
\[\frac{k}{N}\frac{1+{{(1-2p)}^{k-1}}}{2}\le \left( L-1 \right)\frac{1-{{(1-2p)}^{k}}}{2}.\]
Since ${{(1-2p)}^{k-1}}<1$, it suffices to show that
\[\frac{k}{N}\le \left( L-1 \right)\frac{1-{{(1-2p)}^{k}}}{2},\]
or equivalently,
\begin{equation}\label{kkey}
  \frac{2k}{1-{{(1-2p)}^{k}}}\le N\left( L-1 \right).
\end{equation}
We know that
\[1-{{(1-2p)}^{k}}\ge 2kp-C_{k}^{2}{{\left( 2p \right)}^{2}}=2kp-2k(k-1){{p}^{2}}=2kp\left[ 1-p(k-1) \right]\ge 2kp(1-kp).\]
Thus, when $kp\le \frac{1}{2}$, $1-{{(1-2p)}^{k}}\ge 2kp(1-kp)\ge kp$ and
\[\frac{2k}{1-{{(1-2p)}^{k}}}\le \frac{2k}{kp}=\frac{2N}{c\log N}\le 2N,\] when $c\log N>1$.
When $kp>\frac{1}{2}$, ${{(1-2p)}^{k}}\le {{(1-2p)}^{\frac{1}{2p}}}\le \frac{1}{e}$ and
\[\frac{2k}{1-{{(1-2p)}^{k}}}\le \frac{2k}{1-1/e}\le \frac{2N}{1-1/e}.\]
Thus, as long as $L\ge 1+\frac{2}{1-1/e}$, \eqref{kkey} holds.
Jointly considering~\eqref{Derive_2}, we get
\[\Pr[A_1^{(i)}\cup A_2^{(i)}]\le\frac{1+(1-2p)^k}{2}+Lp_{\text{ch}}\frac{1-(1-2p)^k}{2}.\]
Combining~\eqref{Derive_1}, we finally arrive at
\[\begin{split}
\Pr[A_1^{(i)}\cup A_2^{(i)}\cup A_3^{(i)}]\le& \epsilon +(1-\epsilon )\left[ \frac{1+{{(1-2p)}^{k}}}{2}+L{{p}_{\text{ch}}}\frac{1-{{(1-2p)}^{k}}}{2} \right] \\
 & =\epsilon +(1-\epsilon )\left[ 1-\left( 1-L{{p}_{\text{ch}}} \right)\frac{1-{{(1-2p)}^{k}}}{2} \right] \\
 & =1-(1-\epsilon )\left( 1-L{{p}_{\text{ch}}} \right)\frac{1-{{(1-2p)}^{k}}}{2} \\
 & <1-(1-\epsilon -L{{p}_{\text{ch}}})\frac{1-{{(1-2p)}^{k}}}{2} \\
 & =1-(1-\epsilon -L{{p}_{\text{ch}}})\left[ 1-\frac{1+{{(1-2p)}^{k}}}{2} \right] \\
 & =\epsilon +L{{p}_{\text{ch}}}+(1-\epsilon -L{{p}_{\text{ch}}})\frac{1+{{(1-2p)}^{k}}}{2} \\
 & ={{\varepsilon }_{0}}+(1-{{\varepsilon }_{0}})\frac{1+{{(1-2p)}^{k}}}{2},
\end{split}\]
where $\varepsilon_0=Lp_{\text{ch}}+\epsilon$.
\section{Proof of Theorem~\ref{Theorem1}}\label{PofT1}
We will prove that for any $\delta>0$, it holds that
\begin{equation}\label{error_exponent_proof}
  P_e^{(N)} \le {(1 - b_{\delta})^N}{\rm{ + }}\delta e \epsilon \frac{{{N^{2 - c(1 - {\varepsilon _0})(1 - c\delta )}}}}{{\log N}}.
\end{equation}
As shown in what follows, we bound the right hand side of~\eqref{Error_Middle} with two different methods for different $k$'s. First, when $k$ satisfies
\begin{equation}\label{small_k}
  1\le k < \delta \frac{N}{{\log N}},
\end{equation}
define
\begin{equation}\label{u}
  u = N(1 - {\varepsilon _0})\frac{{1 - {{(1 - 2p)}^k}}}{2}
\end{equation}
Then, based on the inequality
\begin{equation}\label{exponential_ineq}
  {(1 - \frac{1}{x})^x} \le {e^{ - 1}},\forall x \in (0,1],
\end{equation}
we have
\begin{equation}\label{k_small_drv1}
\begin{split}
{[{\varepsilon _0} + (1 - {\varepsilon _0})\frac{{1 + {{(1 - 2p)}^k}}}{2}]^N} = &{(1 - \frac{u}{N})^N} = {[{(1 - \frac{u}{N})^{\frac{N}{u}}}]^u} \le {e^{ - u}}.
\end{split}
\end{equation}
From the Taylor's expansion, we get
\[{(1 - 2p)^k} = 1 - 2pk + \frac{{k(k - 1)}}{2}{\theta ^2},\theta  \in [0,2p].\]
By applying the equation above to~\eqref{u}, we get
\[u = N(1 - {\varepsilon _0})[kp - \frac{{k(k - 1)}}{4}{\theta ^2}].\]
Therefore, we have
\[\begin{split}
{e^{ - u}} = &{e^{ - k(1 - {\varepsilon _0}) \cdot c\log N}}\exp \{ N(1 - {\varepsilon _0})\frac{{k(k - 1)}}{4}{\theta ^2}\}\\
\le &{\left( {\frac{1}{N}} \right)^{ck(1 - {\varepsilon _0})}}\exp \{ N(1 - {\varepsilon _0})\frac{{k(k - 1)}}{4}\frac{{4{c^2}{{\log }^2}N}}{{{N^2}}}\}\\
 = &{\left( {\frac{1}{N}} \right)^{ck(1 - {\varepsilon _0})}}{N^{(1 - {\varepsilon _0}) \cdot \frac{{{c^2}k(k - 1)\log N}}{N}}}.
\end{split}\]
Plugging the above inequality into~\eqref{k_small_drv1}, we get
\begin{equation}
\begin{split}\label{k_small_drv2}
&\binom{N}{k}{\epsilon^k}{[{\varepsilon _0} + (1 - {\varepsilon _0})\frac{{1 + {{(1 - 2p)}^k}}}{2}]^N}\\
\le& {\left( {\frac{{Ne}}{k}} \right)^k}{\epsilon^k}{\left( {\frac{1}{N}} \right)^{ck(1 - {\varepsilon _0})}}{N^{(1 - {\varepsilon _0}) \cdot \frac{{{c^2}k(k - 1)\log N}}{N}}}\\
= &{\left( {\frac{e}{k}\epsilon{N^{1 - c(1 - {\varepsilon _0})[1 - \frac{{c(k - 1)\log N}}{N}]}}} \right)^k}<{\left( {\frac{e}{k}\epsilon{N^{1 - c(1 - {\varepsilon _0})(1 - c\delta )}}} \right)^k},
\end{split}
\end{equation}
where the last inequality follows from~\eqref{small_k}.

Second, when $k$ satisfies
\begin{equation}\label{big_k}
  k > \delta \frac{N}{{\log N}},
\end{equation}
we can directly write
\[{(1 - 2p)^k} = {[{(1 - 2p)^{\frac{1}{{2p}}}}]^{2pk}} \le {e^{ - 2pk}} <{e^{ - 2c\delta }}.\]
Therefore, it holds that
\[\begin{split}
&\sum\limits_{k > \delta \frac{N}{{\log N}}} {\binom{N}{k}{{\epsilon}^k}{{[{\varepsilon _0} + (1 - {\varepsilon _0})\frac{{1 + {{(1 - 2p)}^k}}}{2}]^N}}}\\
\le &\sum\limits_{k > \delta \frac{N}{{\log N}}} {\binom{N}{k}{{\epsilon}^k}{{[{\varepsilon _0} + (1 - {\varepsilon _0})\frac{{1 + {e^{ - 2c\delta }}}}{2}]^N}}}\\
\le &{[{\varepsilon _0} + (1 - {\varepsilon _0})\frac{{1 + {e^{ - 2c\delta }}}}{2}]^N}\sum\limits_{k = 0}^N {\binom{N}{k}{{\epsilon}^k}}\\
= &{[{\varepsilon _0} + (1 - {\varepsilon _0})\frac{{1 + {e^{ - 2c\delta }}}}{2}]^N}{(1 +\epsilon)^N}\\
= &{[(1 - (1 - {\varepsilon _0})\frac{{1 - {e^{ - 2c\delta }}}}{2})(1 +\epsilon)]^N}\\
\le &{\{ 1 - [(1 - {\varepsilon _0})(1 - \frac{{1 - {e^{ - 2c\delta }}}}{2}) - \epsilon]\} ^N}\\
=&{\{ 1 - (2b_\delta - \epsilon)\} ^N}.
\end{split}\]
When~\eqref{N_big_enough} holds, we have
\begin{equation}\label{sum_big_k}
\begin{split}
  &\sum\limits_{k > \delta \frac{N}{{\log N}}} {\binom{N}{k}{{(\frac{{{p_{\text{ch}}}}}{{c\log N}})}^k}{{[{\varepsilon _0} + (1 - {\varepsilon _0})\frac{{1 + {{(1 - 2p)}^k}}}{2}]^N}}}
  < {(1 - b_\delta)^N}.
  \end{split}
\end{equation}
Combining~\eqref{Error_Middle} and~\eqref{k_small_drv2}, we get
\[\begin{split}
P_e^{(N)}&\le {(1 - b_\delta)^N}{\rm{+}}\sum\limits_{k < \delta \frac{N}{{\log N}}} {\binom{N}{k}{{\epsilon}^k}{{[{\varepsilon _0} + (1 - {\varepsilon _0})\frac{{1 + {{(1 - 2p)}^k}}}{2}]^N}}} \\
&\le{(1 - b_\delta)^N}+\sum\limits_{k < \delta \frac{N}{{\log N}}} {\left( {\frac{e}{k}\epsilon{N^{1 - c(1 - {\varepsilon _0})(1 - c\delta )}}} \right)^k}\\
&\le{(1 - b_\delta)^N}+ \delta \frac{N}{{\log N}}\frac{e}{k}\epsilon{N^{1 - c(1 - {\varepsilon _0})(1 - c\delta )}}\\
&\le {(1 - b_\delta)^N}{\rm{ + }}\delta e \epsilon \frac{{{N^{2 - c(1 - {\varepsilon _0})(1 - c\delta )}}}}{{\log N}}.
\end{split}\]
When $2 < c(1 - {\varepsilon _0})(1 - c\delta )$, the right hand side decreases polynomially with $N$.

\bibliographystyle{ieeetr}
\bibliography{rough}

\begin{thebibliography}{10}

\bibitem{Gal_TIT_88}
R.~Gallager, ``Finding parity in a simple broadcast network,'' {\em IEEE
  Transactions on Information Theory}, vol.~34, pp.~176--180, Mar 1988.

\bibitem{Kara_TIT_11}
N.~Karamchandani, R.~Appuswamy, and M.~Franceschetti, ``Time and energy
  complexity of function computation over networks,'' {\em IEEE Transactions on
  Information Theory}, vol.~57, pp.~7671--7684, Dec 2011.

\bibitem{Gir_CM_06}
A.~Giridhar and P.~Kumar, ``Toward a theory of in-network computation in
  wireless sensor networks,'' {\em IEEE Communications Magazine}, vol.~44,
  pp.~98--107, April 2006.

\bibitem{Dim_PI_10}
A.~Dimakis, S.~Kar, J.~Moura, M.~Rabbat, and A.~Scaglione, ``Gossip algorithms
  for distributed signal processing,'' {\em Proceedings of the IEEE}, vol.~98,
  pp.~1847--1864, Nov 2010.

\bibitem{Goy_SJC_08}
N.~Goyal, G.~Kindler, and M.~Saks, ``Lower bounds for the noisy broadcast
  problem,'' {\em SIAM Journal on Computing}, vol.~37, no.~6, pp.~1806--1841,
  2008.

\bibitem{Li_TMC_13}
C.~Li and H.~Dai, ``Efficient in-network computing with noisy wireless
  channels,'' {\em IEEE Transactions on Mobile Computing}, vol.~12,
  pp.~2167--2177, Nov 2013.

\bibitem{Zheng_INFOCOM_07}
R.~Zheng and R.~Barton, ``Toward optimal data aggregation in random wireless
  sensor networks,'' in {\em Proceedings of the 26th IEEE International
  Conference on Computer Communications}, pp.~249--257, May 2007.

\bibitem{Mos_IPSN_07}
T.~Moscibroda, ``The worst-case capacity of wireless sensor networks,'' in {\em
  Proceedings of the 6th International Symposium on Information Processing in
  Sensor Networks (ISPN 2007)}, pp.~1--10, April 2007.

\bibitem{Joo_Allerton_08}
C.~Joo and N.~Shroff, ``On the delay performance of in-network aggregation in
  lossy wireless sensor networks,'' {\em Networking, IEEE/ACM Transactions on},
  vol.~22, pp.~662--673, April 2014.

\bibitem{Mar_IPSN_03}
D.~Marco, E.~J. Duarte-Melo, M.~Liu, and D.~L. Neuhoff, ``On the many-to-one
  transport capacity of a dense wireless sensor network and the compressibility
  of its data,'' in {\em Proceedings of the International Symposium on
  Information Processing in Sensor Networks (ISPN 2003)}, pp.~1--16, Springer,
  2003.

\bibitem{Zan_ITJ_14}
A.~Zanella, N.~Bui, A.~Castellani, L.~Vangelista, and M.~Zorzi, ``Internet of
  things for smart cities,'' {\em IEEE Internet of Things Journal}, vol.~1,
  pp.~22--32, Feb 2014.

\bibitem{Kush_AC_97}
E.~Kushilevitz, ``Communication complexity,'' {\em Advances in Computers},
  vol.~44, pp.~331--360, 1997.

\bibitem{Newman_CC_04}
I.~Newman, ``Computing in fault tolerance broadcast networks,'' in {\em
  Proceedings of 19th IEEE Annual Conference on Computational Complexity},
  pp.~113--122, June 2004.

\bibitem{Kush_ACM_98}
E.~Kushilevitz and Y.~Mansour, ``Computation in noisy radio networks.,'' in
  {\em Proceedings of the 9th annual ACM-SIAM symposium on Discrete Algorithms,
  Society for Industrial and Applied Mathematics}, vol.~98, pp.~236--243, 1998.

\bibitem{Wang_ISIT_13}
C.-Y. Wang, S.-W. Jeon, and M.~Gastpar, ``Multi-round computation of
  type-threshold functions in collocated {G}aussian networks,'' in {\em
  Proceedings of 2013 IEEE International Symposium on Information Theory
  (ISIT)}, pp.~2154--2158, July 2013.

\bibitem{Ying_TIT_07}
L.~Ying, R.~Srikant, and G.~Dullerud, ``Distributed symmetric function
  computation in noisy wireless sensor networks,'' {\em IEEE Transactions on
  Information Theory}, vol.~53, pp.~4826--4833, Dec 2007.

\bibitem{Dutta_ACM_08}
C.~Dutta, Y.~Kanoria, D.~Manjunath, and J.~Radhakrishnan, ``A tight lower bound
  for parity in noisy communication networks,'' in {\em Proceedings of the 19th
  Annual ACM-SIAM Symposium on Discrete Algorithms}, SODA '08, (Philadelphia,
  PA, USA), pp.~1056--1065, Society for Industrial and Applied Mathematics,
  2008.

\bibitem{Kano_ISIT_07}
Y.~Kanoria and D.~Manjunath, ``On distributed computation in noisy random
  planar networks,'' in {\em Proceedings of the 2007 IEEE International
  Symposium on Information Theory (ISIT)}, pp.~626--630, 2007.

\bibitem{Kamath_TIT_14}
S.~Kamath, D.~Manjunath, and R.~Mazumdar, ``On distributed function computation
  in structure-free random wireless networks,'' {\em IEEE Transactions on
  Information Theory}, vol.~60, pp.~432--442, Jan 2014.

\bibitem{App_TIT_14}
R.~Appuswamy and M.~Franceschetti, ``Computing linear functions by linear
  coding over networks,'' {\em IEEE Transactions on Information Theory},
  vol.~60, pp.~422--431, Jan 2014.

\bibitem{Kow_TIT_12}
H.~Kowshik and P.~Kumar, ``Optimal function computation in directed and
  undirected graphs,'' {\em IEEE Transactions on Information Theory}, vol.~58,
  pp.~3407--3418, June 2012.

\bibitem{Jeon_ISIT_13}
S.-W. Jeon, C.-Y. Wang, and M.~Gastpar, ``Computation over gaussian networks
  with orthogonal components,'' {\em IEEE Transactions on Information Theory},
  vol.~60, pp.~7841--7861, Dec 2014.

\bibitem{Khu_TMC_08}
N.~Khude, A.~Kumar, and A.~Karnik, ``Time and energy complexity of distributed
  computation of a class of functions in wireless sensor networks,'' {\em IEEE
  Transactions on Mobile Computing}, vol.~7, pp.~617--632, May 2008.

\bibitem{Banerjee_ISIT_11}
S.~Banerjee, P.~Gupta, and S.~Shakkottai, ``Towards a queueing-based framework
  for in-network function computation,'' {\em Queueing Systems}, vol.~72,
  no.~3-4, pp.~219--250, 2012.

\bibitem{Bol_Spr_98}
B.~Bollob$\acute{a}$s, ``Random graphs,'' in {\em Modern Graph Theory},
  vol.~184 of {\em Graduate Texts in Mathematics}, pp.~215--252, Springer New
  York, 1998.

\bibitem{Pip_FOC_85}
N.~Pippenger, ``On networks of noisy gates,'' in {\em 26th IEEE Annual
  Symposium on Foundations of Computer Science}, pp.~30--38, IEEE, 1985.

\bibitem{Pip_TIT_91}
N.~Pippenger, G.~Stamoulis, and J.~Tsitsiklis, ``On a lower bound for the
  redundancy of reliable networks with noisy gates,'' {\em IEEE Transactions on
  Information Theory}, vol.~37, pp.~639--643, May 1991.

\bibitem{Lub_FOCS_02}
M.~Luby, ``L{T} codes,'' in {\em Proceedings of the 2002 Annual Symposium on
  Foundations of Computer Science (FOCS)}, pp.~271--280, 2002.

\bibitem{Maz_JSAC_14}
A.~Mazumdar, V.~Chandar, and G.~Wornell, ``Update-efficiency and local
  repairability limits for capacity approaching codes,'' {\em IEEE Journal on
  Selected Areas in Communications}, vol.~32, pp.~976--988, May 2014.

\bibitem{Alf_JSAC_09}
G.~Alfano, M.~Garetto, and E.~Leonardi, ``Capacity scaling of wireless networks
  with inhomogeneous node density: upper bounds,'' {\em IEEE Journal on
  Selected Areas in Communications}, vol.~27, pp.~1147--1157, September 2009.

\bibitem{Alf_ACM_10}
G.~Alfano, M.~Garetto, E.~Leonardi, and V.~Martina, ``Capacity scaling of
  wireless networks with inhomogeneous node density: Lower bounds,'' {\em
  IEEE/ACM Transactions on Networking}, vol.~18, pp.~1624--1636, Oct. 2010.

\bibitem{Dim_TON_06}
A.~G. Dimakis, V.~Prabhakaran, and K.~Ramchandran, ``Decentralized erasure
  codes for distributed networked storage,'' {\em IEEE/ACM Transactions on
  Networking}, vol.~14, pp.~2809--2816, June 2006.

\bibitem{abadi2012capacity}
H.~K. Abadi, P.~Pad, H.~Saeedi, F.~Marvasti, and K.~Alishahi, ``Capacity
  achieving linear codes with random binary sparse generating matrices over the
  binary symmetric channel,'' in {\em 2012 IEEE International Symposium on
  Information Theory}, pp.~621--625, IEEE, 2012.

\bibitem{Yang_All_14}
Y.~Yang, P.~Grover, and S.~Kar, ``Can a noisy encoder be used to communicate
  reliably?,'' in {\em Proceedings of the 52nd Allerton Conference on Control,
  Communication and Computing}, pp.~659--666, Sept 2014.

\bibitem{ahlswede2000network}
R.~Ahlswede, N.~Cai, S.-Y. Li, and R.~W. Yeung, ``Network information flow,''
  {\em IEEE Transactions on information theory}, vol.~46, no.~4,
  pp.~1204--1216, 2000.

\bibitem{koetter2008coding}
R.~Koetter and F.~R. Kschischang, ``Coding for errors and erasures in random
  network coding,'' {\em IEEE Transactions on Information Theory}, vol.~54,
  no.~8, pp.~3579--3591, 2008.

\bibitem{lim2011noisy}
S.~H. Lim, Y.-H. Kim, A.~El~Gamal, and S.-Y. Chung, ``Noisy network coding,''
  {\em IEEE Transactions on Information Theory}, vol.~57, no.~5,
  pp.~3132--3152, 2011.

\bibitem{jaggi2008resilient}
S.~Jaggi, M.~Langberg, S.~Katti, T.~Ho, D.~Katabi, M.~M{\'e}dard, and
  M.~Effros, ``Resilient network coding in the presence of byzantine
  adversaries,'' {\em IEEE Transactions on Information Theory}, vol.~54, no.~6,
  pp.~2596--2603, 2008.

\bibitem{silva2008rank}
D.~Silva, F.~R. Kschischang, and R.~Koetter, ``A rank-metric approach to error
  control in random network coding,'' {\em IEEE transactions on information
  theory}, vol.~54, no.~9, pp.~3951--3967, 2008.

\bibitem{Gal_Wil_68}
R.~G. Gallager, {\em Information theory and reliable communication}.
\newblock John Wiley \& Sons, 1968.

\bibitem{Wald_AMS_48}
A.~Wald and J.~Wolfowitz, ``Optimum character of the sequential probability
  ratio test,'' {\em The Annals of Mathematical Statistics}, vol.~19, no.~3,
  pp.~pp. 326--339, 1948.

\bibitem{Gup_TIT_00}
P.~Gupta and P.~Kumar, ``The capacity of wireless networks,'' {\em IEEE
  Transactions on Information Theory}, vol.~46, pp.~388--404, Mar 2000.

\bibitem{Cor_MIT_90}
T.~H. Cormen, {\em Introduction to algorithms}.
\newblock MIT press, 2009.

\bibitem{Cover_Wiley_06}
T.~M. Cover and J.~A. Thomas, {\em Elements of Information Theory, 2nd
  Edition}.
\newblock John Wiley \& Sons, 2006.

\bibitem{Che_AMS_52}
H.~Chernoff, ``A measure of asymptotic efficiency for tests of a hypothesis
  based on the sum of observations,'' {\em The Annals of Mathematical
  Statistics}, pp.~493--507, 1952.

\bibitem{7541796}
Y.~Yang, S.~Kar, and P.~Grover, ``Energy efficient distributed coding for data
  collection in a noisy sparse network,'' in {\em 2016 IEEE International
  Symposium on Information Theory (ISIT)}, pp.~2734--2738, July 2016.

\end{thebibliography}

\end{document}